\documentclass{aa}  
\usepackage{natbib}
\usepackage{float}
\usepackage{placeins}
\bibpunct{(}{)}{;}{a}{}{,}

\usepackage{graphicx}
\graphicspath{{./}{figure/}}
\usepackage{txfonts}
\usepackage{hyperref}
\begin{document}

   \title{J-PLUS: Support vector machine applied to STAR-GALAXY-QSO classification}

   \author{Cunshi Wang \inst{1,2} 
   \and Yu Bai \inst{1}
   \and C.~L\'opez-Sanjuan\inst{\ref{CEFCA}}
   \and Haibo Yuan \inst{3}
   \and Song Wang \inst{1} 
   \and Jifeng Liu \inst{1,2} 
   \and David Sobral \inst{4} 
   \and P. O. Baqui \inst{5} 
   \and E.\ L.\ Mart\'in \inst{\ref{IAC},\ref{ULL},\ref{CSIC}}
   \and Carlos Andres Galarza \inst{\ref{ON}}
\and J.~Alcaniz\inst{\ref{ON}}
\and R.~E.~Angulo\inst{\ref{DIPC},\ref{ikerbasque}}
\and A.~J.~Cenarro\inst{\ref{CEFCA}}
\and D.~Crist\'obal-Hornillos\inst{\ref{CEFCA}}
\and R.~A.~Dupke\inst{\ref{ON},\ref{MU},\ref{Alabama}}
\and A.~Ederoclite\inst{\ref{USP}}
\and C.~Hern\'andez-Monteagudo\inst{\ref{IAC},\ref{ULL}}
\and A.~Mar\'{\i}n-Franch\inst{\ref{CEFCA}}
\and M.~Moles\inst{\ref{CEFCA}}
\and L.~Sodr\'e Jr.\inst{\ref{USP}}
\and H.~V\'azquez Rami\'o\inst{\ref{CEFCA}}
\and J.~Varela\inst{\ref{CEFCA}}
}

   \institute{Key Laboratory of Optical Astronomy, National Astronomical Observatories,               Chinese Academy of Sciences, \\
            20A Datun Road, Chaoyang District, Beijing 100012,People's Republic of China
            \and
             College of Astronomy and Space Sciences, University of Chinese Academy of Sciences, Beijing 100049, China
             \and 
             Department of Astronomy, Beijing Normal University, Beijing 100875, People's Republic of China
             \and
             Department of Physics, Lancaster University, Lancaster LA1 4YB, UK
             \and
             PPGFis \& Núcleo de Astrofísica e Cosmologia (Cosmo-ufes), Universidade Federal do Espírito Santo, 29075-910 Vitória, ES, Brazil
             \and
             Departamento de Astrof\'isica, Universidad de La Laguna (ULL), 
             E-38206 La Laguna, Tenerife, Spain \label{ULL}
             \and
        Consejo Superior de Investigaciones Cient\'ificas (CSIC), E-28006 
Madrid, Spain \label{CSIC}
             \and
             Consejo Superior de Investigaciones Cient\'ificas (CSIC), E-28006 
             Madrid, Spain Centro de Estudios de F\'{\i}sica del Cosmos de Arag\'on (CEFCA), Unidad Asociada al CSIC, Plaza San Juan 1, 44001 Teruel, Spain\label{CEFCA}
        \and
         Instituto de Astronomia, Geof\'{\i}sica e Ci\^encias 
Atmosf\'ericas, Universidade de S\~ao Paulo, 05508-090 S\~ao Paulo, 
Brazil\label{USP}
         \and
         Observat\'orio Nacional - MCTI (ON), Rua Gal. Jos\'e Cristino 
77, S\~ao Crist\'ov\~ao, 20921-400 Rio de Janeiro, Brazil\label{ON}
         \and
         Donostia International Physics Centre (DIPC), Paseo Manuel de 
Lardizabal 4, 20018 Donostia-San Sebastián, Spain\label{DIPC}
            \and
         IKERBASQUE, Basque Foundation for Science, 48013, Bilbao, 
Spain\label{ikerbasque}
         \and
         University of Michigan, Department of Astronomy, 1085 South 
University Ave., Ann Arbor, MI 48109, USA\label{MU}
         \and
         University of Alabama, Department of Physics and Astronomy, 
Gallalee Hall, Tuscaloosa, AL 35401, USA\label{Alabama}
         \and
         Instituto de Astrof\'{\i}sica de Canarias, La Laguna, 38205, 
Tenerife, Spain\label{IAC}
             }

 
  \abstract
   {In modern astronomy, machine learning has proved to be efficient and effective in mining big data from the newest telescopes.}
   {In this study, we construct a supervised machine-learning algorithm to classify the objects in the Javalambre Photometric Local Universe Survey first data release (J-PLUS DR1).}
   {The sample set is featured with 12-waveband photometry and labeled with spectrum-based catalogs, including Sloan Digital Sky Survey (SDSS) spectroscopic data, the Large Sky Area Multi-Object Fiber Spectroscopic Telescope (LAMOST), and VERONCAT - the Veron Catalog of Quasars \& AGN (VV13). The performance of the classifier is presented with the applications of blind test validations based on RAdial Velocity Extension (RAVE), the Kepler Input Catalog (KIC), the 2 MASS (the Two Micron All Sky Survey) Redshift Survey (2MRS), and the UV-bright Quasar Survey (UVQS). A new algorithm was applied to constrain the potential extrapolation that could decrease the performance of the machine-learning classifier.}
   {The accuracies of the classifier are 96.5\% in the blind test and 97.0\% in training cross-validation. The $F_1$-scores for each class are presented to show the balance between the precision and the recall of the classifier. We also discuss different methods to constrain the potential extrapolation.}
   {}

   \maketitle

%

\section{Introduction} \label{intro}

Developments in computer science and the technological applications have changed the ways of data processing and knowledge management. Especially, as a growing realm of technology, machine learning has gained worldwide popularity due to its powerful ability to manage large amounts of data. Machine-learning algorithms can reveal potential patterns and physical meanings that are otherwise indistinguishable by traditional methods. Furthermore, machine learning enables us to construct the structure of each observed quantity and to reveal its manner of working.

In modern astronomy, the newest telescopes now produce large amounts of unprocessed data. The Javalambre Photometric Local Universe Survey (J-PLUS, \citealt{jplus2019}) is designed to observe several thousand square degrees in the optical bands. It has been designed to observe more than 13 million objects with the Javalambre Auxiliary Survey Telescope (JAST80) at the Sierra de Javalambre in Spain, and to enhance knowledge from the Solar System to cosmology \footnote{\url{http://j-plus.es/survey/science}}, such as the Coma cluster \citep{jim19}, low-metallicity stars \citep{whit19}, and galaxy formation \citep{nog19}. 

The current classification of sources detected by J-PLUS is morphological, this is to say it aims to distinguish between point-like and extended sources \citep{jplus2019,lop19}. It is therefore not able to differentiate stars from quasi-stellar objects (QSOs), and it does not include valuable color information from the 12 optical J-PLUS bands. This paper presents the spectrum-based classification for the J-PLUS first data release (DR1) with machine-learning algorithms. The input catalog is a modified version of the J-PLUS data set which has been recalibrated by \citet{yuanp}. This version includes 13,265,168 objects with magnitudes obtained with 12 different filters (Sect. \ref{jplus}). From Sect. \ref{sdss} to \ref{vv10}, we label the data set as STAR, GALAXY, and QSO based on the spectroscopy surveys, including the Sloan Digital Sky Survey (SDSS), the Large Sky Area Multi-Object Fiber Spectroscopy Telescope (LAMOST), and VERONCAT - the Veron Catalog of Quasars \& AGN (VV13).

Several machine-learning algorithms have been applied for the classification (Sect. \ref{method}), including the Support Vector Machine (SVM,\citealt{cortes95}), linear discrimination, the $k$-nearest neighbor ($k-$NN,\citealt{knn67,knn77}), Bayesian, and decision trees \citep{quin86}. In the pretraining, we adopted the algorithm with the highest accuracy (Sect. \ref{pretrain}). In Sect. \ref{svm}, we present the processes to test the parameters of the algorithms and to train the classifier. We also provide the blind test and a new method to constrain potential extrapolation (Sect. \ref{validation}) in our prediction.

We present our result in Sect. \ref{result}, including our result catalogs (Sect. \ref{catalog}), considerations about ambiguous objects from the classification probabilities (Sect. \ref{abmig}), and a comparison between the J-PLUS parameter (Sect. \ref{classstar}). In Sect. \ref{discuss}, we discuss different methods to constrain the extrapolation. The classifier is compared with other published classifiers, and the difference is analyzed in detail (Sect. \ref{cdm}). Section \ref{outlook} gives an outlook of the Javalambre Physics of the Accelerating Universe Astrophysical Survey (J-PAS, \citealt{jpas,minijpas}) and our future work.

\section{Data} 
\label{data}
The rapid advance in telescopes and detectors has led to a significant data explosion in modern astronomy. New technologies help us accelerate information acquisition from the huge datasets. Several studies have focused on developing classifiers, and they have proved that  spectral-based methods are more reliable than those only based on photometric data \citep{bai19, ball06}.

\begin{table*}
\centering
\caption{Constitution of a sample set \label{sampleset}}
\begin{tabular}{lcccc}
\hline\hline \noalign{\smallskip}
Catalog & STAR &  GALAXY & QSO  & Total \\ 
\noalign{\smallskip}
  \hline
  \noalign{\smallskip}
  SDSS DR16     & 45,350 & 68,381 & 44,745 & 158,476 \\
  SDSS APOGEE & 13,749 & 0 & 0 & 13,749 \\
  LAMOST DR7 & 299,907 & 16,004 & 4,758 & 345,975  \\
  LAMOST A-, F-, G- and K- type stars & 212,114 & 0 & 0 & 212,114 \\
  LAMOST A- stars & 5,145 & 0 & 0 & 5,145\\
  LAMOST M- stars & 25,604 & 0 & 0 & 25,604 \\
  VV13 & 0 & 0 & 4,744 & 4,744 \\
  \hline
  \end{tabular}
  \tablefoot{The numbers reveal how many objects there are that correspond to each catalog and class, after crossing with the J-PLUS catalog. The sample contains 468,685 objects with a full 12 magnitudes from the catalogs, with 348,085 STAR, 74,701 GALAXY, and 45,899 QSO objects. There are repeated objects in different catalogs, which cause the inequality of the sum. This table presents all objects for training and testing. See \ref{BT} for a blind test.} 
\end{table*}

\subsection{J-PLUS}
\label{jplus}
J-PLUS\footnote{\url{www.j-plus.es}} is being conducted from the Observatorio Astrof\'{\i}sico de Javalambre (OAJ, Teruel, Spain; \citealt{oaj}) using the 83\,cm JAST80 and T80Cam, a panoramic camera of 9.2k $\times$ 9.2k pixels that provides a $2\deg^2$ field of view (FoV) with a pixel scale of 0.55 arcsec pix$^{-1}$ \citep{t80cam}. The J-PLUS filter system is composed of 12 passbands, including five broad and seven medium bands from 3000 to 9000 \AA. The J-PLUS observational strategy, image reduction, and main scientific goals are presented in \citet{jplus2019}. J-PLUS DR1 covers a sky area of 1,022 $\rm deg^2$, and the limiting magnitudes are in the range $21-22$. For different kinds of objects, the magnitudes of these $12$ bands exhibit different distributions \footnote{\url{http://j-plus.es/datareleases/data_release_dr1}}, and such a difference gives us a theoretical foundation for object classification.

Compared to other catalogs, the J-PLUS catalog is an ideal data set for classification owing to its characteristic of both large amounts and multiple wavebands. Multiple bands could provide more information for a single object. In machine learning, these 12-band magnitudes lead to a more expanding training instance space and a smoother training structure. We adopted the $12$ band magnitudes as training features, which are $u$, $J0378$, $J0395$, $J0410$, $J0430$, $g$, $J0515$, $r$, $J0660$, $i$, $J0861$, and $z$. We name them mag1 through to mag12.

Recently, \citet{yuanp} recalibrated the J-PLUS catalog and increased the accuracy of photometric calibration by using the method of stellar color regression (SCR), similar to the method in \citet{yuan15}. The catalog in \citet{yuanp} contains 13,265,168 objects, including 4,126,928 objects with all 12 valid magnitudes. 

\subsection{SDSS}
\label{sdss}
The observation of SDSS has covered one-third of the sky and yielded more than 3 million spectra. We explore the spectroscopy survey sets in data release 16 (DR16; \citealt{sdssdr16}). With the help of SDSS Catalog Archive Server Jobs\footnote{\url{http://skyserver.sdss.org/casjobs/}}, the objects with $zWarning=0$ were chosen to label the J-PLUS data as “STAR”, “GALAXY”, and “QSO”.

The Apache Point Observatory Galactic Evolution Experiment (APOGEE) has observed more than 100,000 stars in the Milky Way, with reliable spectral information including stellar parameters and radial velocities \citep{zas13}. We adopted the APOGEE catalog to enlarge the training set.

We cross-matched the J-PLUS catalog with SDSS DR16 using Tool for OPerations on Catalogues And Tables (Topcat, \citealt{topcat}) \footnote{\url{http://www.star.bris.ac.uk/~mbt/topcat/}} with a tolerance of one arcsec, and we obtained 45,350 stars, 68,381 galaxies, and 44,745 QSOs from the general catalog, as well as 13,749 stars from APOGEE. After cross-matching with other catalogs, APOGEE contributes 6,147 independent stars.

\subsection{LAMOST}
\label{lamost}
LAMOST \citep{cui12, luo12, zhao12, wang96, su04} is located at the Xinglong Observatory in China, which is able to observe 4,000 objects in $20\deg^2$ simultaneously. LAMOST has many scientific projects, and two of them aim to understand the structure of the  Milky Way \citep{deng12} and external galaxies. The low-resolution spectra of LAMOST have a limiting magnitude of about $20$mag in the $g$ band for a resolution R=500. Data release 7 (DR7) was adopted to label the sample. We also adopted information from stellar catalogs from DR7, including the A-, F-, G-, and K-type star catalog, as well as the A- and M-star catalogs.

The A-, F-, G-, and K-type star catalog has stars with a $g$ band signal-to-noise ratio higher than 6 in dark nights or 15 in bright nights. The A- and M-star catalogs contain all A and M stars from the pilot and general surveys. For overlapping stars, we followed the priority of the star catalogs and the general catalog.

In the LAMOST DR7 catalog, the cross-match yields 299,907 stars, 16,004 galaxies, and 4,758 QSOs. There are 212,114 matched stars in the A-, F-, G-, and K-type star catalog and 5,145 and 25,604 stars in the A- and M-star catalogs, respectively. Nearly all of the stars (except only one star) from the star catalogs are covered in the DR7 general catalog.

\begin{figure}
\includegraphics[width=0.45\textwidth]{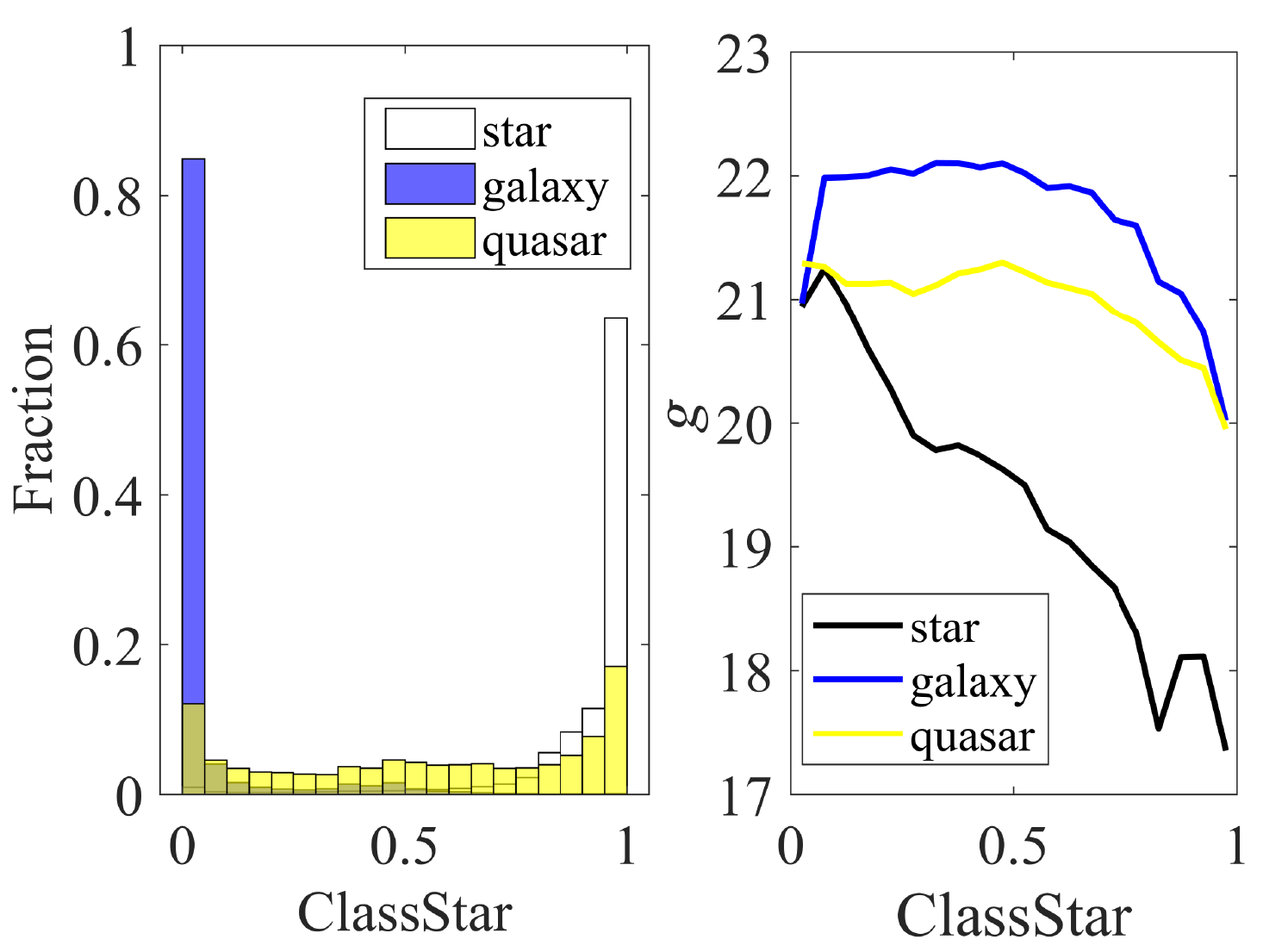}
\caption{Comparison between the class in the sample set and the J-PLUS "\texttt{CLASS\_STAR}" parameter. The panel on the left-hand side shows the normalized distributions of \texttt{CLASS\_STAR}. The panel on the right-hand side shows the relation between the average magnitudes in the \textit{g} band corresponding to each bin (the left panel) of \texttt{CLASS\_STAR}. The white box and black line denote denotes the stellar objects, blue stands for the galaxies, and yellow is for the QSOs. \label{jplusclass}}
\end{figure}

\subsection{QSO catalog}
\label{vv10}
Quasars in VV13 (VERONCAT - Veron Catalog of Quasars \& AGN, the 13th edition) were also employed to enlarge our QSO samples. The catalog contains AGN objects with spectroscopic parameters (including redshift; \citealt{vv13}). The VV13 contains 4,744 QSOs after a one arcsec tolerance cross identification with J-PLUS, 4,593 QSOs are included in SDSS DR16, and 1,339 QSOs are in LAMOST DR7. The VV13 catalog provides 108 additional QSOs.

\subsection{Sample construction} 
The machine-learning sample is made up of SDSS, LAMOST, and VV13 (Table \ref{sampleset}, see more in Appendix \ref{appc}, and magnitude distributions are in Appendix \ref{appb}). There are 468,685 unique objects with 12 valid magnitudes, including 74,701 galaxies, 45,899 QSOs, and 348,085 stars. These 468,685 objects were all put in training with a 10-fold validation. The blind test set was carried out with 2,853 objects in other catalogs, see \ref{BT}. 

J-PLUS DR1 contains the stellar probability \texttt{CLASS\_STAR}, estimated by \texttt{SExtractor} \citep{sextractor} with an artificial neural network (ANN). We present the comparison between the probability and the classification of the sample in Fig. \ref{jplusclass}. In our sample set, about 20\% of the QSOs have a stellar probability of more than 95\%, and more than 10\% of the QSOs have a stellar probability of less than 5\%. In the right panel, the \texttt{CLASS\_STAR} roughly increases as the g-band magnitude becomes dimmer, because the stars in the sample set are brighter than galaxies (see Appendix \ref{appb}). The magnitude - \texttt{CLASS\_STAR} relation is not significant for quasars.

\section{Methodology}
\label{method}
Machine learning has developed many algorithms that are able to deal with big data effectively. Three of them, that is to say decision trees, SVM, and $k$-NN, are the most popular ones.

\subsection{Pretraining}
\label{pretrain}
A pretraining process with 10-fold validation was adopted in order to determine which algorithm fits our problem best. The No-Free-Lunch theorem \citep{shai14} tells us that a perfect learning algorithm that can fit every problem does not exist. In the pretraining, we considered the accuracy to be the most important factor of the training performance. The accuracies of the pretraining are shown in Table \ref{TFM}. 

In the $k$-NN algorithm, the label of each data point is defined by its neighborhood. By introducing a metric function, the algorithm can calculate the distance between every two objects. For each object, the nearest $k$-objects are determined, and its label is defined. This process continues until the labels of all objects are stable. The $k$-NN gives a reasonable result for a nonlinear or discrete training set, and it has good performance when extrapolating a prediction and separating for outliers. However, the $k$-NN algorithm cannot present reliable results for unbalance data that are dominated by objects in one or two classes \citep{shai14}. This is one reason why we precluded the algorithm. In our test, we adopted a 10-NN algorithm with a Euclid norm, and no hyperparameters or weights were involved.

Decision tree is a nonparametric supervised learning method. The tree in the algorithm is built by the threshold calculated from the sample. For each node of a tree, a gain function defines the loss of the prediction \citep{quin86}. If the loss function is low, the node is split. This procedure continues until all objects in the training set are labeled. The time cost of decision tree is low \citep{shai14}, but the gain function may lead to a bias or overfitting for the unbalanced data set. Random forest (RF, \citealt{rf01}) and bagging tree are enhanced decision tree algorithms that can decrease overfitting.

In our work, we tested three tree algorithms without hyperparameters. The decision tree algorithm is based on the Gini index \citep{quin86}, and the maximum split is 100. The RF (\citealt{rf01}) algorithm also contains 30 learners and maximum splits to 468,684. The AdaBoost algorithm \citep{ada} contains 30 learners with a learning rate of 0.1, and it has a maximum split of 20.

For each model, we examined the accuracy and training time (Table \ref{TFM}), and the SVM algorithm provides the highest validation accuracy. Since the model accuracy is the primary factor in our consideration, we decided to adopt the SVM algorithm even if it needs a relatively long training time. The training time becomes significant in other situations, such as transient detection.

\begin{table}
\caption{Accuracy and time cost for algorithms\label{TFM}}
\centering
\begin{tabular}{lcc}
\hline \hline
\noalign{\smallskip}
 Algorithm & Accuracy & Time Cost \\
 \noalign{\smallskip}
 \hline
 \noalign{\smallskip}
  Decision Tree  & $92.6\%$ & $96s$ \\
  Linear Discrimination  & $86.9\%$ & $26s$ \\
  Bayesian & $74.3\%$ & $10s$ \\
  SVM & $96.4\%$ & $90m$ \\
  $k-$NN & $95.7\%$ & $23m$ \\
  AdaBoost & $92.0\%$ & $3m$ \\
  Random Forest & $96.2\%$ & $7m$\\
  \hline
\end{tabular}
  \tablefoot{The last column is the rough training time cost for the training sample. The "s" stands for second and "m" is for minute. See \citet{fisher} for details about the linear discrimination algorithm.}
\end{table}

\subsection{SVM}
\label{svm}
SVM is a binary classification method (see \citet{cortes95} and \citet{svm1992} for details). The theory of SVM is presented in \citet{crist00} and \citet{shai14}.

In brief, the SVM algorithm generates a super surface in the instance space by maximizing the margin. The margin is defined by the smallest distance between the object and the super surface. Given a super surface, the algorithm divides the instance space into two parts and labels the object in each part. The algorithm then compares each label with the sample and calculates the loss function. The margin is maximized when the loss function reaches its minimum. For our classification problem, there are 12 dimensions in the instance space. 

\begin{figure}
\includegraphics[width=0.45\textwidth]{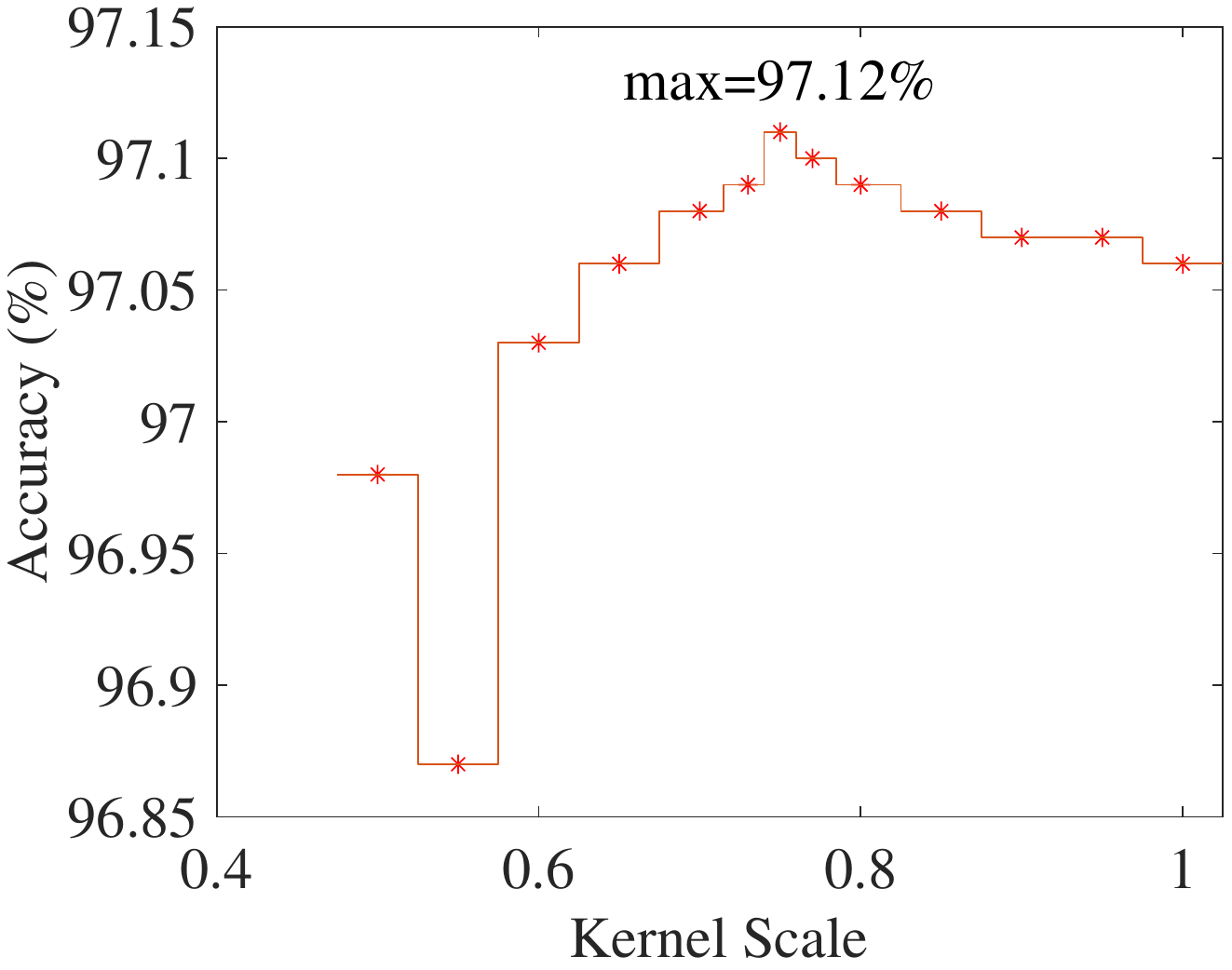}
\caption{Different kernel scales and their corresponding accuracies. The maximum is at 0.75.\label{kernelscale}}
\end{figure}

SVM is a binary classification algorithm, while we are facing a multi-classification algorithm. The coding method can change a multi-classification problem into several binary classifications, such as one-versus-one coding and one-versus-all coding. For a $k$-classification problem, one-versus-one coding finds all binary combinations of the labels. After making a democratic decision, the algorithm produces the predicted label. One-versus-one coding needs $\frac{k(k-1)}{2} $ binary classifications to reach the aim. One-versus-all coding singly picks one label out and defines it as a positive class, and the rest $(k-1)$ of the labels are negative. After $k$ times binary classifications, the one-versuss-all coding presents the labels by democratic decision. One-versus-one coding has a higher accuracy in our classification.

The Gaussian kernel, also known as the radial basis function (RBF) kernel, is an important parameter in the SVM algorithm construction. It can accelerate the optimization of the margin in the SVM algorithm. In the Gaussian kernel, a kernel scale is an adjustable parameter that measures the distance to the half-space. A small kernel scale constrains the kernel function in low variation, and further parameterizes the margin exquisitely. The farther the data points are located from the margin, the less they weigh. In order to find the best kernel scale, we tested the scale from 0.5 to 1, with a step size of 0.05. For each kernel scale, we trained a classifier and calculated its accuracy. Finally, we conclude that 0.75 is the best kernel scale (Fig. \ref{kernelscale}).

\begin{figure}
\includegraphics[width=0.5\textwidth]{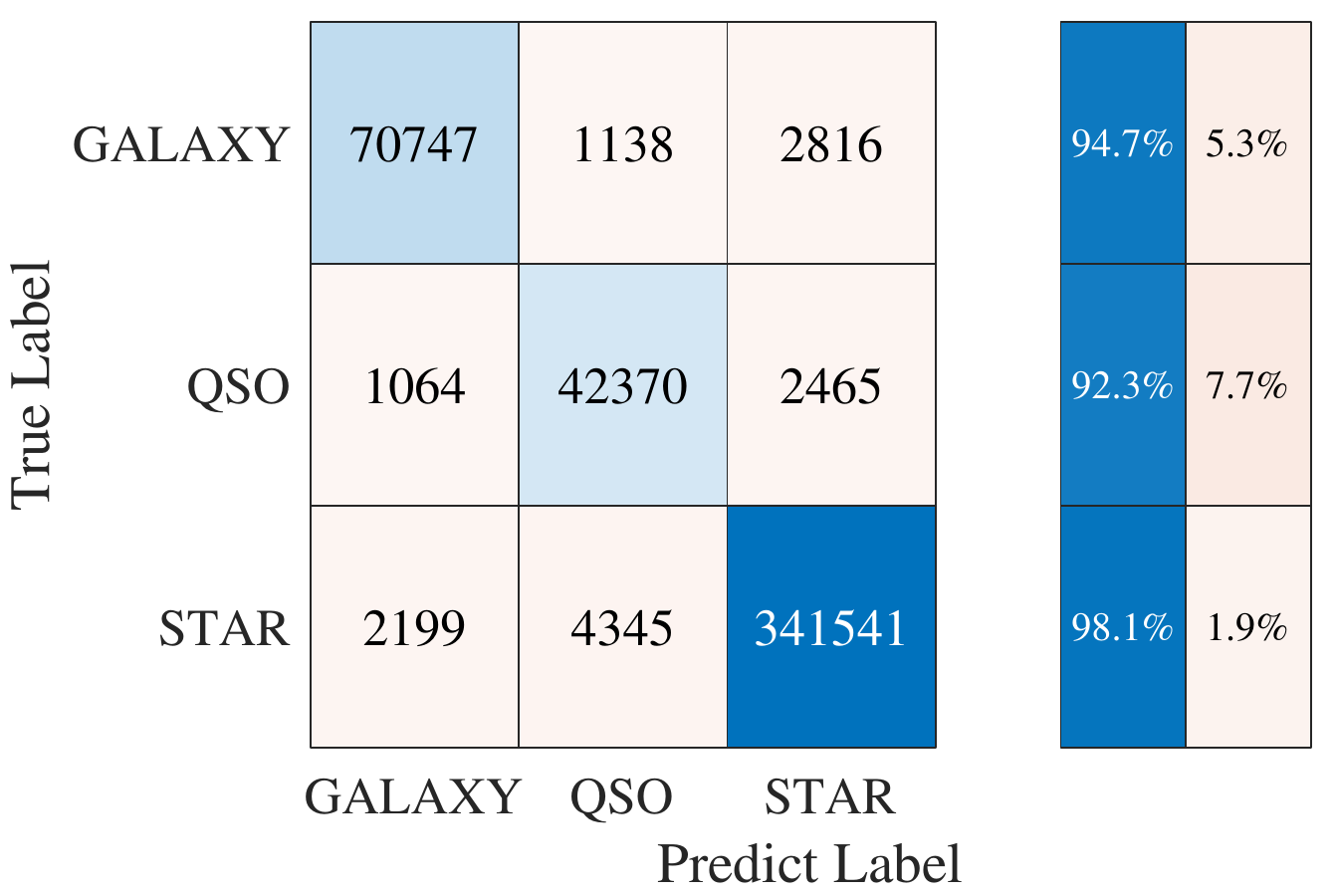}
\caption{Training confusion matrix. The blue rectangles show the correct labels, while the pink ones represent the error labels. \label{FCF}}
\end{figure}

The magnitude uncertainties in J-PLUS DR1 \citep{yuanp} depend on the observing condition and the photometric calibration. In our training process, we employed uncertainties  as the training weight to describe the reliability of the data. 

The confusion matrix is shown in Fig. \ref{FCF}. The total cross-validation accuracy is $97\%$. The low accuracy of QSO may be due to its relatively small sample size.

\subsection{Validation}
\label{validation}
Model validation has been designed to show the effectiveness and to avoid potential overfitting. 
Extrapolation is significant in model validation. It has been proved that the prediction accuracy might decrease when extrapolating outside the feature space region of training samples \citep{wang20}.
The other validating procedure is a blind test, which can reveal the potential overfitting of the classifier. Moreover, the appropriate training data size would be implied by comparing the training and blind test accuracy.

\subsubsection{Extrapolation}
\label{outlier}

Applying any extrapolation may cause low accuracy due to the nonrepresentativeness between training data and predicting data \citep{wang20}. Here, we use the density contour of the training sample to define the potential extrapolation. A dozen three-dimensional density contour surfaces were generated based on the distribution of training data. These surfaces were used as the boundary of the potential extrapolation. The magnitude combinations are (mag1, mag2, mag3), (mag2, mag3, mag4), … , and (mag12, mag1, mag2), and an example is shown in Fig. \ref{mag13}.  We present all the contour surfaces in Appendix \ref{app}. We then define the potential extrapolation with these 12 contour surfaces for the prediction. There are 3,496,867 ($84.73\%$) objects of J-PLUS DR1 located inside these contours.

\begin{figure}
\includegraphics[width=0.45\textwidth]{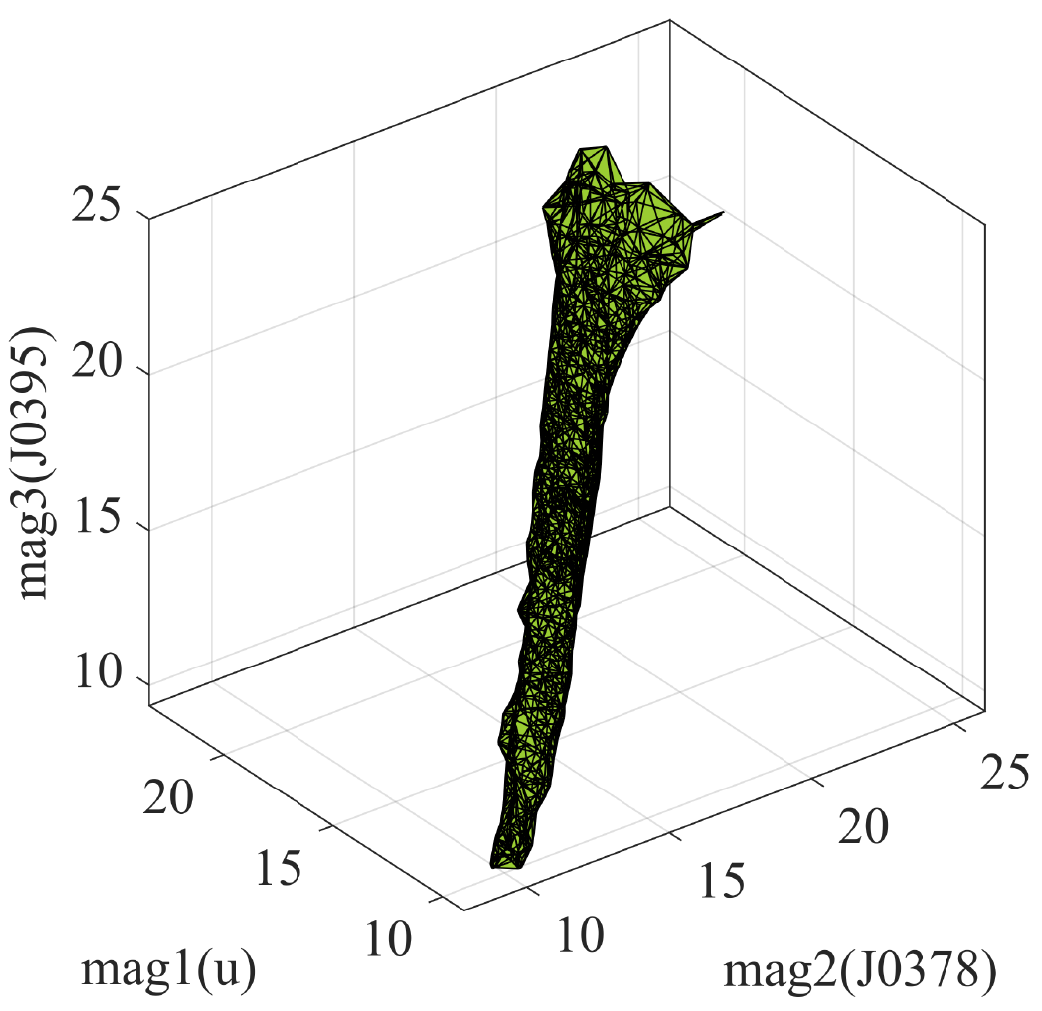}
\caption{Density contour of the first three magnitudes of the training data set. The contour stands for the three-dimensional density of 5\% of the training data.  \label{mag13}}
\end{figure}

\subsubsection{Blind test}
\label{BT}

\begin{figure}
\includegraphics[width=0.45\textwidth]{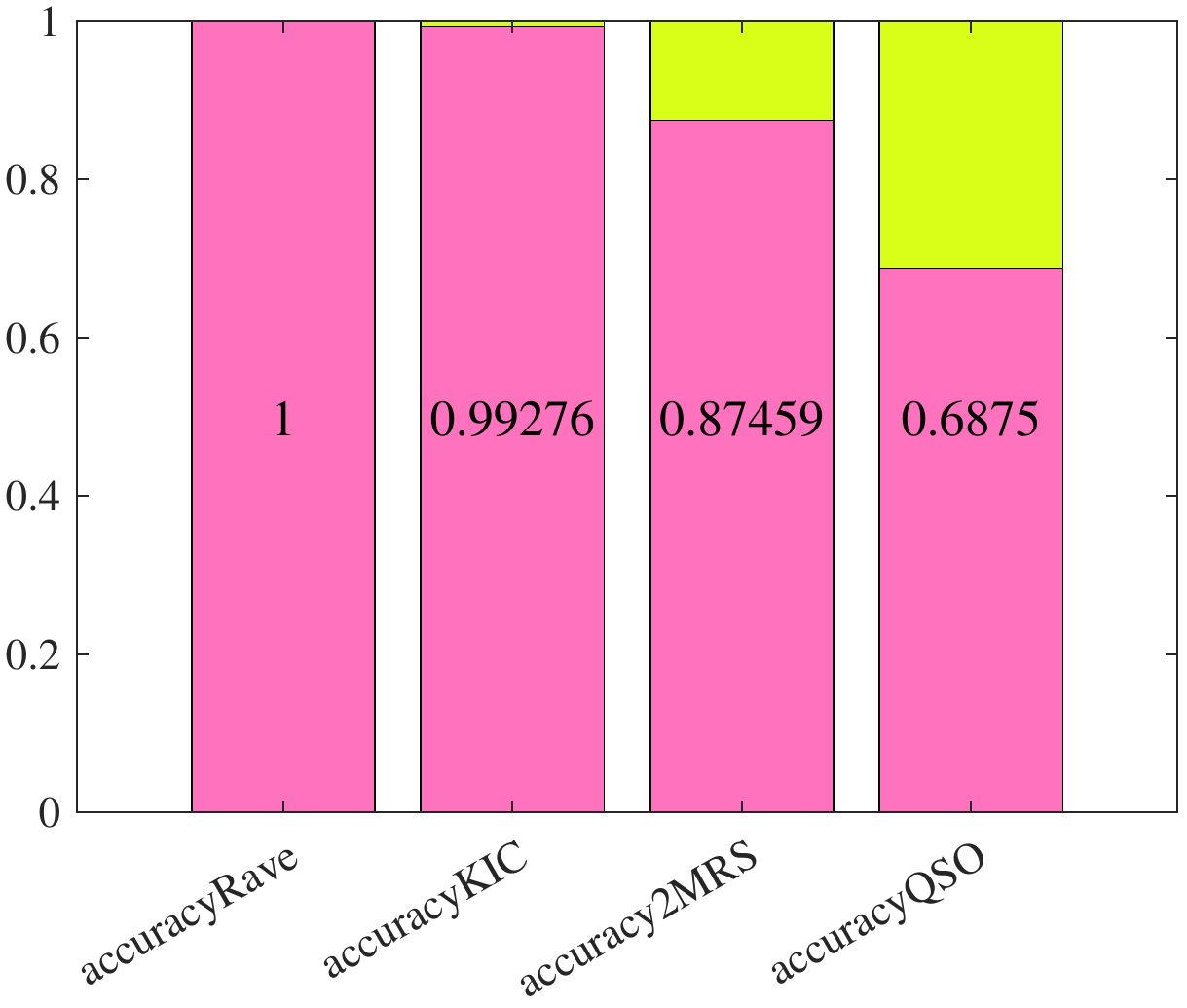}
\caption{Accuracy distribution for different interpolating data blind  sets. The red bars show the correct objects, while the yellow bars show the incorrect ones. The numbers are the accuracies. \label{blindtestplot}}
\end{figure}

\begin{figure}
\includegraphics[width=0.5\textwidth]{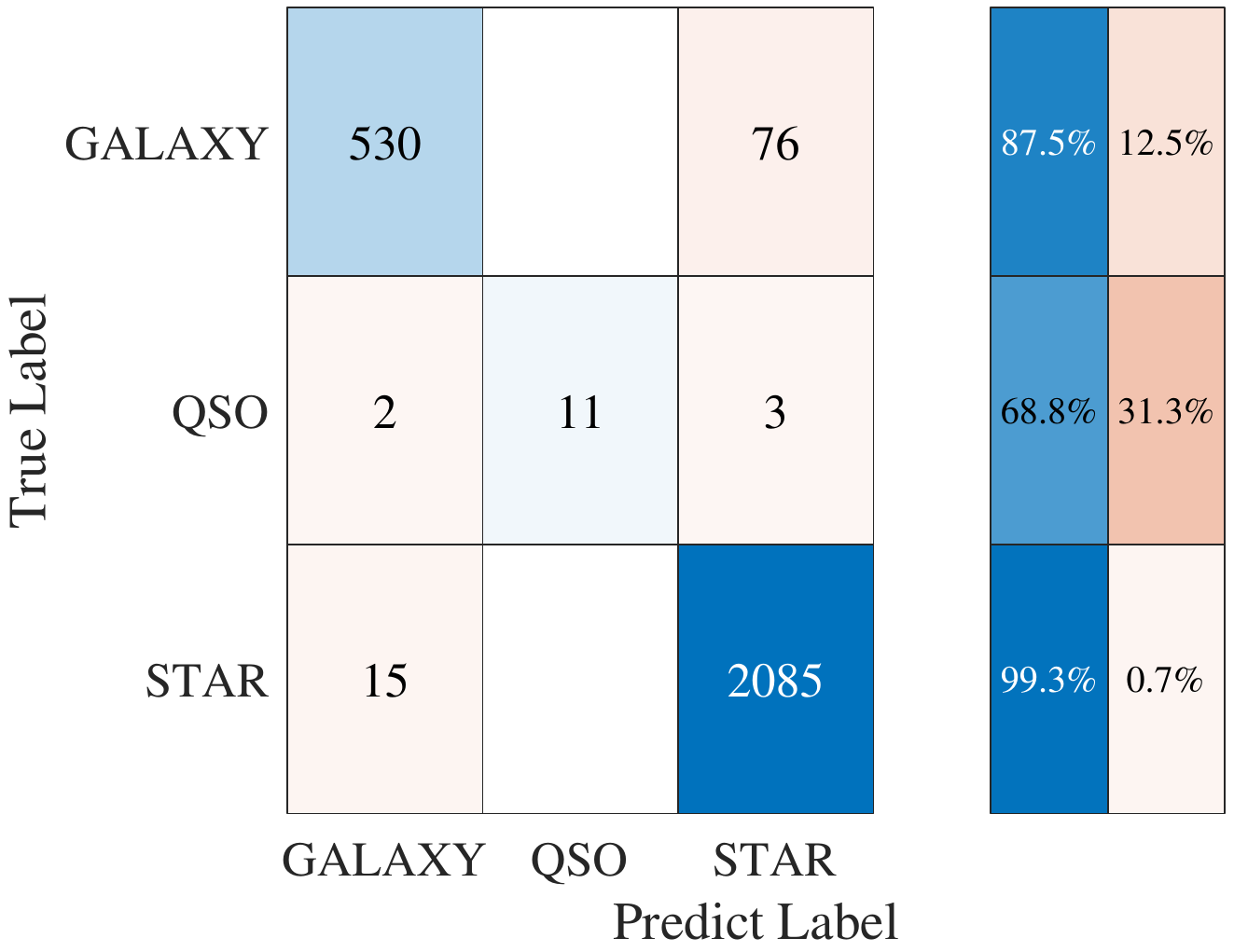}
\caption{Confusion matrix for interpolating the blind test, and the accuracy is 96.5\%. The colors are the same as in Fig. \ref{FCF}. \label{blindtestconf}}
\end{figure}

We applied a blind test to reveal the classifier's validation and its potential overfitting of the training data. The blind test data set (Table \ref{btset}) was built by stars from the RAdial Velocity Experiment (RAVE) and Kepler Input Catalog (KIC), galaxies from the 2 MASS (Two Micron All Sky Survey) Redshift Survey (2MRS) and QSOs from the UV-bright Quasar Survey (UVQS). The accuracy distribution and the confusion matrix of the blind test are shown in Fig. \ref{blindtestplot} and \ref{blindtestconf}. 

\begin{table}
\caption{Constitution of a blind test set \label{btset}}
\centering
\begin{tabular}{lcccc}
\hline \hline \noalign{\smallskip}
  Catalog & Interpolation &  Extrapolation & Total \\
  \noalign{\smallskip}
  \hline
  \noalign{\smallskip}
  RAVE  & 29 & 3 & 32 \\
  KIC & 2,071 & 64 & 2,135 \\
  2MRS & 606 & 46 & 652  \\
  UVQS & 16 & 18 & 34 \\
  \hline
  \noalign{\smallskip}
  Total & 2,722 & 131 & 2,853\\
  \hline
\end{tabular}
  \tablefoot{Every catalog is crossed with J-PLUS and all 12 magnitudes are available. The extrapolation stands for the objects suffering from potential extrapolating, and the others are interpolations.}
\end{table}

RAVE is a stellar survey that focuses on obtaining stellar radial velocities \citep{rave2020}. It provides precise spectroscopic parameters of stars. We obtained only 70 stars by cross-matching with J-PLUS with a one arcsec tolerance after removing the stars in the sample set. There are three stars suffering from potential extrapolating. The number of stars is too small to validate our algorithm, so the KIC catalog was adopted to enlarge the blind test set. The KIC catalog contains 2,135 stars of which 64 are extrapolations.

For galaxies in the blind test, we adopted the 2MRS catalog from \citet{2mrs}. It is a redshift sky survey based on the 2 MASS database, including galaxies with high redshift. There are 652 galaxies and 46 extrapolating ones. These objects are independent of the sample set.

We used the UVQS catalog \citep{uvqs} for QSO-blind testing and obtained 34 objects after cross-matching with J-PLUS. There are 18 objects that have fallen into the extrapolating region. UVQS contains UV bright QSOs, while the observation wavelength of VV13 is mainly in optical bands. This difference may cause a bias between training and testing, and further result in misclassifications.

\begin{figure}
\includegraphics[width=0.45\textwidth]{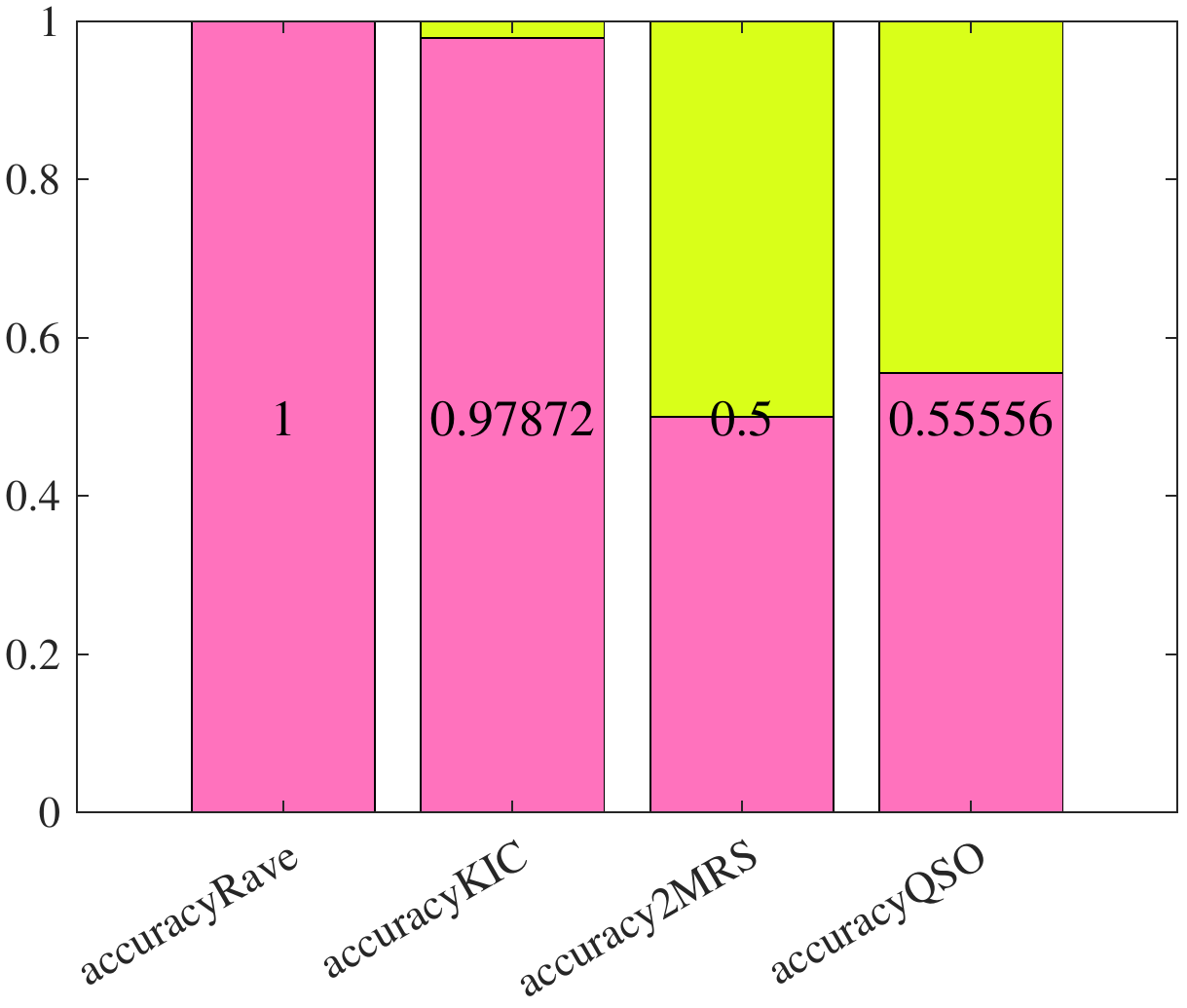}
\caption{Accuracy distribution for different extrapolating  data blind  sets. The colors are the same as in Fig. \ref{blindtestplot}. \label{extraacc}}
\end{figure}

\begin{figure}
\includegraphics[width=0.5\textwidth]{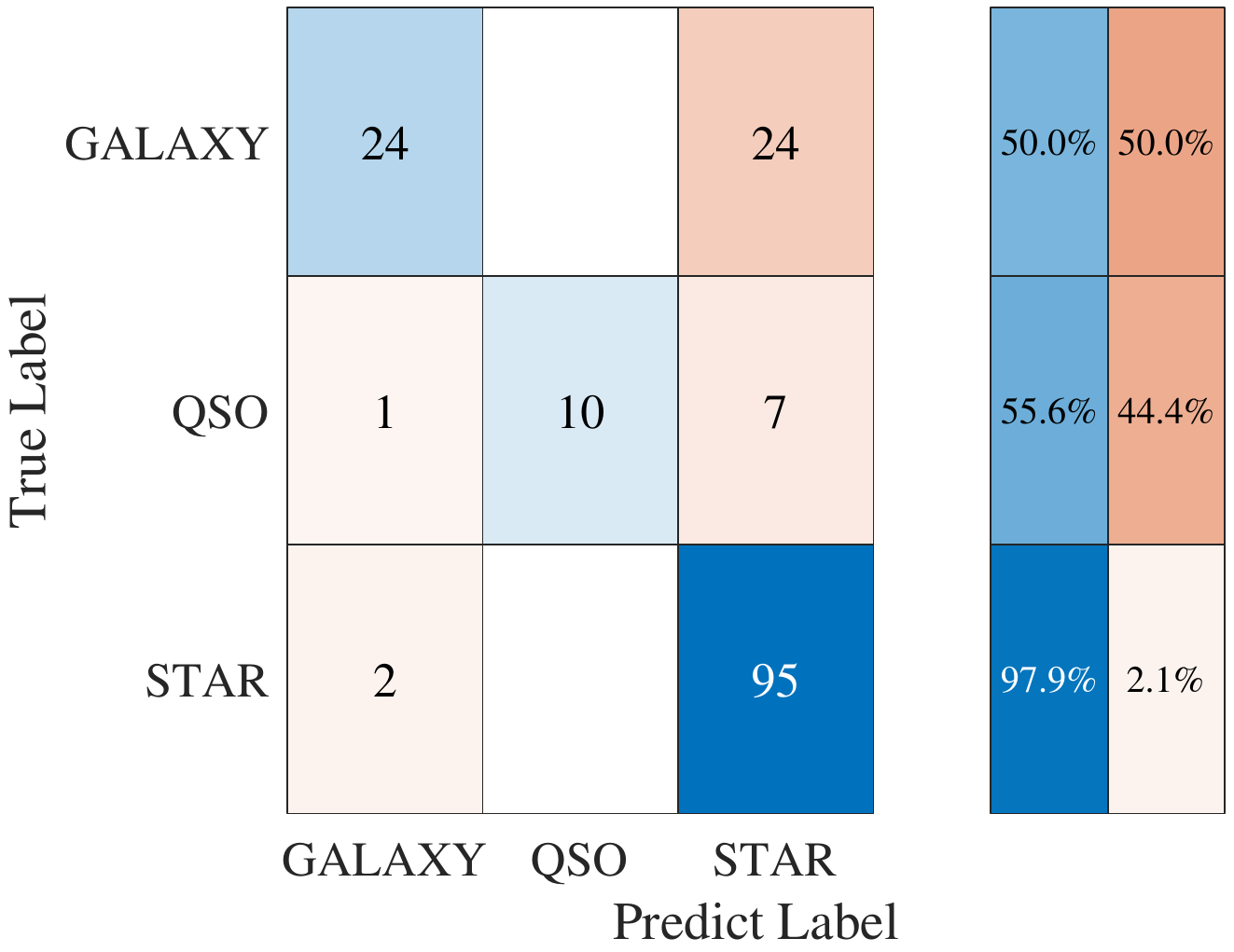}
\caption{Confusion matrix for the extrapolating blind test, and the accuracy is 79.1\% . \label{extraconf}}
\end{figure}

The blind test set was constructed by the independent objects from the four catalogs. We then separated the testing data into the interpolation and extrapolation samples.

We also adopted some other parameters to describe the classifier: recall, precision, and \textit{$F_1$}-score. We first define true positives (TPs), false positives (FPs), and false negatives (FNs) to demonstrate these parameters. TP is the number that both the blind test labels and the predicted labels are positive. FP is the number that blind test labels are negative while the predicted labels are positive, FN is the number that blind test labels are positive, while the predicted labels are negative. Readers should recall that $= \frac{\rm TP}{\rm TP+ \rm FN}$ shows the fraction of right prediction for a label. Precision $= \frac{\rm TP}{\rm TP+ \rm FP}$ shows the fraction of right prediction, and ${F_1-\rm score} = \frac{2}{\frac{1}{\rm precision}+\frac{1}{\rm recall}}=\frac{2\rm TP}{2\rm TP+ \rm FP+ \rm FN}$ shows the harmonic mean of the precision and the recall. 

The total accuracy is $96.5\%$ for the interpolating sample (Fig. \ref{blindtestplot} and \ref{blindtestconf}), and the parameters are shown in Table \ref{btf}. We present the accuracy distribution corresponding to the magnitudes as well (Fig. \ref{fmag}). See more in Appendix \ref{appb}. The blind test indicates a high reliability of the classifier. For the rest of the sample, the total accuracy is 79.1\%  (Fig. \ref{extraacc} and \ref{extraconf}), which is much lower than the interpolating sample, and the test parameters are shown in Table \ref{btef}. This indicates that it is significant and effective to constrain extrapolation in prediction.

\begin{table}
\caption{Parameters for interpolating the blind test\label{btf}}
\centering  
\begin{tabular}{lccc}
\hline \hline \noalign{\smallskip}
  Parameters & STAR &  GALAXY  & QSO \\
  \noalign{\smallskip}
\hline
\noalign{\smallskip}
  Recall & 99.4$\%$ & 88.5$\%$ & 76.9$\%$ \\
  Precision & 96.1$\%$ & 97.7$\%$ & 1 \\
  $F_1$-score & 95.0$\%$ & 92.9$\%$ & 87.0$\%$  \\
  \hline
\end{tabular}
\end{table}

\begin{table}
\caption{Parameters for extrapolating the blind test\label{btef}}
\centering  
\begin{tabular}{lccc}
\hline \hline \noalign{\smallskip}
  Parameters & STAR &  GALAXY  & QSO \\
  \noalign{\smallskip}
  \hline
  \noalign{\smallskip}
  Recall & 99.0$\%$ & 61.0$\%$ & 68.3$\%$ \\
  Precision & 83.3$\%$ & 90.4$\%$ & 1 \\
  $F_1$-score & 90.5$\%$ & 72.9$\%$ & 81.2$\%$  \\
  \hline
\end{tabular}
\end{table}

\begin{figure}
 
\includegraphics[width=0.45\textwidth]{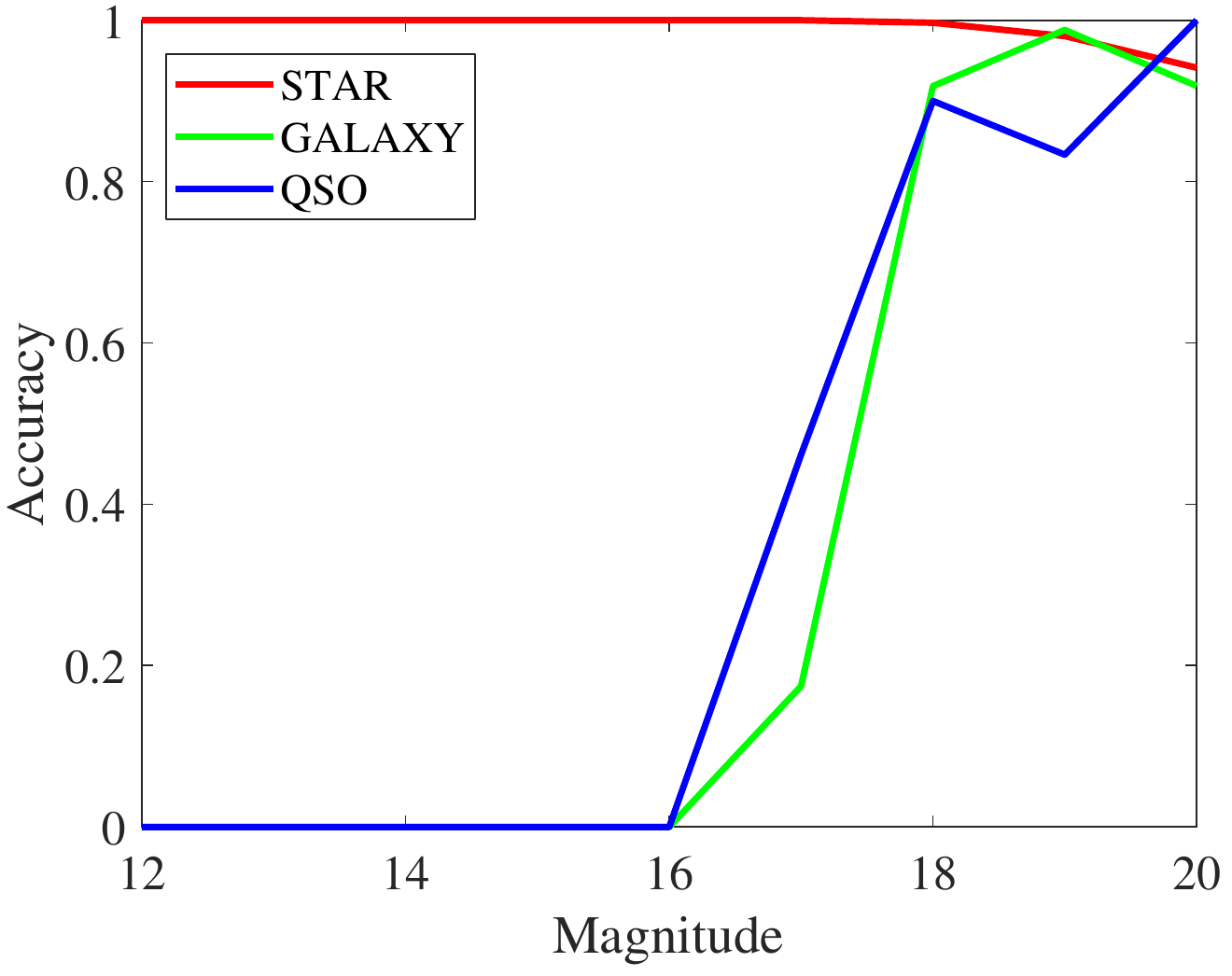}
\includegraphics[width=0.5\textwidth]{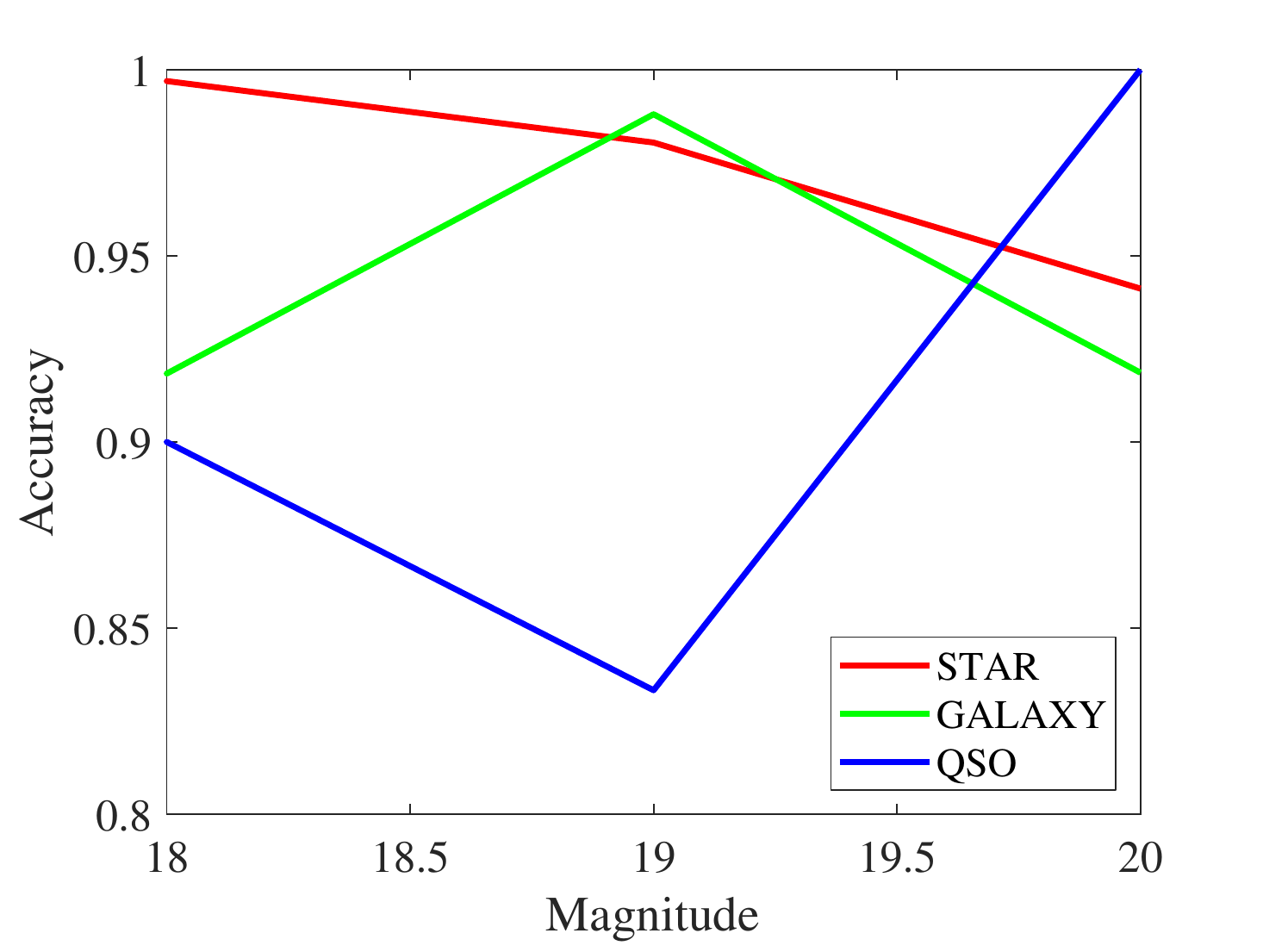}
\caption{Accuracy distribution of the blind test corresponds to mag6 (\textit{g} band, top panel). The bottom panel shows the detail of the upper figure from 18 mag to 20 mag. The zero accuracies of bright GALAXY and QSO are caused by sample insufficiency. \label{fmag}}
\end{figure}

\section{Results}
\label{result}

\subsection{Classification catalogs}
\label{catalog}

The total number of objects in the J-PLUS data set is 13,265,168, and there are 4,126,928 objects with valid 12 magnitudes. We obtained a classifier using the 12-band magnitudes and their corresponding errors to classify objects into STAR, GALAXY, and QSO categories. The classifier was constructed with a SVM algorithm based on the data from J-PLUS, SDSS, LAMSOT, and VV13. We present a new classification catalog in Table \ref{table1}. In order to avoid potential extrapolation, we set up 12 contours and there are 3,496,867 objects located inside. 

\begin{table*}
\caption{J-PLUS classification \label{table1}}
\centering  
\begin{tabular}{cccccc}
\hline \hline \noalign{\smallskip}
ID & R.A. & Dec & class$\_$star & PredictClass & Probability\\
\noalign{\smallskip}
\hline
\noalign{\smallskip}
26016-5 & 255.46835 & 22.76001 & 0.9990 & STAR & 92.96\% \\
26016-15 & 255.30506 & 22.76083 & 0.9990 & STAR & 99.91\% \\
26016-16 & 255.36753 & 22.76113 & 0.9990 & STAR & 99.20\% \\
26016-22 & 255.27928 & 22.76202 & 0.9536 & GALAXY & 79.71\% \\
26016-29 & 255.45243 & 22.76157 & 0.9814 & GALAXY & 76.31\% \\
26016-33 & 255.39194 & 22.76157 & 0.9780 & GALAXY & 93.95\% \\
26016-50 & 255.16394 & 22.76439 & 0.0460 & STAR & 64.35\% \\
26016-55 & 255.66306 & 22.76168 & 1.0000 & STAR & 99.49\% \\
26016-63 & 254.86508 & 22.76506 & 0.7656 & QSO & 86.82\% \\
26016-64 & 254.75625 & 22.76496 & 0.9663 & STAR & 84.58\% \\
\hline
\end{tabular}
\tablefoot{This table contains the top ten objects located inside the 12 contours. ID is the object identity from J-PLUS \footnote{\url{http://archive.cefca.es/catalogues/jplus-dr1/navigator.html}}. R.A. and Dec are the right ascension and declination of the objects. The class$\_$star column comes from the J-PLUS catalog, denoting the probability of stars. The PredictClass column presents our classification. The last column provides the probabilities of the predicted class. The sum of the probabilities for the three classes is equal to $100\%$. In this work, we developed a SVM algorithm with a one-versus-one strategy and a Gaussian kernel equal to 0.75. The table is uploaded with the paper. }
\end{table*}

\begin{figure}
 
\includegraphics[width=0.5\textwidth]{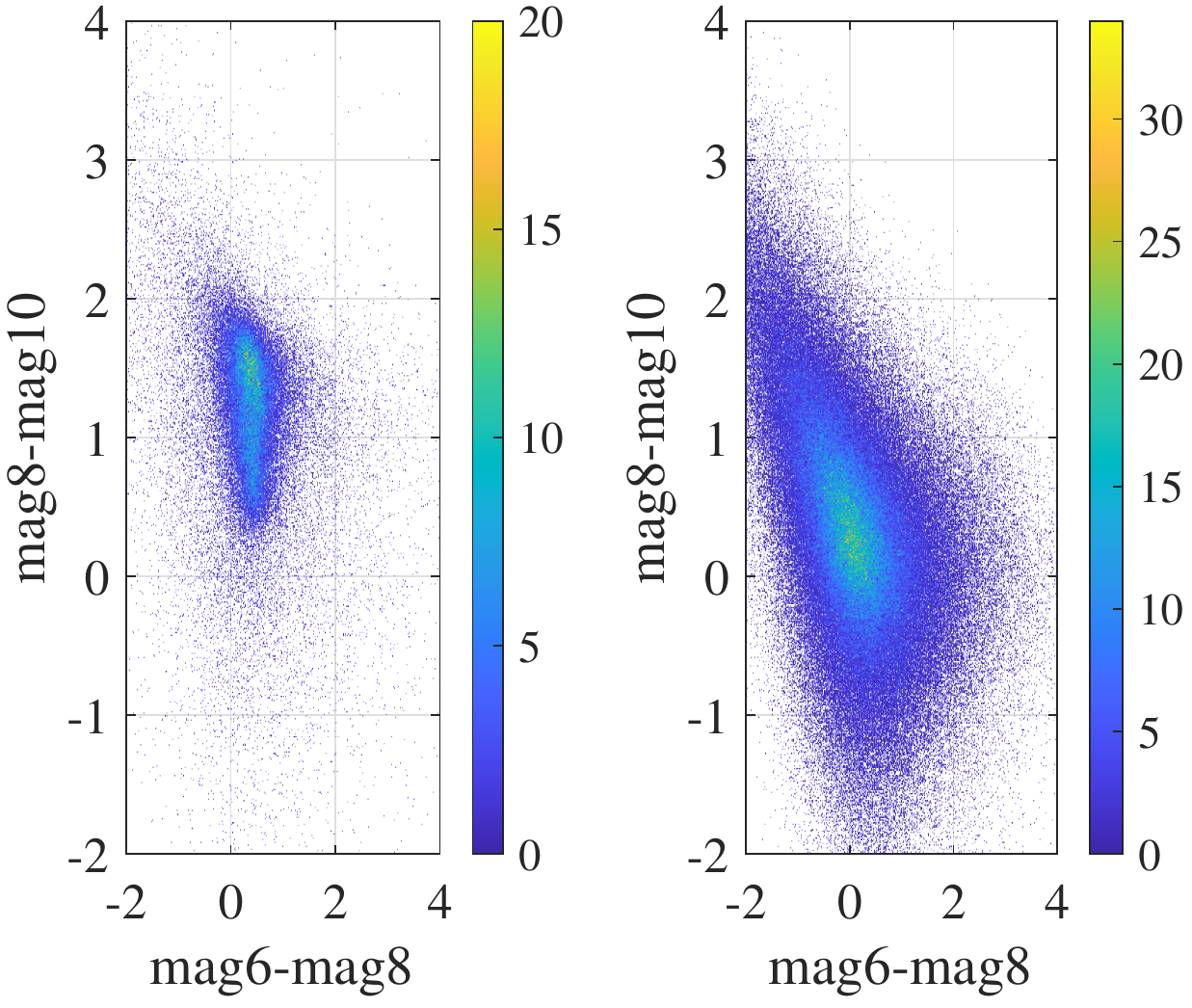}
\caption{Color-color diagram of GALAXYs. The left panel is the sample, and the right panel is the interpolation set. The color is the density of the sample, with a color bin of 0.01mag$^2$. \label{colorgalaxy}}
\end{figure}

\begin{figure}
 
\includegraphics[width=0.5\textwidth]{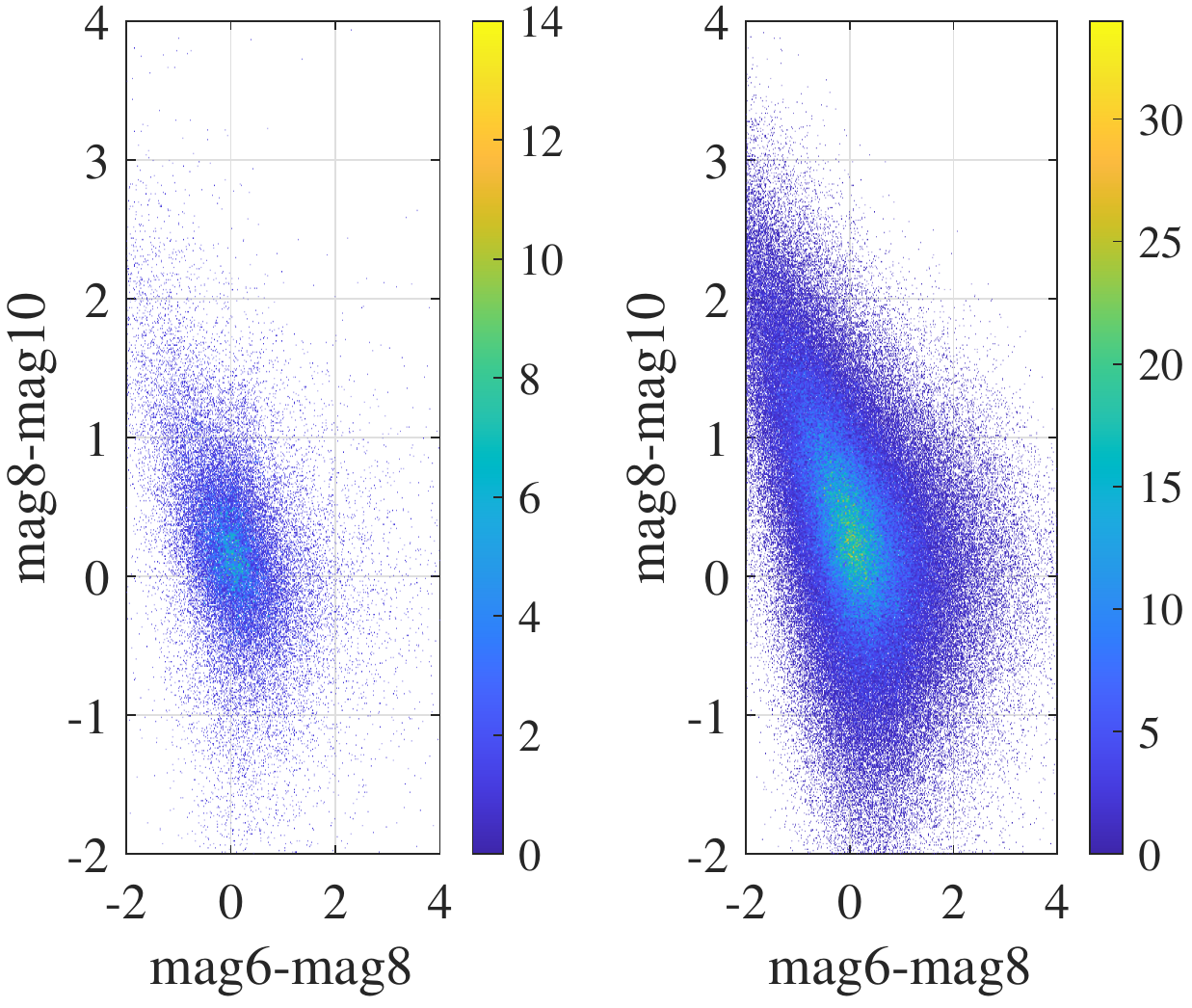}
\caption{Color-color diagram of QSOs (similar to Fig. \ref{colorgalaxy}). \label{colorqso}}
\end{figure}

\begin{figure}
 
\includegraphics[width=0.5\textwidth]{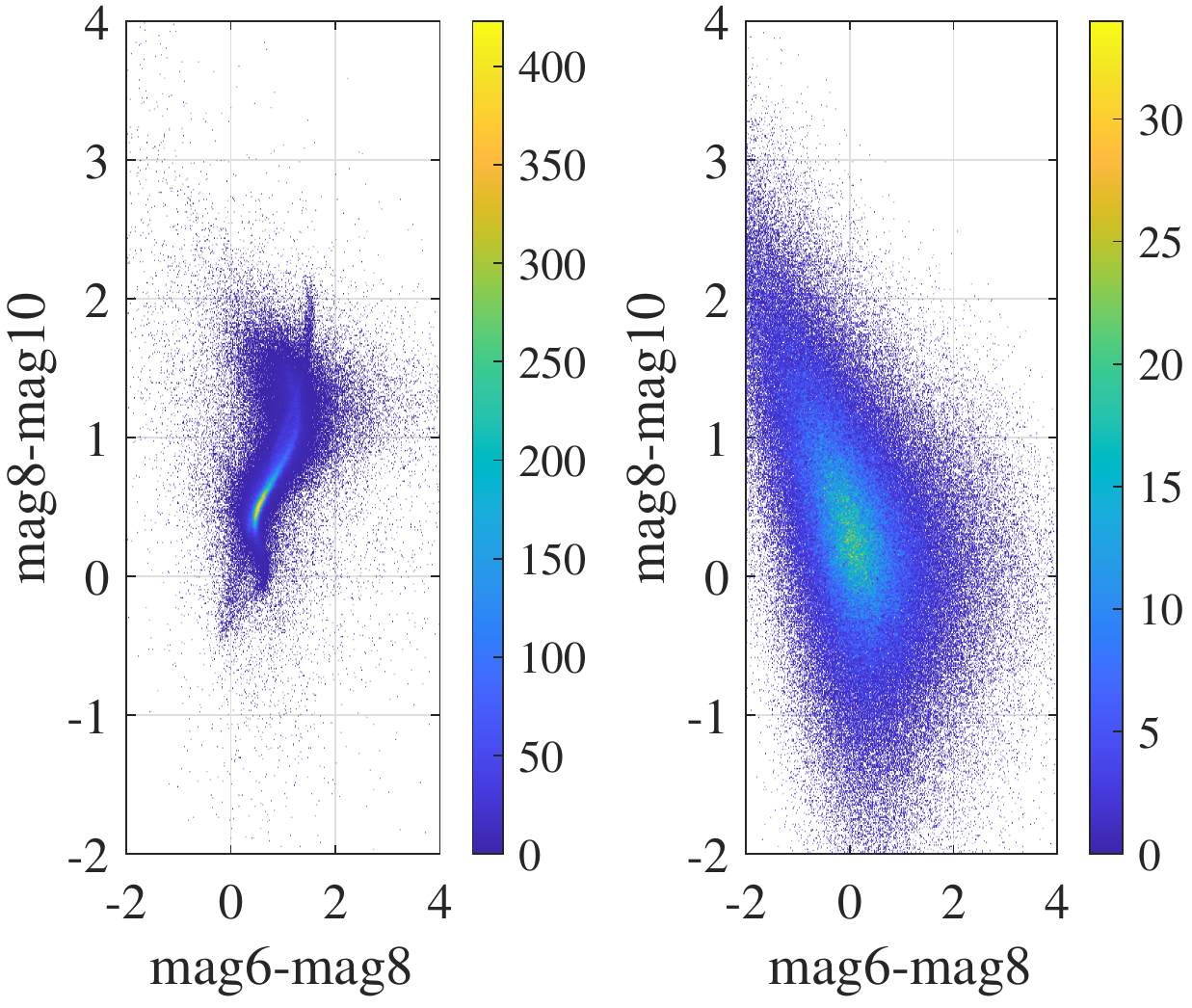}
\caption{Color-color diagram of STARs (similar to Fig. \ref{colorgalaxy}). \label{colorstar}}
\end{figure}

We have 2,493,424 stars, 613,686 galaxies, and 389,757 QSOs. The average probability is $95.63\%$ for STAR, $86.62\%$ for GALAXY, and $79.04\%$ for QSO. We also present the color-color plot of these interpolating objects (Fig. \ref{colorgalaxy}, \ref{colorqso}, and \ref{colorstar}). In these plots, we chose mag6$-$mag8 and mag8$-$mag10 ($g-r$ and $r-i$) to show the spread of interpolation objects. We also provide the magnitude distributions of each class in Appendix \ref{appb}. The objects suffering from potential extrapolation are shown in Table \ref{tableextra}, including 223,924 stars, 239,616 galaxies, and 166,521 QSOs, which is a total of 630,061 objects. 

\begin{table*}
\centering  
\caption{Extrapolation objects \label{tableextra}}
\begin{tabular}{cccccc}
\hline \hline \noalign{\smallskip}
ID & RA & Dec & class$\_$star & PredictClass & Probability \\
\noalign{\smallskip}
\hline
\noalign{\smallskip}
26016-2 & 255.64129 & 22.75862 & 0.9213 & STAR & 58.49\% \\
26016-3 & 255.54013 & 22.75905 & 0.9345 & STAR & 60.68\%\\
26016-9 & 255.39760 & 22.76004 & 0.4218 & QSO & 64.53\%\\
26016-10 & 255.20027 & 22.76122 & 0.9497 & STAR & 64.68\%\\
26016-11 & 255.40765 & 22.76091 & 0.9711 & STAR & 54.15\%\\
26016-13 & 255.38474 & 22.76107 & 0.9604 & STAR & 96.74\%\\
26016-14 & 255.22779 & 22.76128 & 0.9975 & QSO & 89.82\%\\
26016-23 & 255.76778 & 22.76090 & 0.9379 & QSO & 95.45\%\\
26016-34 & 255.61690 & 22.76251 & 0.6918 & GALAXY & 46.79\%\\
26016-37 & 255.37846 & 22.76318 & 0.9785 & STAR & 95.16\%\\
\hline
\end{tabular}
\tablefoot{This table contains the top ten objects located outside the 12 contours. The column labels for this table are the same as in Table \ref{table1}. The table is uploaded with the paper.}
\end{table*}

\begin{table}
\centering  
\caption{Object numbers and criteria \label{table2}}
\begin{tabular}{lr}
\hline \hline \noalign{\smallskip}
Criterion & Object number \\
\noalign{\smallskip}
\hline
\noalign{\smallskip}
$<0.5$  & 117,952 \\
$<0.4$  & 15,583 \\
$<0.35$  & 943  \\
$<0.34$  & 155 \\
\hline
\end{tabular}
\tablefoot{The first column is the upper limit for the highest probability in three classes. The objects were taken from both Table \ref{table1} and \ref{tableextra}.}
\end{table}

\subsection{Ambiguous objects}
\label{abmig}

The classifier also presents the probabilities of three different classes, which enabled us to select ambiguous objects. The ambiguous objects show characteristics that are unlike any of the three classes. When one's three-class probabilities are similar, it is selected as an ambiguous object. Table \ref{table2} shows different criteria and  their corresponding object numbers. The criterion is the upper limit of the highest probability in three classes.  We present 155 objects with three probabilities lower than 0.34 in Table \ref{abnormal}.

In order to find the abnormal objects from the ambiguous samples, we calculated the Mahalanobis distance \citep{mahal1,mahal2}. We then checked whether the objects were far from each label. The objects that have a higher distance to one label than the distance of this label to the other labels were treated as abnormal objects. These objects are not only located outside the region of three classes, but they are also far from all of them. The criteria of the Mahalanobis distance are as follows: 18.4 between STAR to GALAXY, 23.3 between GALAXY to QSO, and 50.3 between QSO to STAR. Table \ref{ab26} presents 26 abnormal objects. 

\begin{table*}
\centering  
\caption{Ambiguous objects \label{abnormal}}
\begin{tabular}{cccccccc}
\hline \hline \noalign{\smallskip}
ID & RA & Dec & class$\_$star & PredictClass & GALAXY & QSO & STAR \\
\noalign{\smallskip}
\hline
\noalign{\smallskip}
26016-20835 & 255.17392 & 23.50692 & 0.0008 & GALAXY & 33.50\% & 32.81\% & 33.68\%\\
26016-32840 & 255.67090 & 24.07627 & 0.5004 & STAR & 33.38\% & 32.90\% & 33.70\%\\
26015-6320 & 256.61032 & 23.07467 & 0.4304 & STAR & 32.54\% & 33.51\% & 33.94\%\\
26010-33609 & 257.27271 & 25.44612 & 0.0078 & GALAXY & 33.71\% & 33.09\% & 33.19\%\\
26012-8040 & 256.82760 & 25.80270 & 0.9121 & GALAXY & 33.95\% & 32.38\% & 33.65\%\\
26012-24063 & 256.34766 & 26.30697 & 0.4980 & GALAXY & 33.25\% & 33.40\% & 33.33\%\\
26028-3296 & 126.56224 & 29.97372 & 0.8632 & STAR & 32.39\% & 33.63\% & 33.97\%\\
26037-3692 & 139.61591 & 29.96809 & 0.0011 & GALAXY & 33.45\% & 33.74\% & 32.80\%\\
26038-17239 & 144.43648 & 30.72382 & 0.0027 & QSO & 33.22\% & 33.51\% & 33.25\%\\
26036-8820 & 146.18454 & 30.19118 & 0.0050 & GALAXY & 33.45\% & 33.22\% & 32.82\%\\
\hline
\end{tabular}
\tablefoot{This table contains the top ten ambiguous objects for criterion 0.34. The first five column labels of this table are the same as in Table \ref{table1}. The last three column labels provide the probability of the corresponding label. The table is uploaded with the paper.}
\end{table*}

\begin{table*}
\centering  
\caption{Abnormal objects\label{ab26}}
\begin{tabular}{ccccccccccc}
\hline \hline \noalign{\smallskip}
ID & RA & Dec. & \texttt{CLASS\_STAR} & PredictClass & GALAXY & Gdis & QSO & Qdis & STAR & Sdis\\
\noalign{\smallskip}
\hline
\noalign{\smallskip}
26016-32840     & 255.67090 & 24.07627 & 0.5005 &       STAR &  $33.39\%$ & 45.85 &       $32.90\%$ &     59.90 & $33.71\%$ & 214.59 \\
26047-19730 & 121.32316 & 31.91350 & 0.0001 & GALAXY &  $32.76\%$ &     78.09 &       $33.94\%$ &     51.34 & $33.30\%$ &  204.47 \\
26091-3622 & 128.45054 & 34.12899 &     0.4170 & GALAXY &       $32.93\%$ &       201.06 &        $33.70\%$ & 62.20 &     $33.38\%$ &     391.73 \\
33209-2012 & 137.46224 & 39.60661 &     0.1433 &        GALAXY &        $33.52\%$ &       55.05 & $32.89\%$ &     61.91 & $33.59\%$ &     224.79 \\
26141-18524 & 169.58419 & 40.46721 &    0.0016 &        STAR &  $33.52\%$ &       73.44 & $32.63\%$ & 52.22 & $33.85\%$ &         185.80 \\
26145-15610 & 284.78880 & 39.72198 &    0.2306 & GALAXY &       $33.87\%$ & 133.31 & $32.82\%$ & 50.71 &  $33.32\%$ &     278.23 \\
33232-7122 & 126.08110 & 41.29631 &     0.1117 & GALAXY &       $33.60\%$ &       98.28 & $33.32\%$ & 67.51 & $33.08\%$ &         251.50 \\
26151-29824 & 273.32033 & 41.61314 & 0.0013     & GALAXY &       $33.74\%$ & 229.37 & $32.77\%$ & 56.60 &   $33.49\%$ & 255.62 \\
26207-8959 & 137.26562 & 52.62772 &     0.0027 & GALAXY &        $32.94\%$ & 66.90 &       $33.75\%$ & 59.96 &     $33.31\%$ &  312.61 \\
26241-25002 & 137.26562 & 52.62772 &    0.5029 & GALAXY &        $33.31\%$ & 129.51 & $33.41\%$    & 70.59 &       $33.28\%$ & 278.00 \\
\hline
\end{tabular}
\tablefoot{This table contains the top ten abnormal objects. The column labels of this table are the same as in Table \ref{abnormal}. The column Gdis, Qdis, and Sdis provide the Mahalanobis distance to the group GALAXY, QSO, and STAR. The table is uploaded with the paper.}
\end{table*}

\begin{figure}
\centering  
\includegraphics[width=0.45\textwidth]{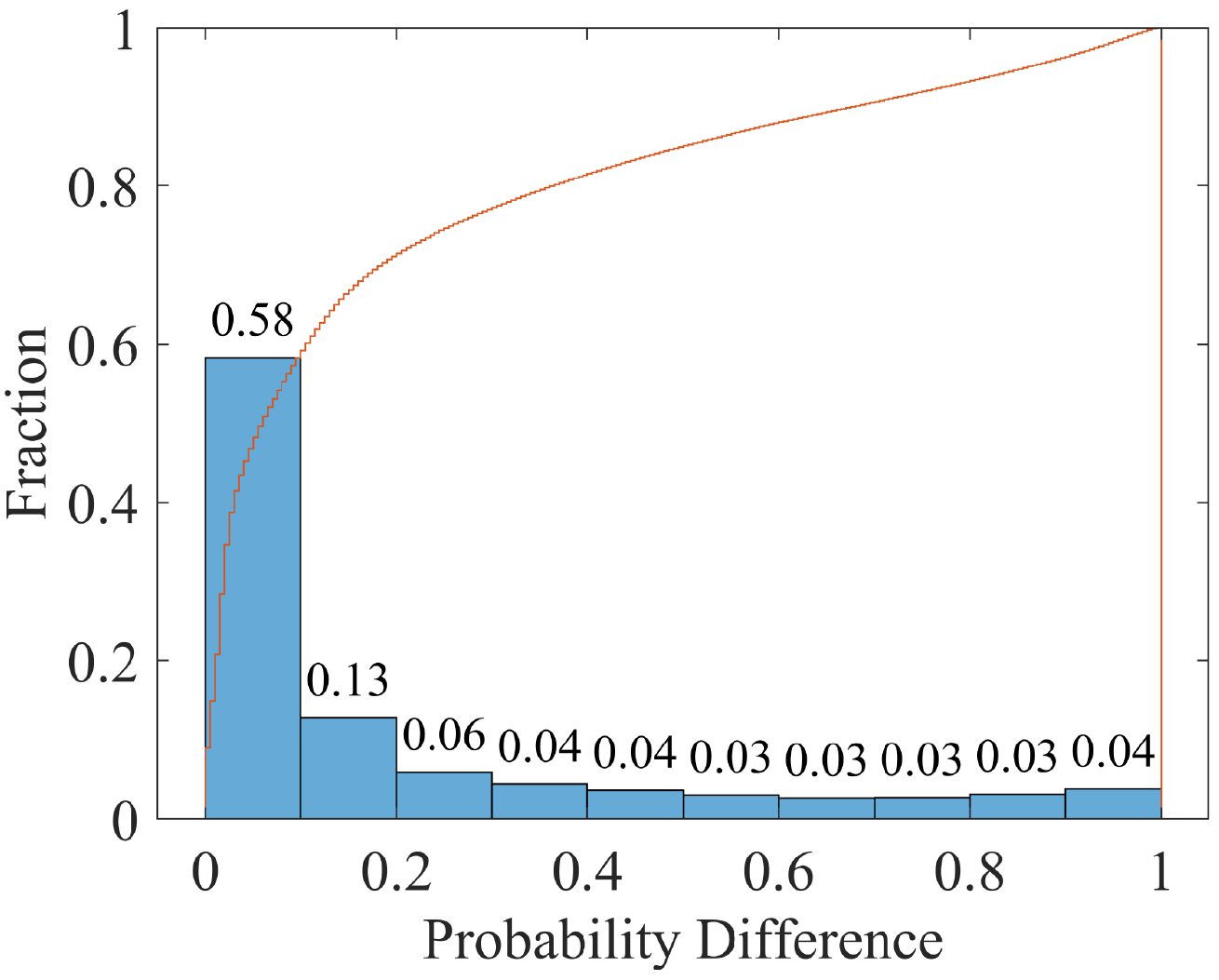}
\caption{Distributions of different probabilities between \texttt{CLASS\_STAR} and our stellar probabilities (blue bar). The red line shows the cumulative distribution function of the difference.\label{Qplot}}
\end{figure}

\subsection{Comparison with \texttt{CLASS\_STAR}}
\label{classstar}

In Fig. \ref{Qplot}, we draw the difference between our results and the \texttt{CLASS\_STAR} in J-PLUS catalog. The figure indicates that there are differences between the J-PLUS \texttt{CLASS\_STAR} and our result. The difference may be caused by the different strategies of classifying the objects: binary classification for J-PLUS based on the point-source detection and our triple classification based on machine learning. The QSOs are probably not distinguished from stars or galaxies with the point-source detection, and such a detection could further result in the difference in Fig. \ref{Qplot}. Therefore, the factor \texttt{CLASS\_STAR} may not be suitable enough for multi-classifications.

\section{Discussion}
\label{discuss}

\subsection{Different ways to constrain extrapolation}
\label{mte}

We constructed three methods to constrain extrapolation, including magnitude cuts and two density-dependence methods. The most straightforward thought is defining intervals based on magnitude distributions of our training sample. We can determine whether an object belongs to the intersection of these intervals or not.

\begin{figure}
 
\includegraphics[width=0.45\textwidth]{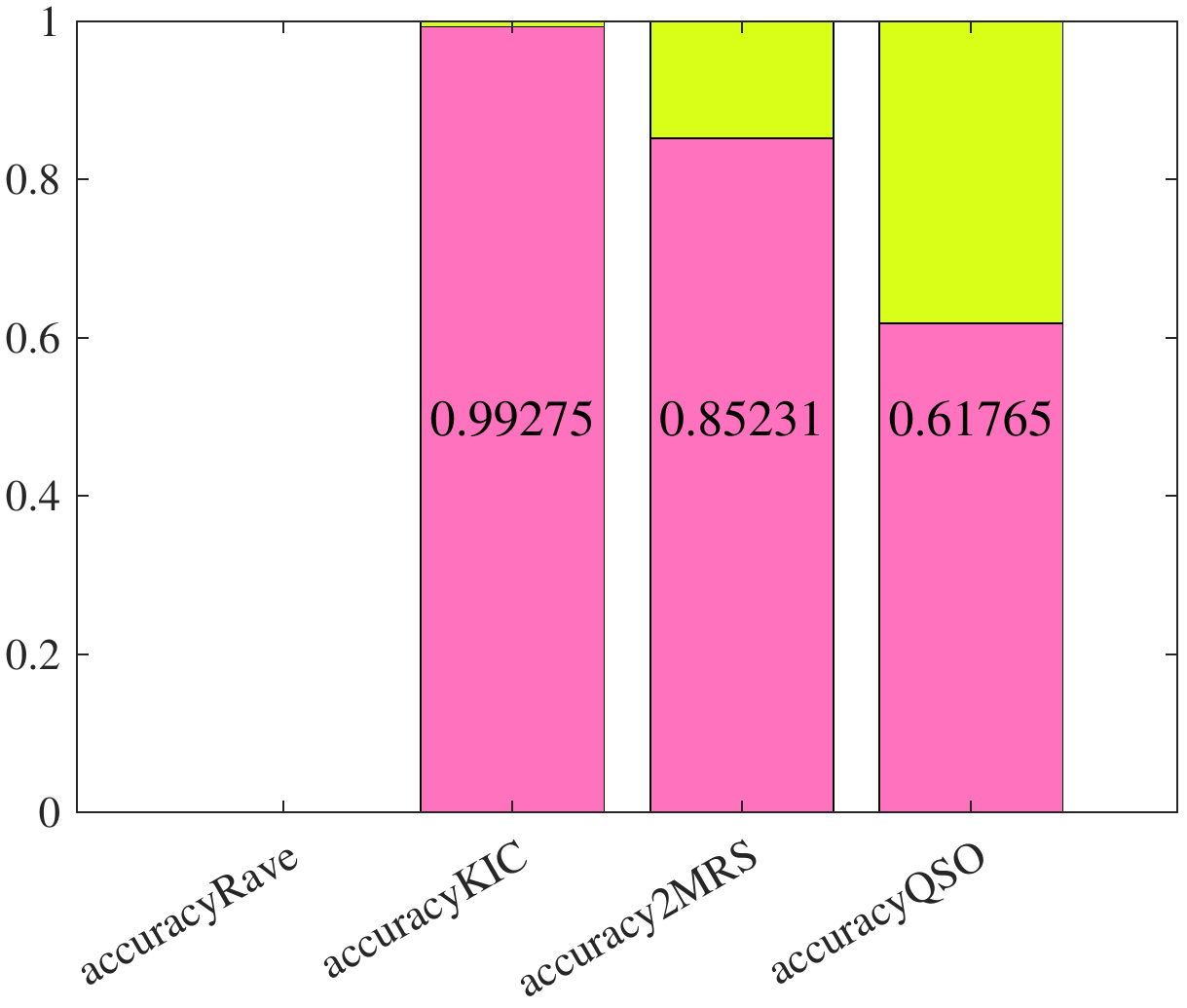}
\caption{Accuracy distribution for each blind test set under the kernel distribution method. This method defines all RAVE objects as extrapolation. \label{keracc}}
\end{figure}
\begin{figure}
 
\includegraphics[width=0.5\textwidth]{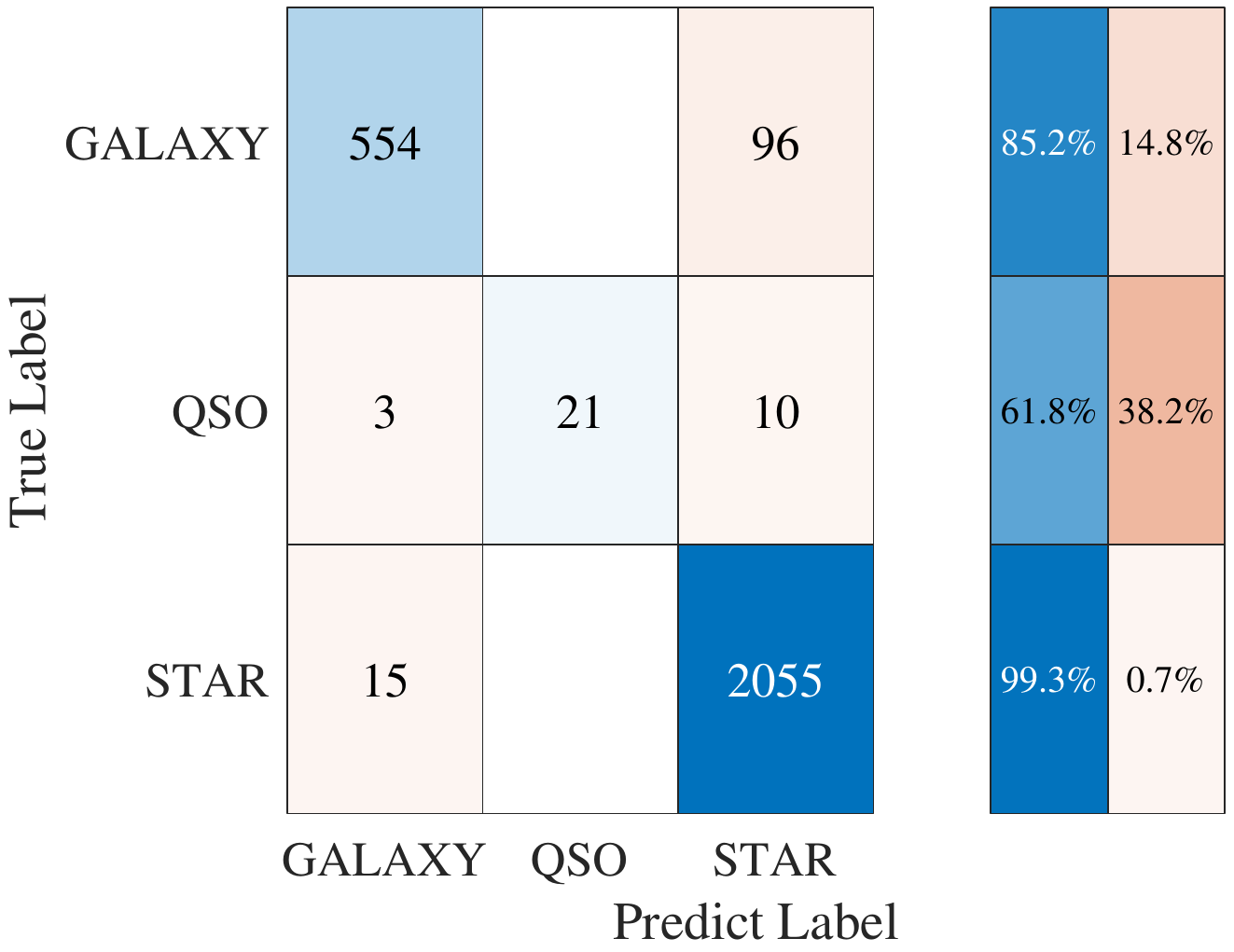}
\caption{Confusion matrix of the blind test by using the kernel distribution method to develop the extrapolations, and the accuracy is $95.5\%$. \label{kerconf}}
\end{figure}

We employed kernel distribution \citep{ker97} to fit the distributions of each dimension in the instance space. The kernel distribution is a kind of probability measure. For each dimension, the objects situated in the middle part of the distribution are defined as interpolations.  By cutting down 0.025 for each side of a magnitude distribution, the intervals were constructed, and the objects could be separated into interpolation or extrapolation. This method results in an accuracy of $95.5\%$  for the blind test. After the selection, 2,749,840 interpolating objects were left, and there was $65.79\%$ of the J-PLUS catalog (Fig. \ref{keracc} and \ref{kerconf}). This method is precluded due to its low accuracy and its unrepresentative of the interpolation boundary.

\begin{figure}
 
\includegraphics[width=0.45\textwidth]{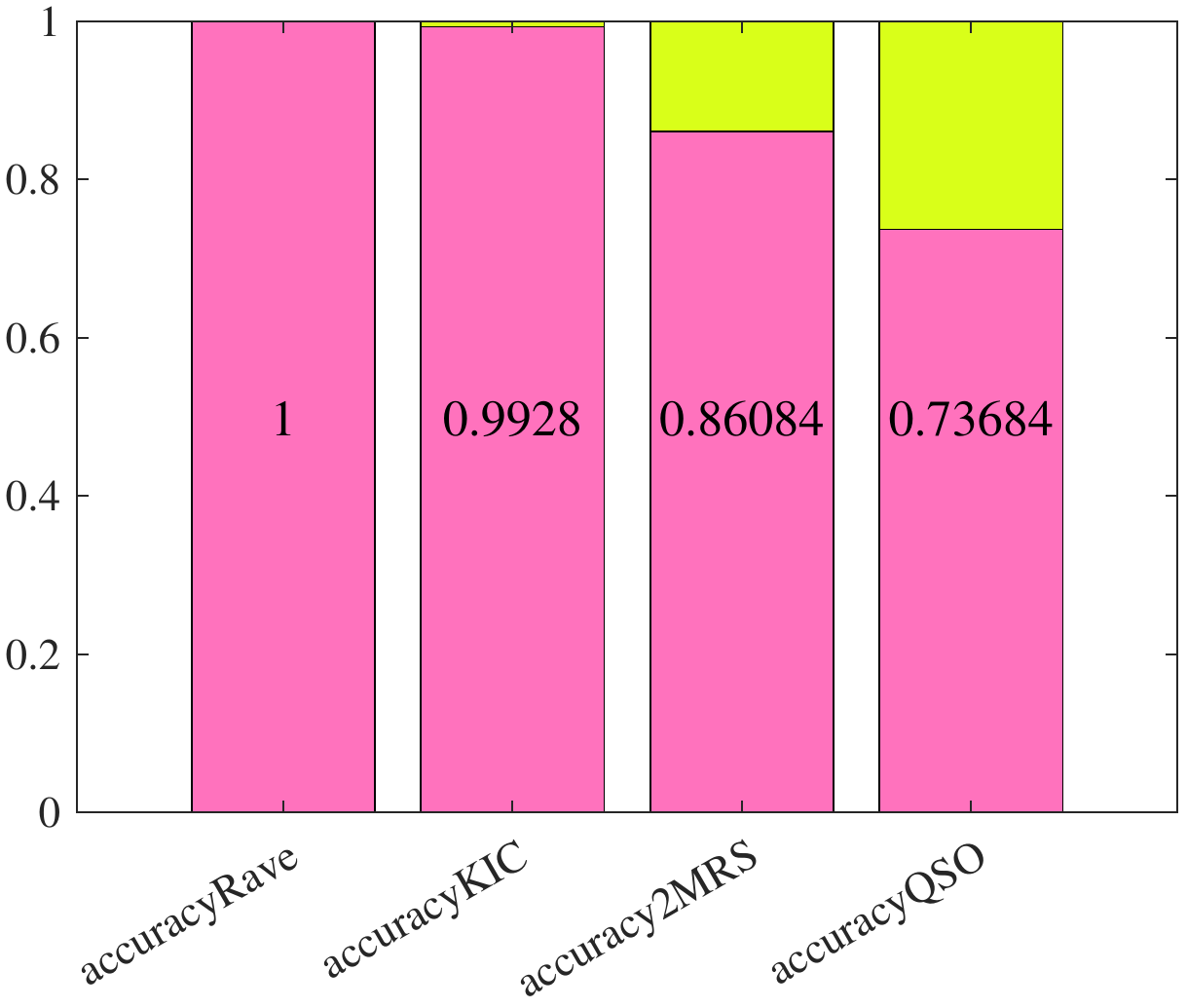}
\caption{Accuracy distribution for each blind test set under the rough contour method. \label{roughacc}}
\end{figure}

\begin{figure}
\includegraphics[width=0.5\textwidth]{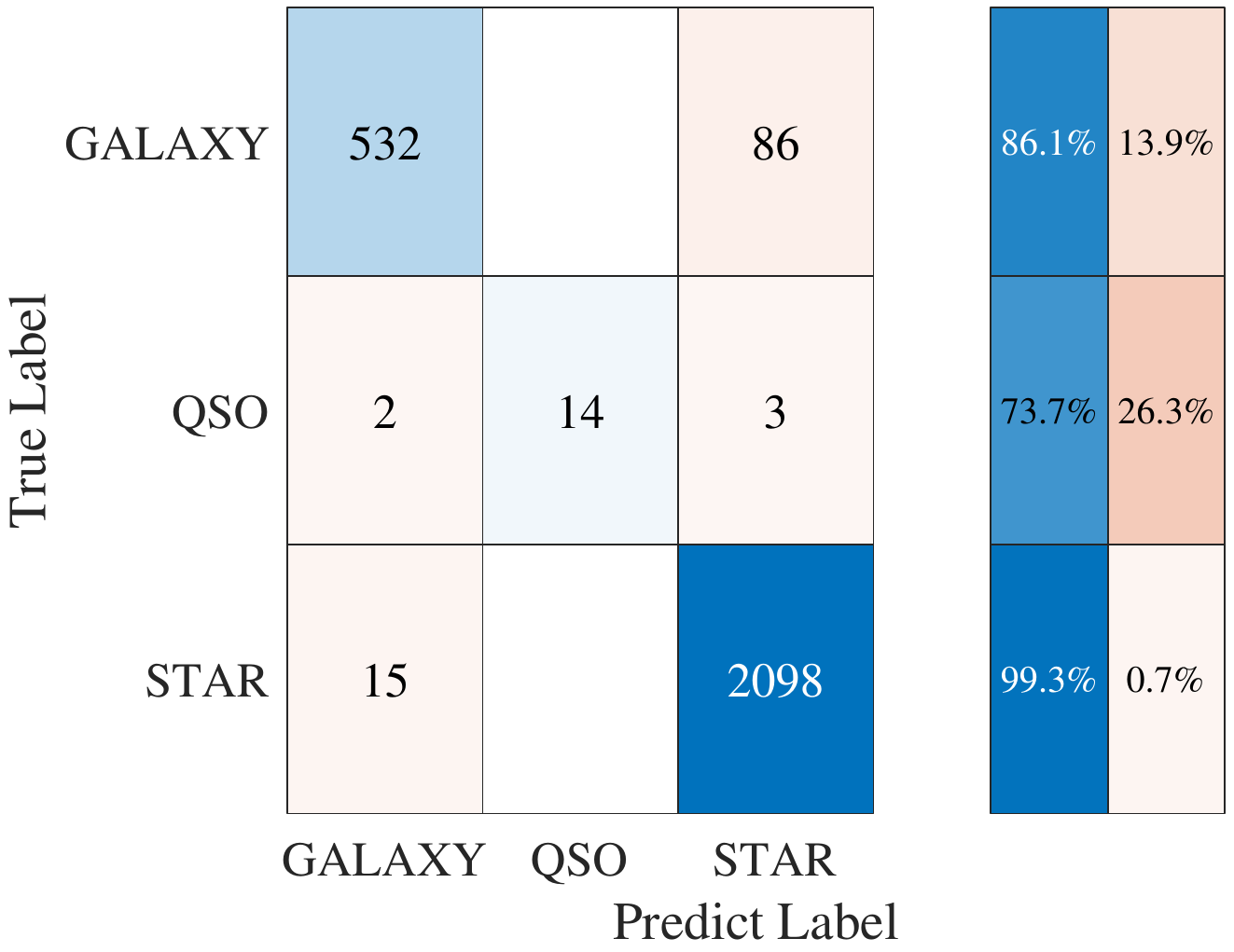}
\caption{Confusion matrix of the blind test by using the rough contour method to develop extrapolation, and the accuracy is 96.1\%. \label{roughconf}}
\end{figure}

The ideal approach is to draw a 12-dimension density contour to select the interpolating sample. The adopted method (Sect. \ref{outlier}) is an approximation of such an ideal approach. The last method is four contours instead of 12 contours, which are (mag1, mag2, mag3), (mag4, mag5, mag6), (mag7, mag8, mag9), and (mag10, mag11, mag12). These rough contours result in an accuracy of $96.1\%$, and they left 3,702,268 interpolating objects (Fig. \ref{roughacc} and \ref{roughconf}). This method is also precluded due to its low accuracy.

\subsection{Comparison of different classifiers}
\label{cdm}
\citet{bai19} used a RF algorithm to gain a classifier with an accuracy of $99\%$. We also tested RF, but its accuracy is lower than SVM. The different results of these two works are probably due to the different sample sizes and wavebands. The accuracy of the blind test is similar to the training accuracy, implying that there is no obvious overfitting in our training process \citep{shai14}.

The sample size may also influence the training accuracy. In our method, the sample size is 468,685, while in \citet{bai19}, the number is 2,973,855. Both SVM or RF have a finite Vapnik-Chervonenkis dimension (VCdim; \citealt{shai14}). If a sample size goes to infinity, the training error and the validation error converge to the approximation error. This implies that there exists a limited accuracy of a classifier. In our work, the training error ($97\%$) is similar to the validation error ($96.5\%$). Therefore, if we enlarge the sample size, the accuracy may not increase significantly. 

\begin{figure*}
\centering  
\includegraphics[width=0.85\textwidth, height=0.55\textwidth]{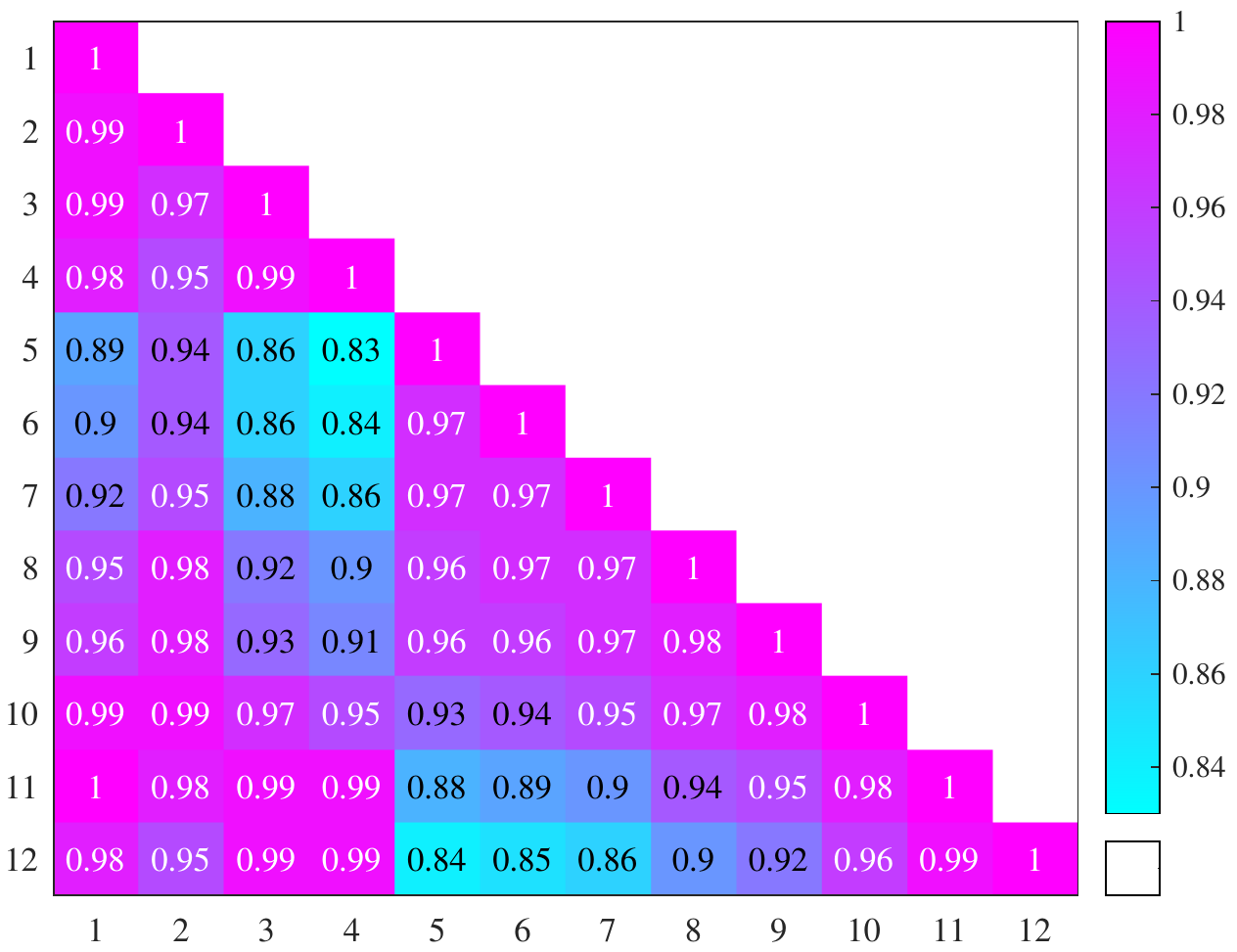}
\caption{Correlation matrix of the 12 bands. The numbers in the axes imply the bands, e.g., 1 corresponds to mag1. The 12 magnitudes are mag1 (\textit{u}), mag2 (J0378), mag3 (J0395), mag4 (J0410), mag5 (J0430), mag6 (\textit{g}), mag7 (J0515), mag8 (\textit{r}), mag9 (J0660), mag10 (\textit{i}), mag11 (J0861), and mag12 (\textit{z}). \label{matrix}}
\end{figure*}

\citet{bai19} applied nine-dimensional color spaces including infrared bands, while we used 12 optical magnitudes. More and broader bands involved in the training would lead to a higher total accuracy. The accuracy in our work is slightly lower. This is probably due to the strong correlation in the 12 bands. We calculated a correlation matrix (Fig. \ref{matrix}) for these 12 bands from their photometric results.

The minimum of the correlation coefficient stands at $(5,4)$, mag4 (J0410), and mag5 (J0430) in Fig. \ref{matrix}. These high correlation values indicate that all of these wavebands are highly correlated. The correlation may be explained by not only the distance of the object that causes a similarity in all magnitudes, but also the overlapping of filter profiles. From the band plot in J-PLUS \footnote{The plot can be found both at \url{http://www.j-plus.es/ancillarydata/dr1_lya_emitting_candidates} and in \citet{jplus2019}}, the filter profile of $u$, $g$, $r$, $i$, $z$ have overlapped with other narrow bands, implying that they are not strongly independent. The wavebands adopted in \citet{bai19} cover a larger range, and the correlations are probably weaker.

The constitution of the data set may also influence the accuracy of a classifier. In \citet{ball06}, a tree algorithm was developed to output the probability of a star, galaxy, and nsng (neither star nor galaxy object). In Bai's and Ball's training samples, there was a significant bias in the sample set. The sample construction of our SVM classifier and the two mentioned classifiers is shown in Table \ref{table3}. \citet{bai19} and \citet{ball06} concluded that the biased sample can also present a training accuracy of better than 95\%.

\begin{table}
\centering 
\caption{Constitution of different algorithms \label{table3}}
\begin{tabular}{lccc}
\hline \hline \noalign{\smallskip}
Class & This work & \citet{bai19} & \citet{ball06} \\
\noalign{\smallskip}
\hline
\noalign{\smallskip}
Galaxy & $15.94\%$ & $27.11\%$ & $75.77\%$\\
QSO(nsng) &  $9.79\%$ &  $1.47\%$ & $11.16\%$\\
Star   & $74.27\%$ &  $71.42\%$ & $13.07\%$\\
\hline
\end{tabular}
\tablefoot{In Bai's paper, the second row is QSO, while in Ball's work, it is nsng.}
\end{table}

\subsection{Future work}
\label{outlook}
The advantages of J-PLUS are 12 optical filters and a large amount of data. The ongoing J-PAS has an all-time system of 56 optical narrowband filters, making it one of the most promising surveys in the world. The way we work on J-PLUS can be copy to J-PAS. The more bands applied, the more precise a classifier could be.

\citet{minij21} developed different classifiers to label the mini-JPAS 
\citep{minijpas}, including RF and Extremely Randomized Trees (ERT). MiniJ-PAS is a previous project to test J-PAS. Their work has gained good performance with Area Under the Curve (AUC) greater than 0.95 in different classifiers. AUC is equal to the positive probability.

SVM is inferior when the instance space has too many dimensions, or when the data set is too large for calculation. More works are required to test the time cost of SVM when we apply larger data with more features. Although SVM is a good algorithm; its performance in J-PAS still needs to be tested considering the computational complexity and no-free-lunch theorem.

\begin{acknowledgements}
This work was supported by the National Natural Science Foundation of China (NSFC) through grants NSFC-11988101/11973054/11933004 and the National Programs on Key Research and Development Project (grant No.2019YFA0405504 and 2019YFA0405000). Strategic Priority Program of the Chinese Academy of Sciences under grant number XDB41000000. 
H, B, Yuan, acknowledge funding from National Natural Science Foundation of China (NSFC) through grants NSFC-12173007.
ELM acknowledges  support  from  the Agencia  Estatal  de  Investigaci\'on  del  Ministerio  de Ciencia e Innovaci\'on (AEI-MCINN) under grant PID2019-109522GB-C53\@. \\

Based on observations made with the JAST80 telescope at the Observatorio Astrofísico de
Javalambre (OAJ), in Teruel, owned, managed, and operated by the Centro de Estudios de Física del Cosmos de Aragón (CEFCA). We acknowledge the OAJ Data Processing and Archiving Unit (UPAD) for reducing the OAJ data used in this work.
Funding for the J-PLUS Project has been provided by the Governments of Spain and Aragón throughthe Fondo de Inversiones de Teruel; the Aragón Government through the Research Groups E96, E103, and E16\_17R; the Spanish Ministry of Science, Innovation and Universities (MCIU/AEI/FEDER, UE) with grants PGC2018-097585-B-C21 and PGC2018-097585-B-C22; the Spanish Ministry of Economy and Competitiveness (MINECO) under AYA2015-66211-C2-1-P, AYA2015-66211-C2-2, AYA2012-30789, and ICTS-2009-14; and European FEDER funding (FCDD10-4E-867, FCDD13-4E-2685). The Brazilian agencies FINEP, FAPESP, and the National Observatory of Brazil have also contributed to this project.

Guoshoujing Telescope (the Large Sky Area Multi-Object Fiber Spectroscopic Telescope LAMOST) is a National Major Scientific Project built by the Chinese Academy of Sciences. Funding for the project has been provided by the National Development and Reform Commission. LAMOST is operated and managed by the National Astronomical Observatories, Chinese Academy of Sciences.
Funding for the Sloan Digital Sky Survey IV has been provided by the Alfred P. Sloan Foundation, the U.S. Department of Energy Office of Science, and the Participating Institutions. SDSS-IV acknowledges support and resources from the Center for High-Performance Computing at the University of Utah. The SDSS website is \url{http://www.sdss.org/}.
SDSS-IV is managed by the Astrophysical Research Consortium for the Participating Institutions of the SDSS Collaboration including the Brazilian Participation Group, the Carnegie Institution for Science, Carnegie Mellon University, the Chilean Participation Group, the French Participation Group, Harvard-Smithsonian Center for Astrophysics, Instituto de Astrofísica de Canarias, The Johns Hopkins University, Kavli Institute for the Physics and Mathematics of the Universe (IPMU)/University of Tokyo, Lawrence Berkeley National Laboratory, Leibniz Institut für Astrophysik Potsdam (AIP), Max-Planck-Institut für Astronomie (MPIA Heidelberg), Max- Planck-Institut für Astrophysik (MPA Garching), Max-Planck-Institut für Extraterrestrische Physik (MPE), National Astronomical Observatories of China, New Mexico State University, New York University, University of Notre Dame, Observatário Nacional/MCTI, The Ohio State University, Pennsylvania State University, Shanghai Astronomical Observatory, United Kingdom Participation Group, Universidad Nacional Autónoma de México, University of Arizona, University of Colorado Boulder, University of Oxford, University of Portsmouth, University of Utah, University of Virginia, University of Washington, University of Wisconsin, Vanderbilt University, and Yale University.
This work is supported by the CSST project on "stellar activity and late evolutionary stage".

\end{acknowledgements}

%
%
\bibliographystyle{aa} 
\bibliography{ms4}

\begin{thebibliography}{42}
\expandafter\ifx\csname natexlab\endcsname\relax\def\natexlab#1{#1}\fi

\bibitem[{{Ahumada} {et~al.}(2020){Ahumada}, {Allende Prieto}, {Almeida},
  {Anders}, {Anderson}, {Andrews}, {Anguiano}, {Arcodia}, {Armengaud},
  {Aubert}, {Avila}, {Avila-Reese}, {Badenes}, {Balland}, {Barger},
  {Barrera-Ballesteros}, {Basu}, {Bautista}, {Beaton}, {Beers}, {Benavides},
  {Bender}, {Bernardi}, {Bershady}, {Beutler}, {Bidin}, {Bird}, {Bizyaev},
  {Blanc}, {Blanton}, {Boquien}, {Borissova}, {Bovy}, {Brandt}, {Brinkmann},
  {Brownstein}, {Bundy}, {Bureau}, {Burgasser}, {Burtin}, {Cano-D{\'\i}az},
  {Capasso}, {Cappellari}, {Carrera}, {Chabanier}, {Chaplin}, {Chapman},
  {Cherinka}, {Chiappini}, {Doohyun Choi}, {Chojnowski}, {Chung}, {Clerc},
  {Coffey}, {Comerford}, {Comparat}, {da Costa}, {Cousinou}, {Covey}, {Crane},
  {Cunha}, {da Silva Ilha}, {Dai}, {Damsted}, {Darling}, {Davidson}, {Davies},
  {Dawson}, {De}, {de la Macorra}, {De Lee}, {de Andrade Queiroz}, {Deconto
  Machado}, {de la Torre}, {Dell'Agli}, {du Mas des Bourboux},
  {Diamond-Stanic}, {Dillon}, {Donor}, {Drory}, {Duckworth}, {Dwelly},
  {Ebelke}, {Eftekharzadeh}, {Eigenbrot}, {Elsworth}, {Eracleous},
  {Erfanianfar}, {Escoffier}, {Fan}, {Farr}, {Fern{\'a}ndez-Trincado},
  {Feuillet}, {Finoguenov}, {Fofie}, {Fraser-McKelvie}, {Frinchaboy},
  {Fromenteau}, {Fu}, {Galbany}, {Garcia}, {Garc{\'\i}a-Hern{\'a}ndez}, {Garma
  Oehmichen}, {Ge}, {Geimba Maia}, {Geisler}, {Gelfand}, {Goddy},
  {Gonzalez-Perez}, {Grabowski}, {Green}, {Grier}, {Guo}, {Guy}, {Harding},
  {Hasselquist}, {Hawken}, {Hayes}, {Hearty}, {Hekker}, {Hogg}, {Holtzman},
  {Horta}, {Hou}, {Hsieh}, {Huber}, {Hunt}, {Ider Chitham}, {Imig}, {Jaber},
  {Jimenez Angel}, {Johnson}, {Jones}, {J{\"o}nsson}, {Jullo}, {Kim},
  {Kinemuchi}, {Kirkpatrick}, {Kite}, {Klaene}, {Kneib}, {Kollmeier}, {Kong},
  {Kounkel}, {Krishnarao}, {Lacerna}, {Lan}, {Lane}, {Law}, {Le Goff}, {Leung},
  {Lewis}, {Li}, {Lian}, {Lin}, {Long}, {Longa-Pe{\~n}a}, {Lundgren}, {Lyke},
  {Ted Mackereth}, {MacLeod}, {Majewski}, {Manchado}, {Maraston}, {Martini},
  {Masseron}, {Masters}, {Mathur}, {McDermid}, {Merloni}, {Merrifield},
  {M{\'e}sz{\'a}ros}, {Miglio}, {Minniti}, {Minsley}, {Miyaji}, {Mohammad},
  {Mosser}, {Mueller}, {Muna}, {Mu{\~n}oz-Guti{\'e}rrez}, {Myers}, {Nadathur},
  {Nair}, {Nandra}, {do Nascimento}, {Nevin}, {Newman}, {Nidever}, {Nitschelm},
  {Noterdaeme}, {O'Connell}, {Olmstead}, {Oravetz}, {Oravetz}, {Osorio},
  {Pace}, {Padilla}, {Palanque-Delabrouille}, {Palicio}, {Pan}, {Pan},
  {Parker}, {Paviot}, {Peirani}, {Pe{\~n}a Ram{\'r}ez}, {Penny}, {Percival},
  {Perez-Fournon}, {P{\'e}rez-R{\`a}fols}, {Petitjean}, {Pieri},
  {Pinsonneault}, {Poovelil}, {Povick}, {Prakash}, {Price-Whelan}, {Raddick},
  {Raichoor}, {Ray}, {Rembold}, {Rezaie}, {Riffel}, {Riffel}, {Rix}, {Robin},
  {Roman-Lopes}, {Rom{\'a}n-Z{\'u}{\~n}iga}, {Rose}, {Ross}, {Rossi},
  {Rowlands}, {Rubin}, {Salvato}, {S{\'a}nchez}, {S{\'a}nchez-Menguiano},
  {S{\'a}nchez-Gallego}, {Sayres}, {Schaefer}, {Schiavon}, {Schimoia},
  {Schlafly}, {Schlegel}, {Schneider}, {Schultheis}, {Schwope}, {Seo},
  {Serenelli}, {Shafieloo}, {Shamsi}, {Shao}, {Shen}, {Shetrone}, {Shirley},
  {Silva Aguirre}, {Simon}, {Skrutskie}, {Slosar}, {Smethurst}, {Sobeck},
  {Sodi}, {Souto}, {Stark}, {Stassun}, {Steinmetz}, {Stello}, {Stermer},
  {Storchi-Bergmann}, {Streblyanska}, {Stringfellow}, {Stutz}, {Su{\'a}rez},
  {Sun}, {Taghizadeh-Popp}, {Talbot}, {Tayar}, {Thakar}, {Theriault}, {Thomas},
  {Thomas}, {Tinker}, {Tojeiro}, {Toledo}, {Tremonti}, {Troup}, {Tuttle},
  {Unda-Sanzana}, {Valentini}, {Vargas-Gonz{\'a}lez}, {Vargas-Maga{\~n}a},
  {V{\'a}zquez-Mata}, {Vivek}, {Wake}, {Wang}, {Weaver}, {Weijmans}, {Wild},
  {Wilson}, {Wilson}, {Wolthuis}, {Wood-Vasey}, {Yan}, {Yang}, {Y{\`e}che},
  {Zamora}, {Zarrouk}, {Zasowski}, {Zhang}, {Zhao}, {Zhao}, {Zheng}, {Zheng},
  {Zhu}, \& {Zou}}]{sdssdr16}
{Ahumada}, R., {Allende Prieto}, C., {Almeida}, A., {et~al.} 2020, \apjs, 249,
  3

\bibitem[{Bai {et~al.}(2018)Bai, Liu, Wang, \& Yang}]{bai19}
Bai, Y., Liu, J., Wang, S., \& Yang, F. 2018, The Astronomical Journal, 157, 9

\bibitem[{Ball {et~al.}(2006)Ball, Brunner, Myers, \& Tcheng}]{ball06}
Ball, N.~M., Brunner, R.~J., Myers, A.~D., \& Tcheng, D. 2006, The
  Astrophysical Journal, 650, 497–509

\bibitem[{{Baqui, P. O.} {et~al.}(2021){Baqui, P. O.}, {Marra, V.}, {Casarini,
  L.}, {Angulo, R.}, {D\'{\i}az-Garc\'{\i}a, L. A.}, {Hern\'andez-Monteagudo,
  C.}, {Lopes, P. A. A.}, {L\'opez-Sanjuan, C.}, {Muniesa, D.}, {Placco, V.
  M.}, {Quartin, M.}, {Queiroz, C.}, {Sobral, D.}, {Solano, E.}, {Tempel, E.},
  {Varela, J.}, {V\'{\i}lchez, J. M.}, {Abramo, R.}, {Alcaniz, J.}, {Benitez,
  N.}, {Bonoli, S.}, {Carneiro, S.}, {Cenarro, A. J.}, {Crist\'obal-Hornillos,
  D.}, {de Amorim, A. L.}, {de Oliveira, C. M.}, {Dupke, R.}, {Ederoclite, A.},
  {Gonz\'alez Delgado, R. M.}, {Mar\'{\i}n-Franch, A.}, {Moles, M.}, {V\'azquez
  Rami\'o, H.}, {Sodr\'e, L.}, \& {Taylor, K.}}]{minij21}
{Baqui, P. O.}, {Marra, V.}, {Casarini, L.}, {et~al.} 2021, A\&A, 645, A87

\bibitem[{{Ben{\'{\i}}tez} {et~al.}(2014){Ben{\'{\i}}tez}, {Dupke}, {Moles},
  {Sodre}, {Cenarro}, {Marin-Franch}, {Taylor}, {Cristobal}, {Fernandez-Soto},
  {Mendes de Oliveira}, {Cepa-Nogue}, {Abramo}, {Alcaniz}, {Overzier},
  {Hernandez-Monteagudo}, {Alfaro}, {Kanaan}, {Carvano}, {Reis}, {Martinez
  Gonzalez}, {Ascaso}, {Ballesteros}, {Xavier}, {Varela}, {Ederoclite},
  {Vazquez Ramio}, {Broadhurst}, {Cypriano}, {Angulo}, {Diego}, {Zandivarez},
  {Diaz}, {Melchior}, {Umetsu}, {Spinelli}, {Zitrin}, {Coe}, {Yepes}, {Vielva},
  {Sahni}, {Marcos-Caballero}, {Shu Kitaura}, {Maroto}, {Masip}, {Tsujikawa},
  {Carneiro}, {Gonzalez Nuevo}, {Carvalho}, {Reboucas}, {Carvalho}, {Abdalla},
  {Bernui}, {Pigozzo}, {Ferreira}, {Chandrachani Devi}, {Bengaly}, {Campista},
  {Amorim}, {Asari}, {Bongiovanni}, {Bonoli}, {Bruzual}, {Cardiel}, {Cava},
  {Cid Fernandes}, {Coelho}, {Cortesi}, {Delgado}, {Diaz Garcia}, {Espinosa},
  {Galliano}, {Gonzalez-Serrano}, {Falcon-Barroso}, {Fritz}, {Fernandes},
  {Gorgas}, {Hoyos}, {Jimenez-Teja}, {Lopez-Aguerri}, {Lopez-San Juan},
  {Mateus}, {Molino}, {Novais}, {OMill}, {Oteo}, {Perez-Gonzalez}, {Poggianti},
  {Proctor}, {Ricciardelli}, {Sanchez-Blazquez}, {Storchi-Bergmann}, {Telles},
  {Schoennell}, {Trujillo}, {Vazdekis}, {Viironen}, {Daflon},
  {Aparicio-Villegas}, {Rocha}, {Ribeiro}, {Borges}, {Martins}, {Marcolino},
  {Martinez-Delgado}, {Perez-Torres}, {Siffert}, {Calvao}, {Sako}, {Kessler},
  {Alvarez-Candal}, {De Pra}, {Roig}, {Lazzaro}, {Gorosabel}, {Lopes de
  Oliveira}, {Lima-Neto}, {Irwin}, {Liu}, {Alvarez}, {Balmes}, {Chueca},
  {Costa-Duarte}, {da Costa}, {Dantas}, {Diaz}, {Fabregat}, {Ferrari},
  {Gavela}, {Gracia}, {Gruel}, {Gutierrez}, {Guzman}, {Hernandez-Fernandez},
  {Herranz}, {Hurtado-Gil}, {Jablonsky}, {Laporte}, {Le Tiran}, {Licandro},
  {Lima}, {Martin}, {Martinez}, {Montero}, {Penteado}, {Pereira}, {Peris},
  {Quilis}, {Sanchez-Portal}, {Soja}, {Solano}, {Torra}, \&
  {Valdivielso}}]{jpas}
{Ben{\'{\i}}tez}, N., {Dupke}, R., {Moles}, M., {et~al.} 2014,
  [ArXiv:1403.5237] [\eprint[arXiv]{1403.5237}]

\bibitem[{{Bertin} \& {Arnouts}(1996)}]{sextractor}
{Bertin}, E. \& {Arnouts}, S. 1996, \aaps, 117, 393

\bibitem[{{Bonoli} {et~al.}(2021){Bonoli}, {Mar{\'\i}n-Franch}, {Varela},
  {V{\'a}zquez Rami{\'o}}, {Abramo}, {Cenarro}, {Dupke}, {V{\'\i}lchez},
  {Crist{\'o}bal-Hornillos}, {Gonz{\'a}lez Delgado},
  {Hern{\'a}ndez-Monteagudo}, {L{\'o}pez-Sanjuan}, {Muniesa}, {Civera},
  {Ederoclite}, {Hern{\'a}n-Caballero}, {Marra}, {Baqui}, {Cortesi},
  {Cypriano}, {Daflon}, {de Amorim}, {D{\'\i}az-Garc{\'\i}a}, {Diego},
  {Mart{\'\i}nez-Solaeche}, {P{\'e}rez}, {Placco}, {Prada}, {Queiroz},
  {Alcaniz}, {Alvarez-Candal}, {Cepa}, {Maroto}, {Roig}, {Siffert}, {Taylor},
  {Benitez}, {Moles}, {Sodr{\'e}}, {Carneiro}, {Mendes de Oliveira}, {Abdalla},
  {Angulo}, {Aparicio Resco}, {Balaguera-Antol{\'\i}nez}, {Ballesteros},
  {Brito-Silva}, {Broadhurst}, {Carrasco}, {Castro}, {Cid Fernandes}, {Coelho},
  {de Melo}, {Doubrawa}, {Fernandez-Soto}, {Ferrari}, {Finoguenov},
  {Garc{\'\i}a-Benito}, {Iglesias-P{\'a}ramo}, {Jim{\'e}nez-Teja}, {Kitaura},
  {Laur}, {Lopes}, {Lucatelli}, {Mart{\'\i}nez}, {Maturi}, {Overzier},
  {Pigozzo}, {Quartin}, {Rodr{\'\i}guez-Mart{\'\i}n}, {Salzano}, {Tamm},
  {Tempel}, {Umetsu}, {Valdivielso}, {von Marttens}, {Zitrin},
  {D{\'\i}az-Mart{\'\i}n}, {L{\'o}pez-Alegre}, {L{\'o}pez-Sainz},
  {Yanes-D{\'\i}az}, {Rueda-Teruel}, {Rueda-Teruel}, {Abril Iba{\~n}ez}, {L
  Ant{\'o}n Bravo}, {Bello Ferrer}, {Bielsa}, {Casino}, {Castillo}, {Chueca},
  {Cuesta}, {Garzar{\'a}n Calderaro}, {Iglesias-Marzoa}, {{\'I}niguez},
  {Lamadrid Gutierrez}, {Lopez-Martinez}, {Lozano-P{\'e}rez}, {Ma{\'\i}cas
  Sacrist{\'a}n}, {Molina-Ib{\'a}{\~n}ez}, {Moreno-Signes}, {Rodr{\'\i}guez
  Llano}, {Royo Navarro}, {Tilve Rua}, {Andrade}, {Alfaro}, {Akras},
  {Arnalte-Mur}, {Ascaso}, {Barbosa}, {Beltr{\'a}n Jim{\'e}nez}, {Benetti},
  {Bengaly}, {Bernui}, {Blanco-Pillado}, {Borges Fernandes}, {Bregman},
  {Bruzual}, {Calderone}, {Carvano}, {Casarini}, {Chaves-Montero},
  {Chies-Santos}, {Coutinho de Carvalho}, {Dimauro}, {Duarte Puertas},
  {Figueruelo}, {Gonz{\'a}lez-Serrano}, {Guerrero}, {Gurung-L{\'o}pez},
  {Herranz}, {Huertas-Company}, {Irwin}, {Izquierdo-Villalba}, {Kanaan},
  {Kehrig}, {Kirkpatrick}, {Lim}, {Lopes}, {Lopes de Oliveira},
  {Marcos-Caballero}, {Mart{\'\i}nez-Delgado}, {Mart{\'\i}nez-Gonz{\'a}lez},
  {Mart{\'\i}nez-Somonte}, {Oliveira}, {Orsi}, {Penna-Lima}, {Reis}, {Spinoso},
  {Tsujikawa}, {Vielva}, {Vitorelli}, {Xia}, {Yuan}, {Arroyo-Polonio},
  {Dantas}, {Galarza}, {Gon{\c{c}}alves}, {Gon{\c{c}}alves}, {Gonzalez},
  {Gonzalez}, {Greisel}, {Jim{\'e}nez-Esteban}, {Landim}, {Lazzaro}, {Magris},
  {Monteiro-Oliveira}, {Pereira}, {Rebou{\c{c}}as}, {Rodriguez-Espinosa},
  {Santos da Costa}, \& {Telles}}]{minijpas}
{Bonoli}, S., {Mar{\'\i}n-Franch}, A., {Varela}, J., {et~al.} 2021, \aap, 653,
  A31

\bibitem[{Boser {et~al.}(1992)Boser, Guyon, \& Vapnik}]{svm1992}
Boser, B.~E., Guyon, I.~M., \& Vapnik, V.~N. 1992, in Proceedings of the Fifth
  Annual Workshop on Computational Learning Theory, COLT '92 (New York, NY,
  USA: Association for Computing Machinery), 144–152

\bibitem[{Bowman \& Azzalini(1997)}]{ker97}
Bowman, A.~W. \& Azzalini, A. 1997, Applied smoothing techniques for data
  analysis: the kernel approach with S-Plus illustrations, Vol.~18 (OUP Oxford)

\bibitem[{Breiman(2001)}]{rf01}
Breiman, L. 2001, Statist. Sci., 16, 199

\bibitem[{Cenarro {et~al.}(2019)Cenarro, Moles, Cristóbal-Hornillos,
  Marín-Franch, Ederoclite, Varela, López-Sanjuan, Hernández-Monteagudo,
  Angulo, Vázquez~Ramió, \& et~al.}]{jplus2019}
Cenarro, A.~J., Moles, M., Cristóbal-Hornillos, D., {et~al.} 2019, Astronomy
  \& Astrophysics, 622, A176

\bibitem[{{Cenarro} {et~al.}(2014){Cenarro}, {Moles}, {Mar{\'{\i}}n-Franch},
  {Crist{\'o}bal-Hornillos}, {Yanes D{\'{\i}}az}, {Ederoclite}, {Varela},
  {V{\'a}zquez-Rami{\'o}}, {Valdivielso}, {Ben{\'{\i}}tez}, {Cepa}, {Dupke},
  {Fern{\'a}ndez-Soto}, {Mendes de Oliveira}, {Sodr{\'e}}, {Taylor},
  {Rueda-Teruel}, {Rueda-Teruel}, {Luis-Simoes}, {Chueca}, {Ant{\'o}n},
  {Bello}, {D{\'{\i}}az-Mart{\'{\i}}n}, {Guill{\'e}n-Civera},
  {Hern{\'a}ndez-Fuertes}, {Iglesias-Marzoa}, {Jim{\'e}nez-Mej{\'{\i}}as},
  {Lasso-Cabrera}, {L{\'o}pez-Alegre}, {L{\'o}pez-Sainz},
  {Rodr{\'{\i}}guez-Hern{\'a}ndez}, {Su{\'a}rez}, {Lamadrid}, {Ma{\'{\i}}cas},
  {Abril-Iba{\~n}ez}, {Tilve}, \& {Rodr{\'{\i}}guez-Llano}}]{oaj}
{Cenarro}, A.~J., {Moles}, M., {Mar{\'{\i}}n-Franch}, A., {et~al.} 2014, in
  \procspie, Vol. 9149, Observatory Operations: Strategies, Processes, and
  Systems V, 91491I

\bibitem[{Cortes \& Vapnik(1995)}]{cortes95}
Cortes, C. \& Vapnik, V. 1995, Machine Learning, 20, 273

\bibitem[{Cover \& Hart(1967)}]{knn67}
Cover, T. \& Hart, P. 1967, IEEE Trans. Inf. Theory, 13, 21

\bibitem[{Cristianini \& Shawe-Taylor(2000)}]{crist00}
Cristianini, N. \& Shawe-Taylor, J. 2000, An Introduction to Support Vector
  Machines and Other Kernel-based Learning Methods (Cambridge University Press)

\bibitem[{{Cui} {et~al.}(2012){Cui}, {Zhao}, {Chu}, {Li}, {Li}, {Zhang}, {Su},
  {Yao}, {Wang}, {Xing}, {Li}, {Zhu}, {Wang}, {Gu}, {Luo}, {Xu}, {Zhang},
  {Liu}, {Zhang}, {Yang}, {Cao}, {Chen}, {Chen}, {Chen}, {Chen}, {Chu}, {Feng},
  {Gong}, {Hou}, {Hu}, {Hu}, {Hu}, {Jia}, {Jiang}, {Jiang}, {Jiang}, {Jin},
  {Li}, {Li}, {Li}, {Liu}, {Liu}, {Lu}, {Mao}, {Men}, {Qi}, {Qi}, {Shi},
  {Tang}, {Tao}, {Wang}, {Wang}, {Wang}, {Wang}, {Wang}, {Wang}, {Wang},
  {Wang}, {Wang}, {Wang}, {Wang}, {Wang}, {Xu}, {Xu}, {Yang}, {Yu}, {Yuan},
  {Yuan}, {Zhai}, {Zhang}, {Zhang}, {Zhang}, {Zhao}, {Zhou}, {Zhou}, {Zhu}, \&
  {Zou}}]{cui12}
{Cui}, X.-Q., {Zhao}, Y.-H., {Chu}, Y.-Q., {et~al.} 2012, Research in Astronomy
  and Astrophysics, 12, 1197

\bibitem[{{De Maesschalck} {et~al.}(2000){De Maesschalck}, Jouan-Rimbaud, \&
  Massart}]{mahal1}
{De Maesschalck}, R., Jouan-Rimbaud, D., \& Massart, D. 2000, Chemometrics and
  Intelligent Laboratory Systems, 50, 1

\bibitem[{{Deng} {et~al.}(2012){Deng}, {Newberg}, {Liu}, {Carlin}, {Beers},
  {Chen}, {Chen}, {Christlieb}, {Grillmair}, {Guhathakurta}, {Han}, {Hou},
  {Lee}, {L{\'e}pine}, {Li}, {Liu}, {Pan}, {Sellwood}, {Wang}, {Wang}, {Yang},
  {Yanny}, {Zhang}, {Zhang}, {Zheng}, \& {Zhu}}]{deng12}
{Deng}, L.-C., {Newberg}, H.~J., {Liu}, C., {et~al.} 2012, Research in
  Astronomy and Astrophysics, 12, 735

\bibitem[{FISHER(1936)}]{fisher}
FISHER, R.~A. 1936, Annals of Eugenics, 7, 179

\bibitem[{Freund \& Schapire(1995)}]{ada}
Freund, Y. \& Schapire, R.~E. 1995, A Decision-Theoretic Generalization of
  on-Line Learning and an Application to Boosting

\bibitem[{guan Wang {et~al.}(1996)guan Wang, qiang Su, quan Chu, Cui, \& nan
  Wang}]{wang96}
guan Wang, S., qiang Su, D., quan Chu, Y., Cui, X., \& nan Wang, Y. 1996, Appl.
  Opt., 35, 5155

\bibitem[{Huchra {et~al.}(2012)Huchra, Macri, Masters, Jarrett, Berlind,
  Calkins, Crook, Cutri, Erdoğdu, Falco, \& et~al.}]{2mrs}
Huchra, J.~P., Macri, L.~M., Masters, K.~L., {et~al.} 2012, The Astrophysical
  Journal Supplement Series, 199, 26

\bibitem[{Jiménez-Teja {et~al.}(2019)Jiménez-Teja, Dupke, Lopes~de Oliveira,
  Xavier, Coelho, Chies-Santos, López-Sanjuan, Alvarez-Candal, Costa-Duarte,
  Telles, \& et~al.}]{jim19}
Jiménez-Teja, Y., Dupke, R.~A., Lopes~de Oliveira, R., {et~al.} 2019,
  Astronomy \& Astrophysics, 622, A183

\bibitem[{{L{\'o}pez-Sanjuan} {et~al.}(2019){L{\'o}pez-Sanjuan}, {V{\'a}zquez
  Rami{\'o}}, {Varela}, {Spinoso}, {Angulo}, {Muniesa}, {Viironen},
  {Crist{\'o}bal-Hornillos}, {Cenarro}, {Ederoclite}, {Mar{\'\i}n-Franch},
  {Moles}, {Ascaso}, {Bonoli}, {Chies-Santos}, {Coelho}, {Costa-Duarte},
  {Cortesi}, {D{\'\i}az-Garc{\'\i}a}, {Dupke}, {Galbany},
  {Hern{\'a}ndez-Monteagudo}, {Logro{\~n}o-Garc{\'\i}a}, {Molino}, {Orsi},
  {Placco}, {Sampedro}, {San Roman}, {Vilella-Rojo}, {Whitten}, {Mendes de
  Oliveira}, \& {Sodr{\'e}}}]{lop19}
{L{\'o}pez-Sanjuan}, C., {V{\'a}zquez Rami{\'o}}, H., {Varela}, J., {et~al.}
  2019, \aap, 622, A177

\bibitem[{Luo {et~al.}(2012)Luo, Zhang, Zhao, Zhao, Cui, Li, Chu, Shi, Wang,
  Zhang, Bai, Chen, Wang, Guo, Chen, Du, Kong, Lei, Li, Song, Wu, Zhang, Zhou,
  Zuo, Du, He, Hou, Dong, Li, Li, Li, Song, Tian, Wang, Wu, Yang, Yuan, Cao,
  Chen, Chen, Chen, Chu, Feng, Gong, Gu, Hou, Huo, Hu, Hu, Hu, Jia, Jiang,
  Jiang, Jiang, Jin, Li, Li, Li, Li, Li, Liu, Liu, Liu, Lu, Lu, Luo, Mao, Men,
  Ni, Qi, Qi, Shi, Su, Sun, Su, Tang, Tao, Tu, Wang, Wang, Wang, Wang, Wang,
  Wang, Wang, Wang, Wang, Wang, Wang, Wang, Wang, Wang, Wei, Xue, Xing, Xu, Xu,
  Xu, Yang, Yang, Yao, Yu, Yuan, Zhai, Zhang, Zhang, Zhang, Zhang, Zhang,
  Zhang, Zhao, Zhou, Zhu, Zhu, \& Zou}]{luo12}
Luo, A.-L., Zhang, H.-T., Zhao, Y.-H., {et~al.} 2012, Research in Astronomy and
  Astrophysics, 12, 1243

\bibitem[{Mahalanobis(1936)}]{mahal2}
Mahalanobis, P.~C. 1936, Proceedings of the National Institute of Sciences
  (Calcutta), 2, 49

\bibitem[{{Mar{\'{\i}}n-Franch} {et~al.}(2015){Mar{\'{\i}}n-Franch}, {Taylor},
  {Cenarro}, {Cristobal-Hornillos}, \& {Moles}}]{t80cam}
{Mar{\'{\i}}n-Franch}, A., {Taylor}, K., {Cenarro}, J., {Cristobal-Hornillos},
  D., \& {Moles}, M. 2015, in IAU General Assembly, Vol.~29, 2257381

\bibitem[{Monroe {et~al.}(2016)Monroe, Prochaska, Tejos, Worseck, Hennawi,
  Schmidt, Tumlinson, \& Shen}]{uvqs}
Monroe, T.~R., Prochaska, J.~X., Tejos, N., {et~al.} 2016, The Astronomical
  Journal, 152, 25

\bibitem[{Nogueira-Cavalcante {et~al.}(2019)Nogueira-Cavalcante, Dupke, Coelho,
  Dantas, Gonçalves, Menéndez-Delmestre, Lopes~de Oliveira, Jiménez-Teja,
  López-Sanjuan, Alcaniz, \& et~al.}]{nog19}
Nogueira-Cavalcante, J.~P., Dupke, R., Coelho, P., {et~al.} 2019, Astronomy \&
  Astrophysics, 630, A88

\bibitem[{Quinlan(1986)}]{quin86}
Quinlan, J.~R. 1986, Machine Learning, 1, 81

\bibitem[{Shalev-Shwartz \& Ben-David(2014)}]{shai14}
Shalev-Shwartz, S. \& Ben-David, S. 2014, Understanding Machine Learning: From
  Theory to Algorithms (Cambridge University Press)

\bibitem[{Steinmetz {et~al.}(2020)Steinmetz, Matijevič, Enke, Zwitter,
  Guiglion, McMillan, Kordopatis, Valentini, Chiappini, Casagrande, \&
  et~al.}]{rave2020}
Steinmetz, M., Matijevič, G., Enke, H., {et~al.} 2020, The Astronomical
  Journal, 160, 82

\bibitem[{Stone(1977)}]{knn77}
Stone, C.~J. 1977, The Annals of Statistics, 5, 595

\bibitem[{Su \& Cui(2004)}]{su04}
Su, D.-Q. \& Cui, X.-Q. 2004, Chinese journal of Astronomy and Astrophysics, 4,
  1

\bibitem[{{Taylor}(2005)}]{topcat}
{Taylor}, M.~B. 2005, in Astronomical Society of the Pacific Conference Series,
  Vol. 347, Astronomical Data Analysis Software and Systems XIV, ed.
  P.~{Shopbell}, M.~{Britton}, \& R.~{Ebert}, 29

\bibitem[{{V{\'e}ron-Cetty} \& {V{\'e}ron}(2010)}]{vv13}
{V{\'e}ron-Cetty}, M.~P. \& {V{\'e}ron}, P. 2010, \aap, 518, A10

\bibitem[{Wang(2021)}]{wang20}
Wang, S. 2021, Private communication

\bibitem[{Whitten {et~al.}(2019)Whitten, Placco, Beers, Chies-Santos, Bonatto,
  Varela, Cristóbal-Hornillos, Ederoclite, Masseron, Lee, \& et~al.}]{whit19}
Whitten, D.~D., Placco, V.~M., Beers, T.~C., {et~al.} 2019, Astronomy \&
  Astrophysics, 622, A182

\bibitem[{Yuan(2021)}]{yuanp}
Yuan, H. 2021, In preparation and private communication

\bibitem[{Yuan {et~al.}(2015)Yuan, Liu, Xiang, Huang, Zhang, \& Chen}]{yuan15}
Yuan, H., Liu, X., Xiang, M., {et~al.} 2015, The Astrophysical Journal, 799,
  133

\bibitem[{Zasowski {et~al.}(2013)Zasowski, Johnson, Frinchaboy, Majewski,
  Nidever, Pinto, Girardi, Andrews, Chojnowski, Cudworth, \& et~al.}]{zas13}
Zasowski, G., Johnson, J.~A., Frinchaboy, P.~M., {et~al.} 2013, The
  Astronomical Journal, 146, 81

\bibitem[{Zhao {et~al.}(2012)Zhao, Zhao, Chu, Jing, \& Deng}]{zhao12}
Zhao, G., Zhao, Y.-H., Chu, Y.-Q., Jing, Y.-P., \& Deng, L.-C. 2012, Research
  in Astronomy and Astrophysics, 12, 723

\end{thebibliography}

\begin{appendix}
\section{Density contours of the sample set}
\label{app}
We present all 12 three-dimensional contours of the predictions.
\newpage
\newpage

\begin{figure*}
    \centering 
    
    \includegraphics[width=0.7\textwidth]{b1.pdf}
    \includegraphics[width=0.7\textwidth]{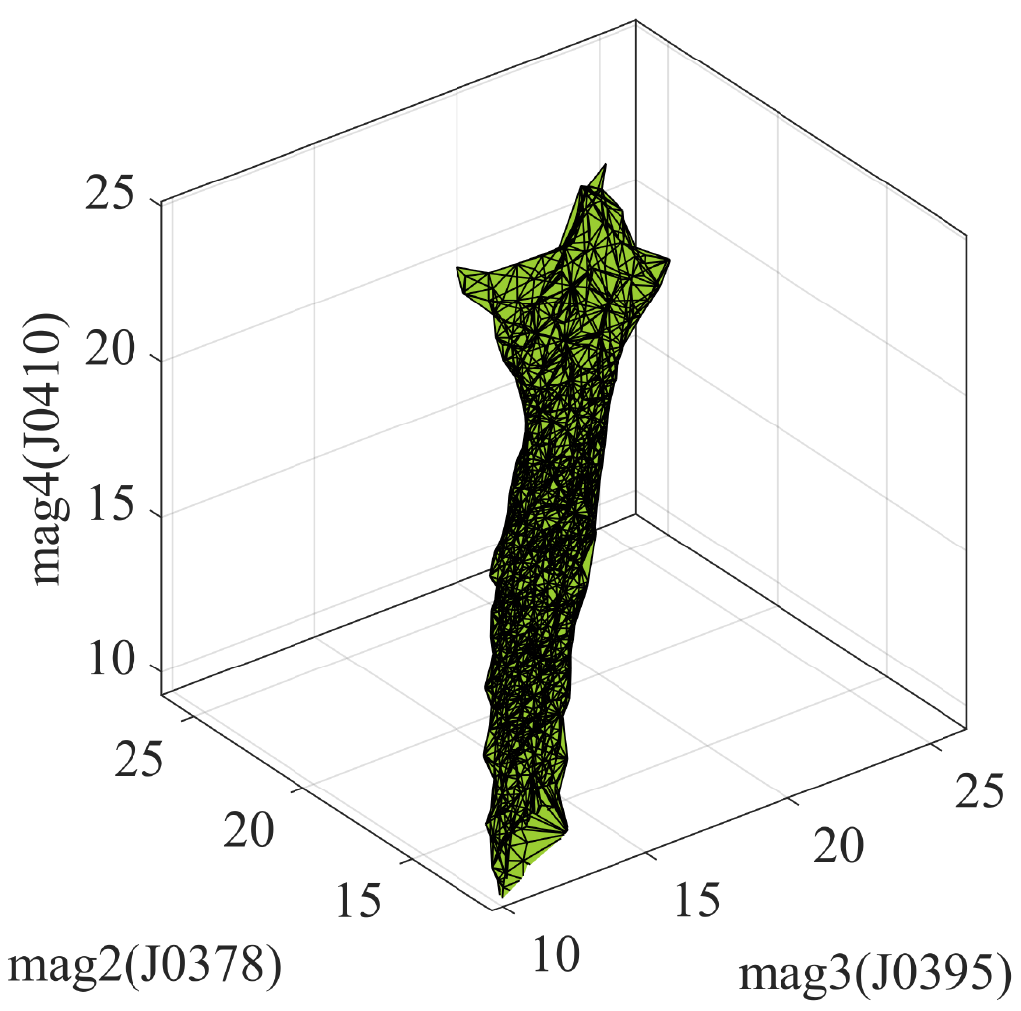}
    \caption{First two contours for extrapolation constraining.}
\end{figure*}

\begin{figure*}
    \centering 
    \includegraphics[width=0.7\textwidth]{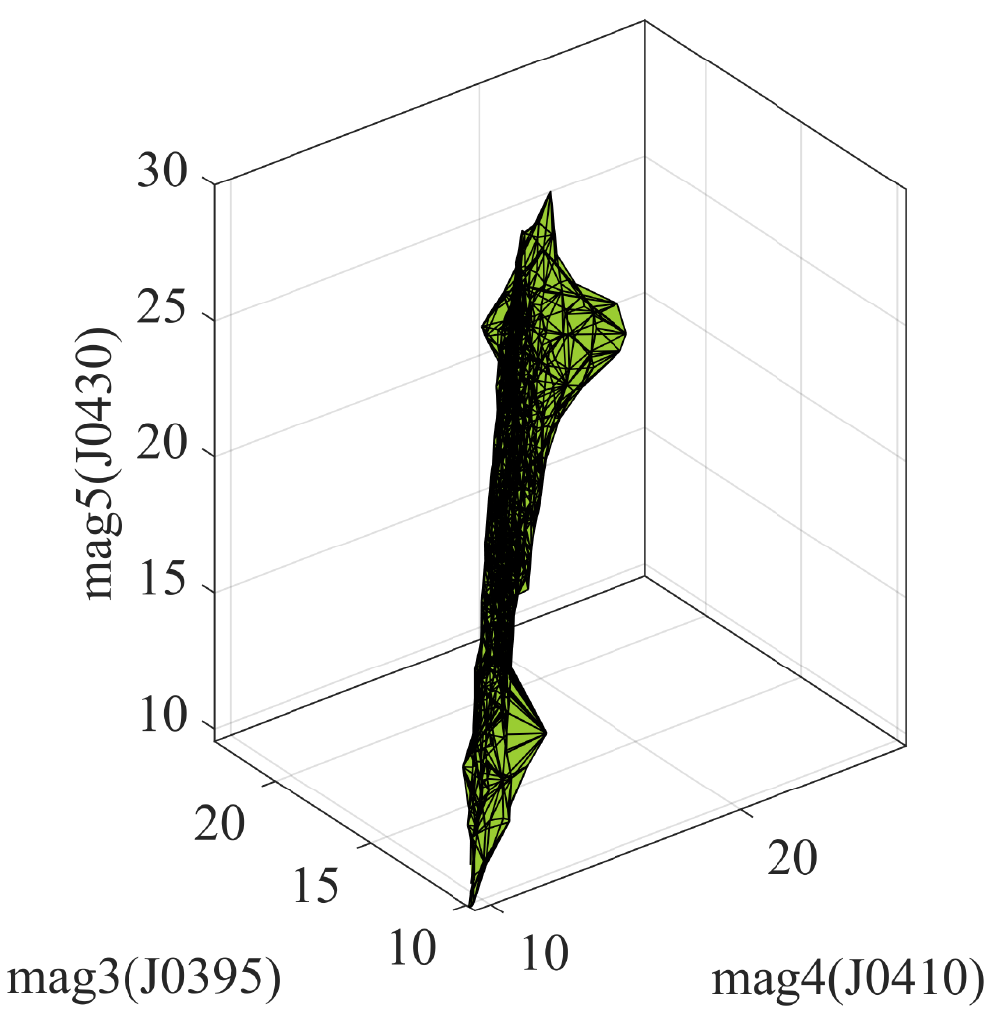}
    \includegraphics[width=0.7\textwidth]{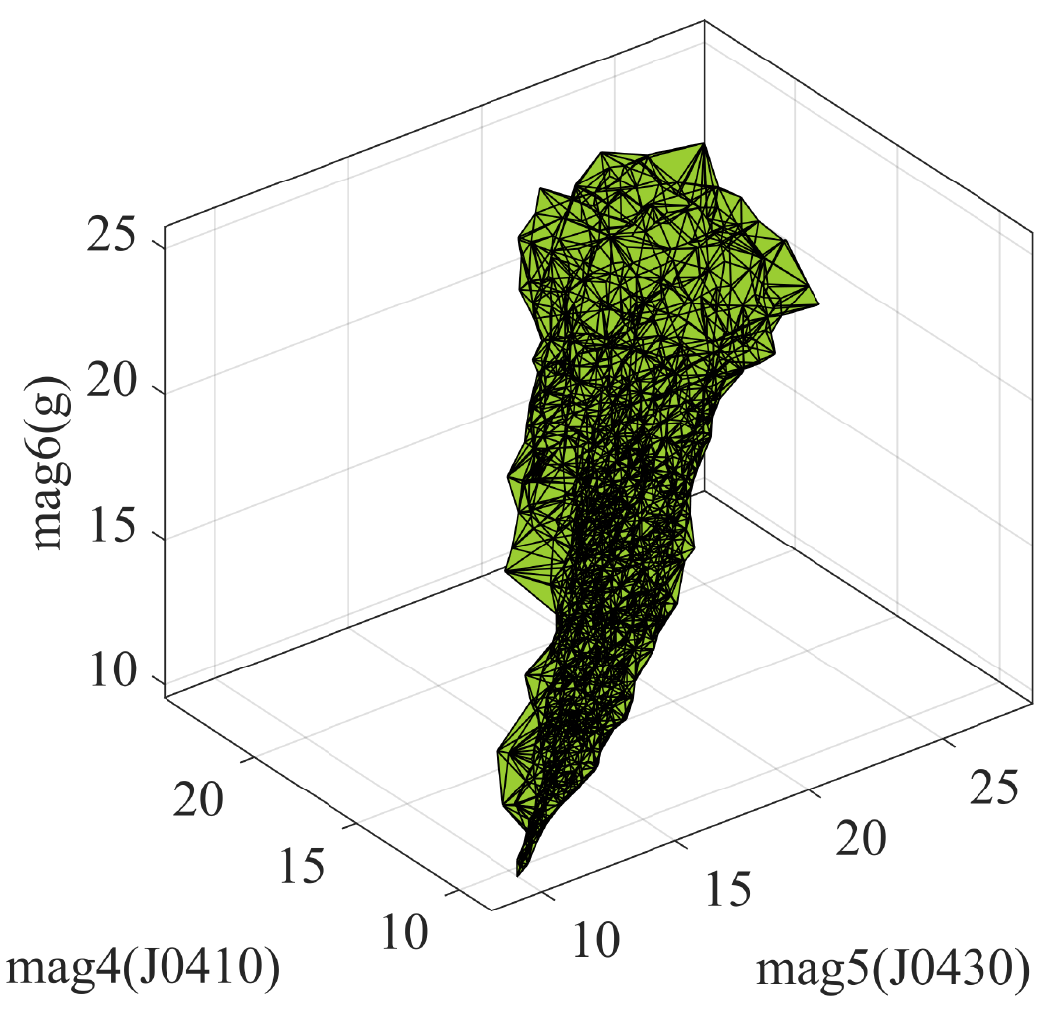}
    \caption{Third and fourth contour for extrapolation constraining.}
\end{figure*}

\begin{figure*}
    \centering 
    \includegraphics[width=0.7\textwidth]{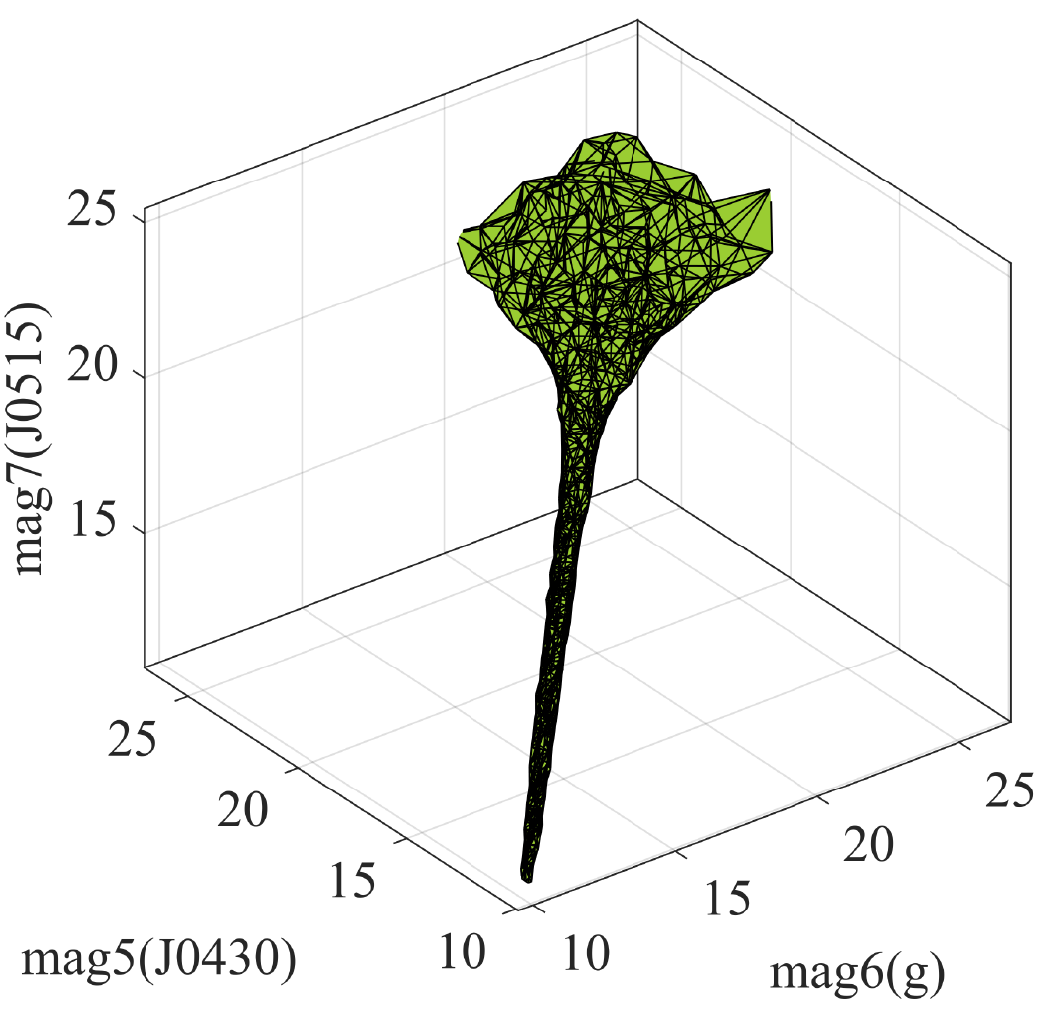}
    \includegraphics[width=0.7\textwidth]{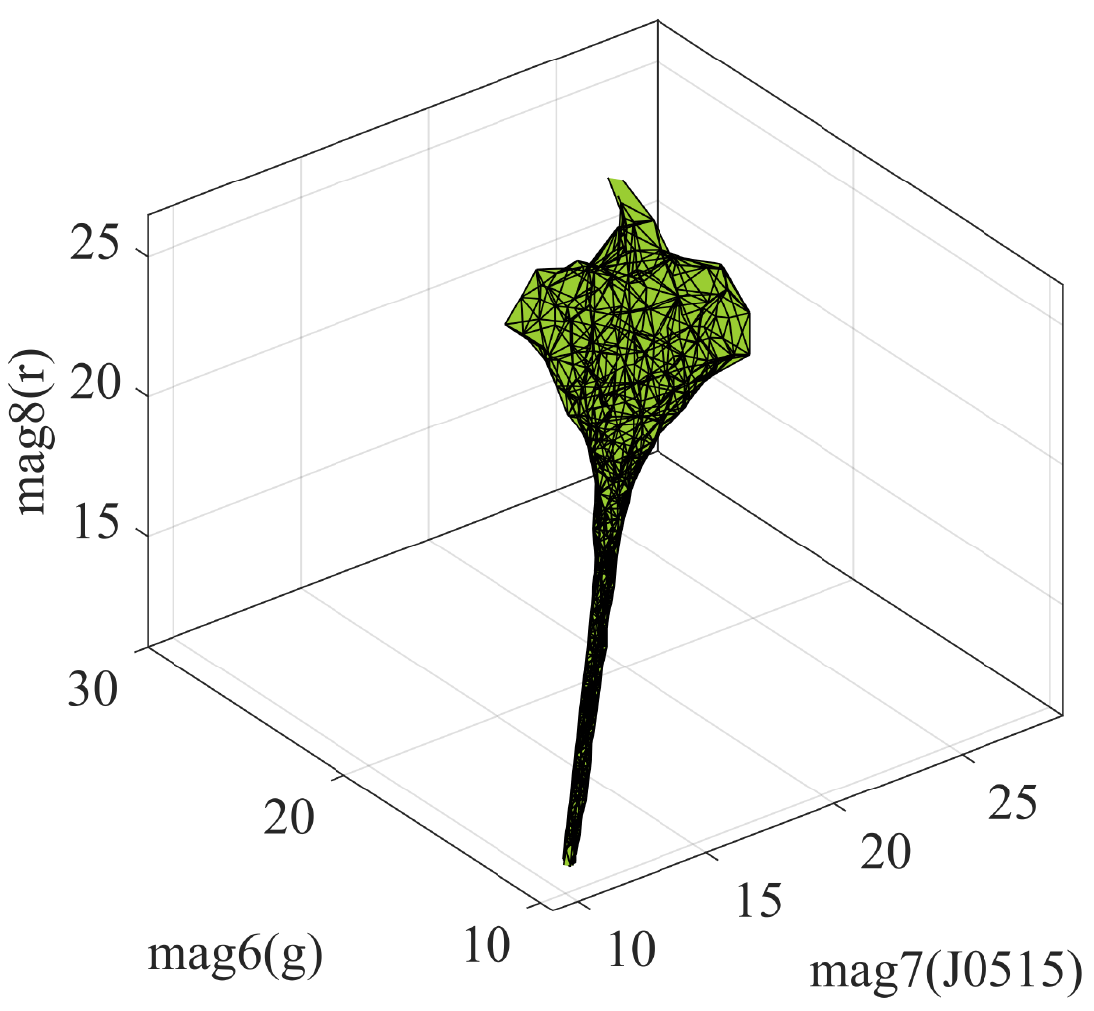}
    \caption{Fifth and sixth contour for extrapolation constraining.}
\end{figure*}

\begin{figure*}
    \centering 
    \includegraphics[width=0.7\textwidth]{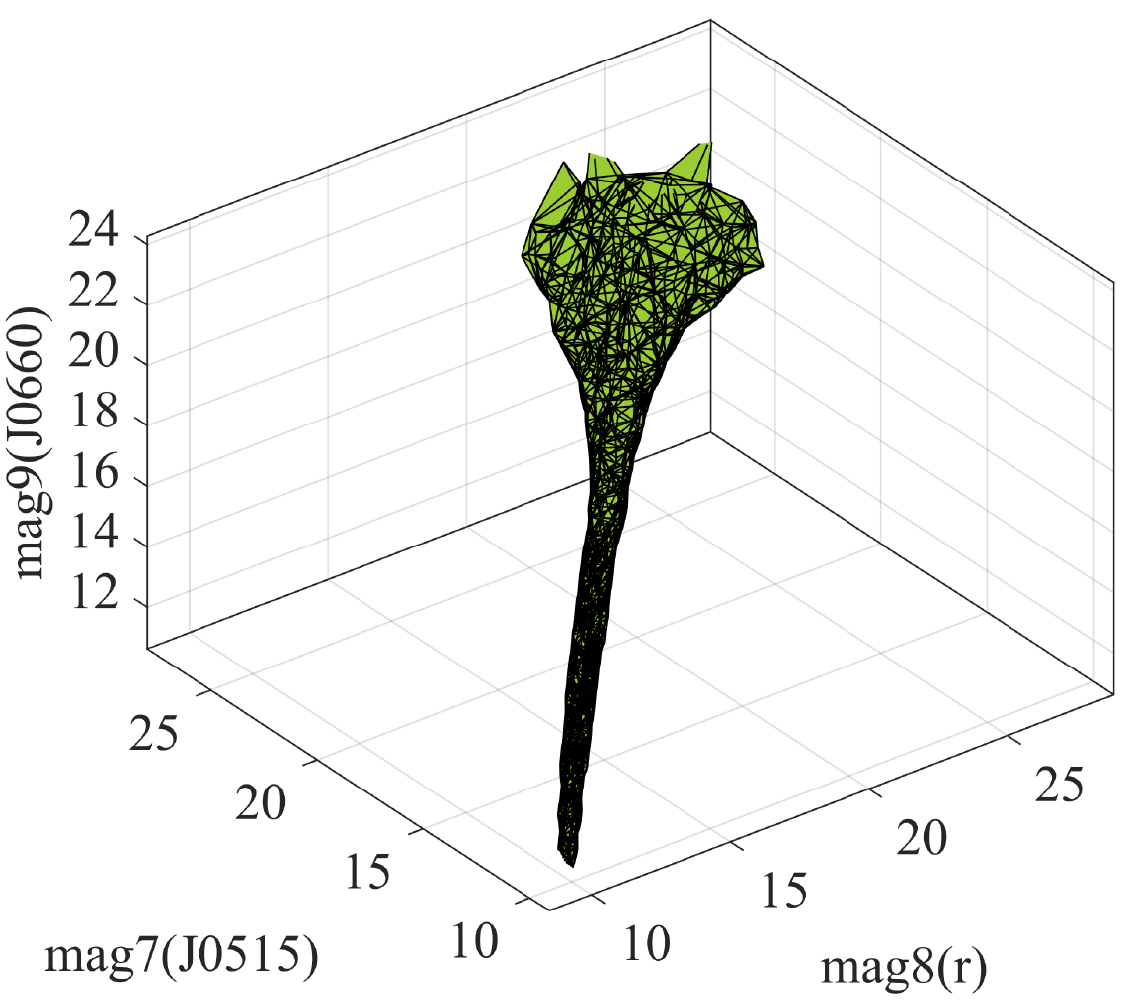}
    \includegraphics[width=0.7\textwidth]{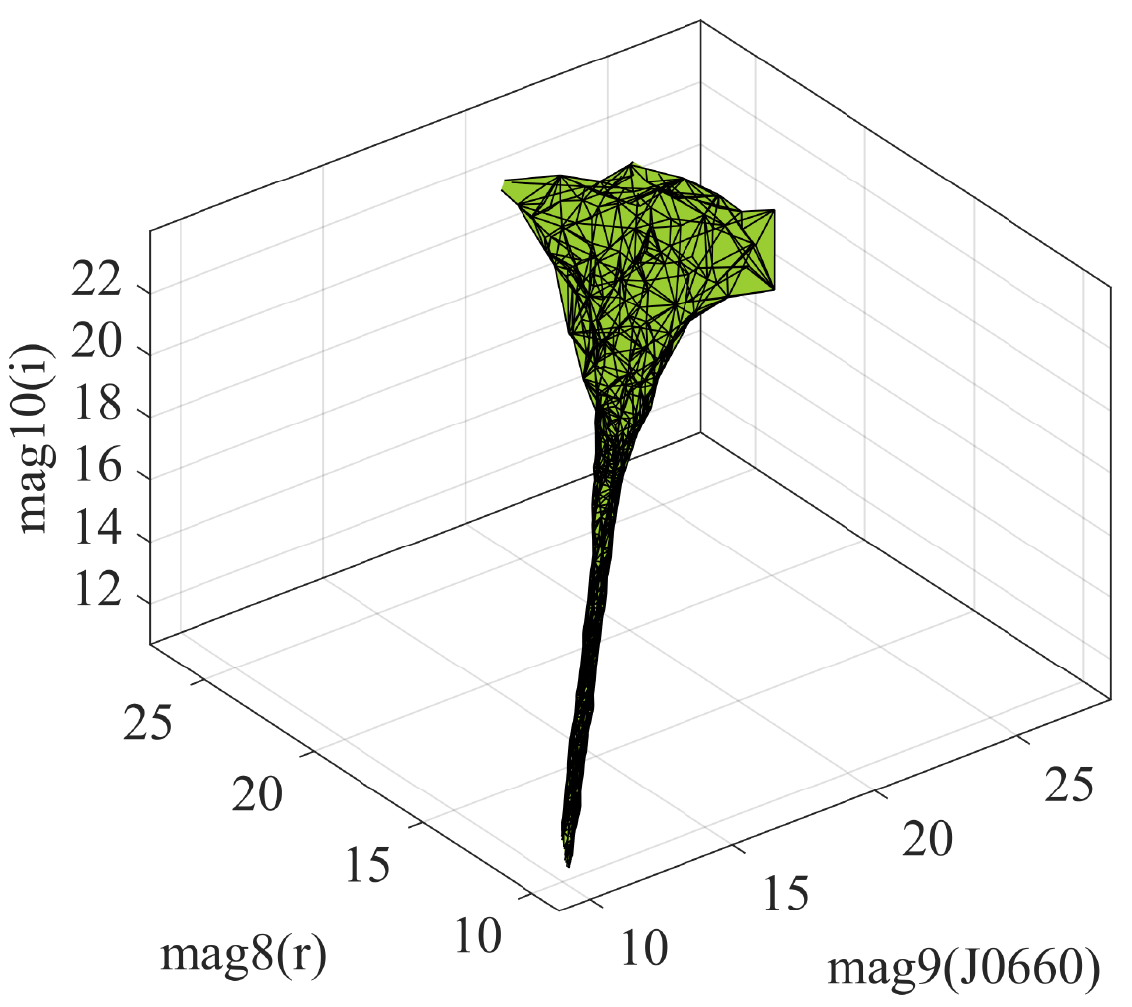}
    \caption{Seventh and eighth contour for extrapolation constraining.}
\end{figure*}

\begin{figure*}
    \centering 
    \includegraphics[width=0.7\textwidth]{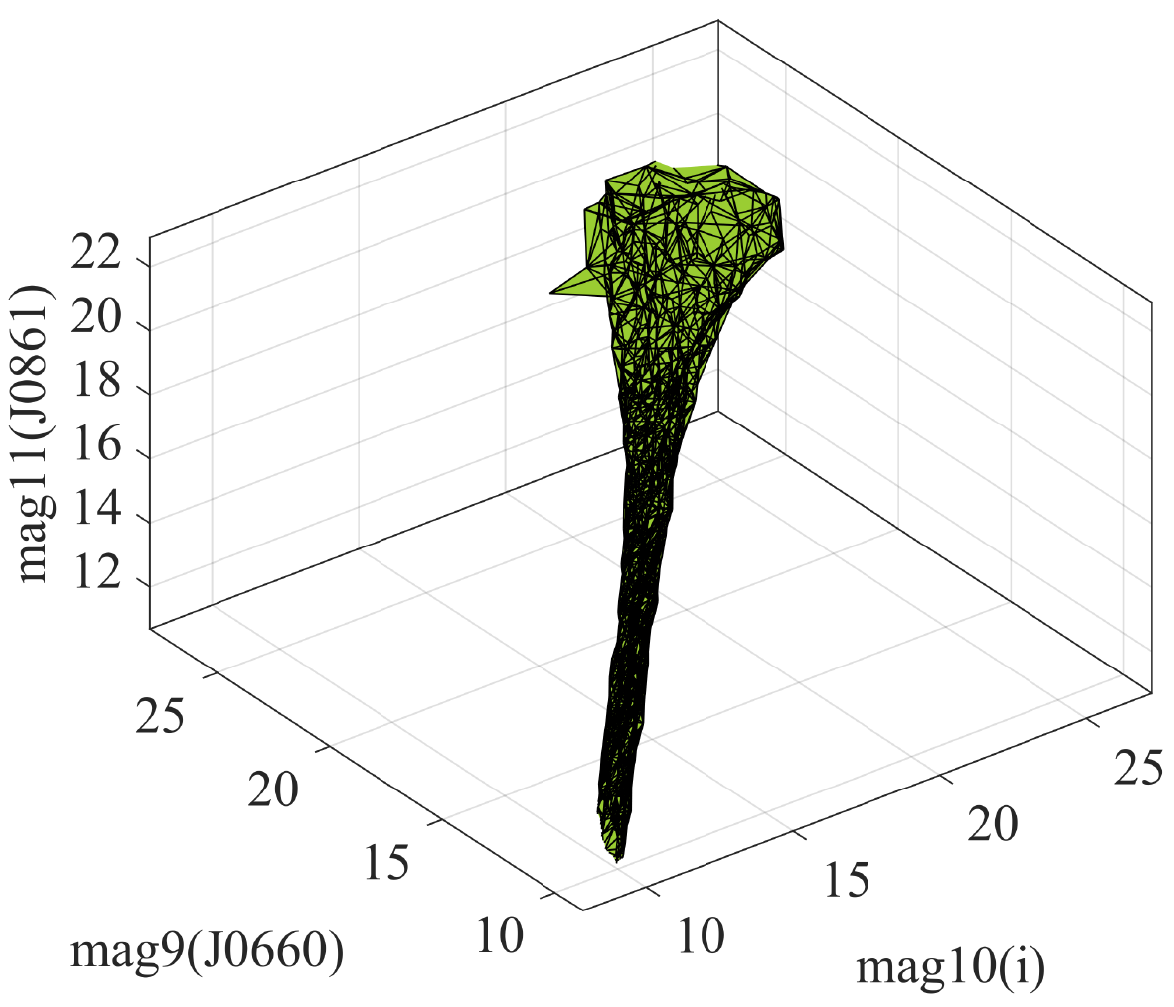}
    \includegraphics[width=0.7\textwidth]{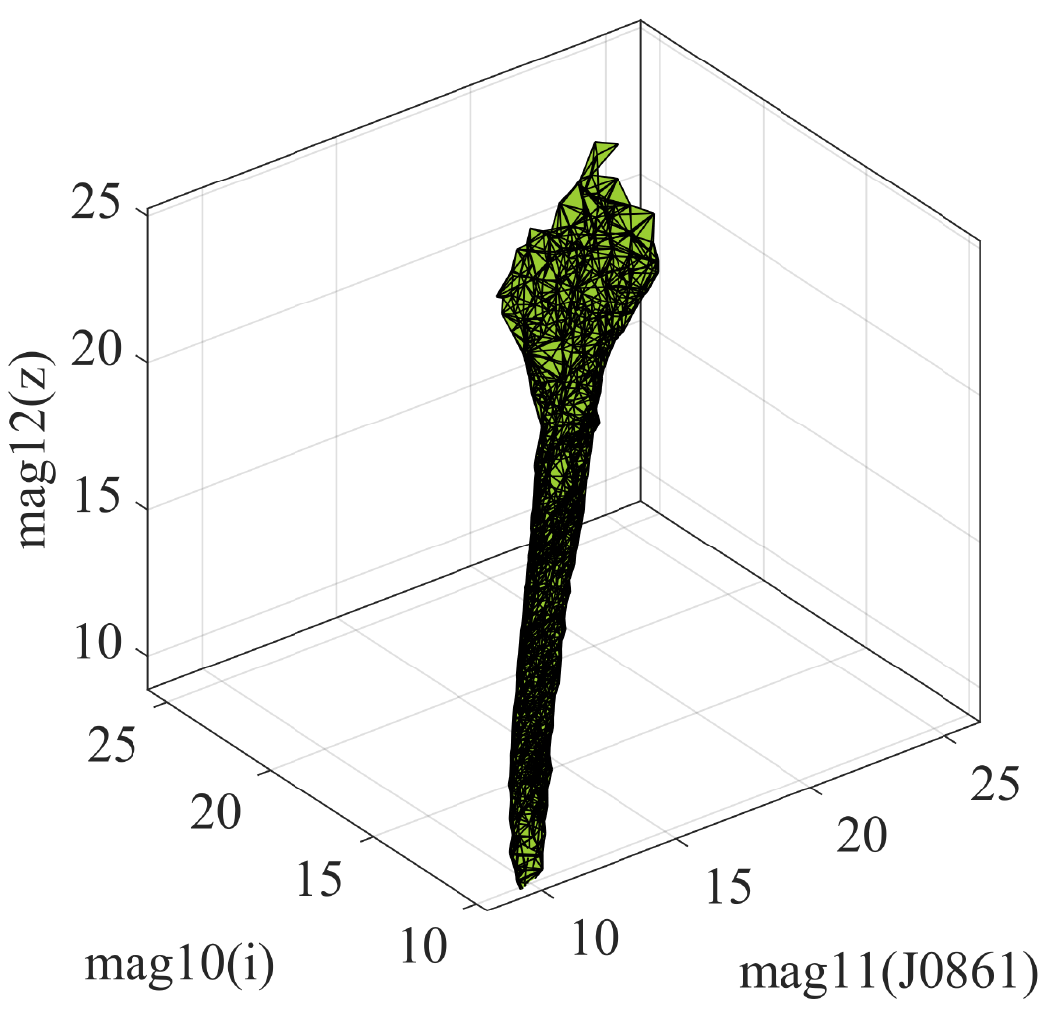}
    \caption{Ninth  and tenth contour for extrapolation constraining.}
\end{figure*}

\begin{figure*}
    \centering 
    \includegraphics[width=0.7\textwidth]{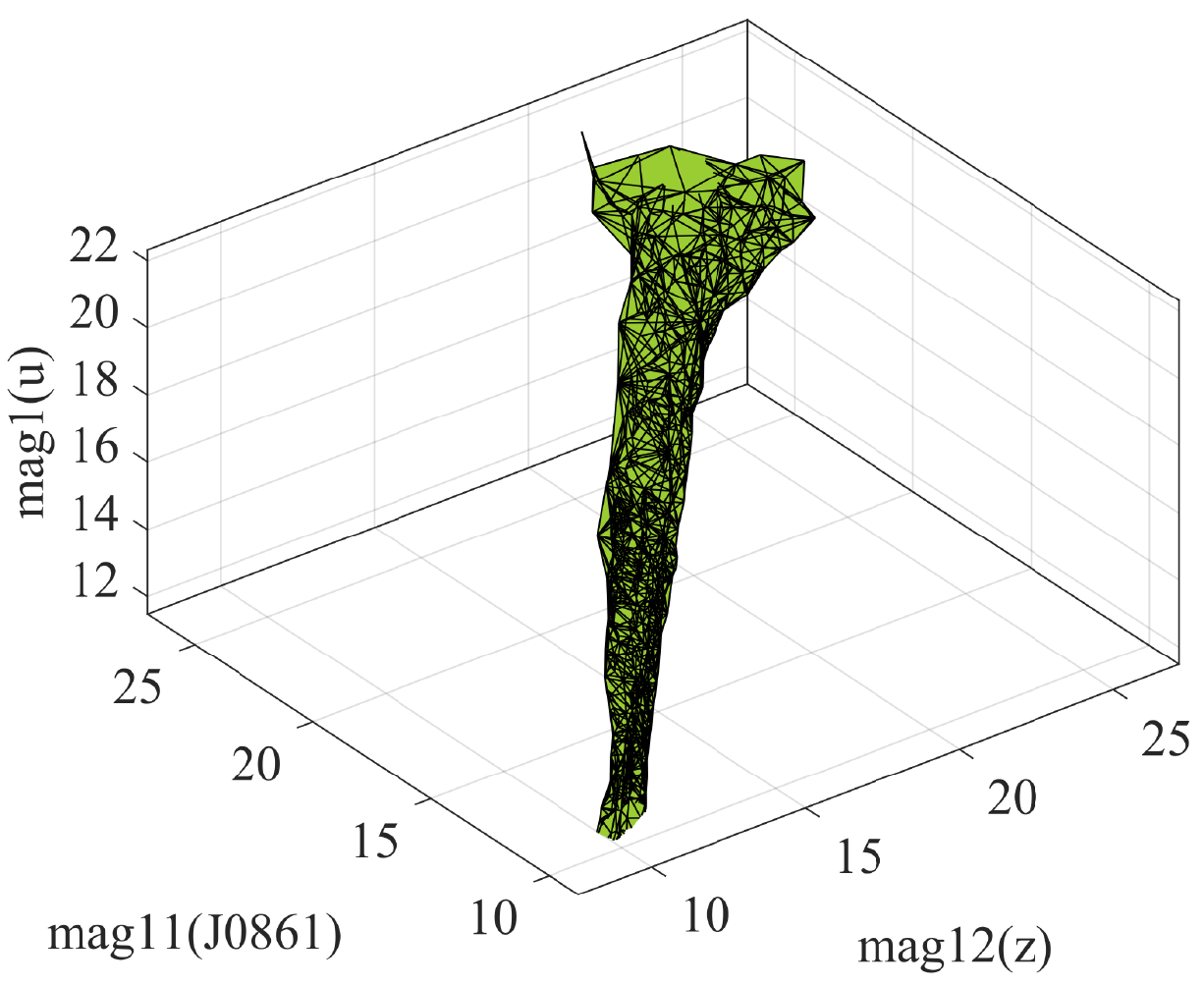}
    \includegraphics[width=0.9\textwidth]{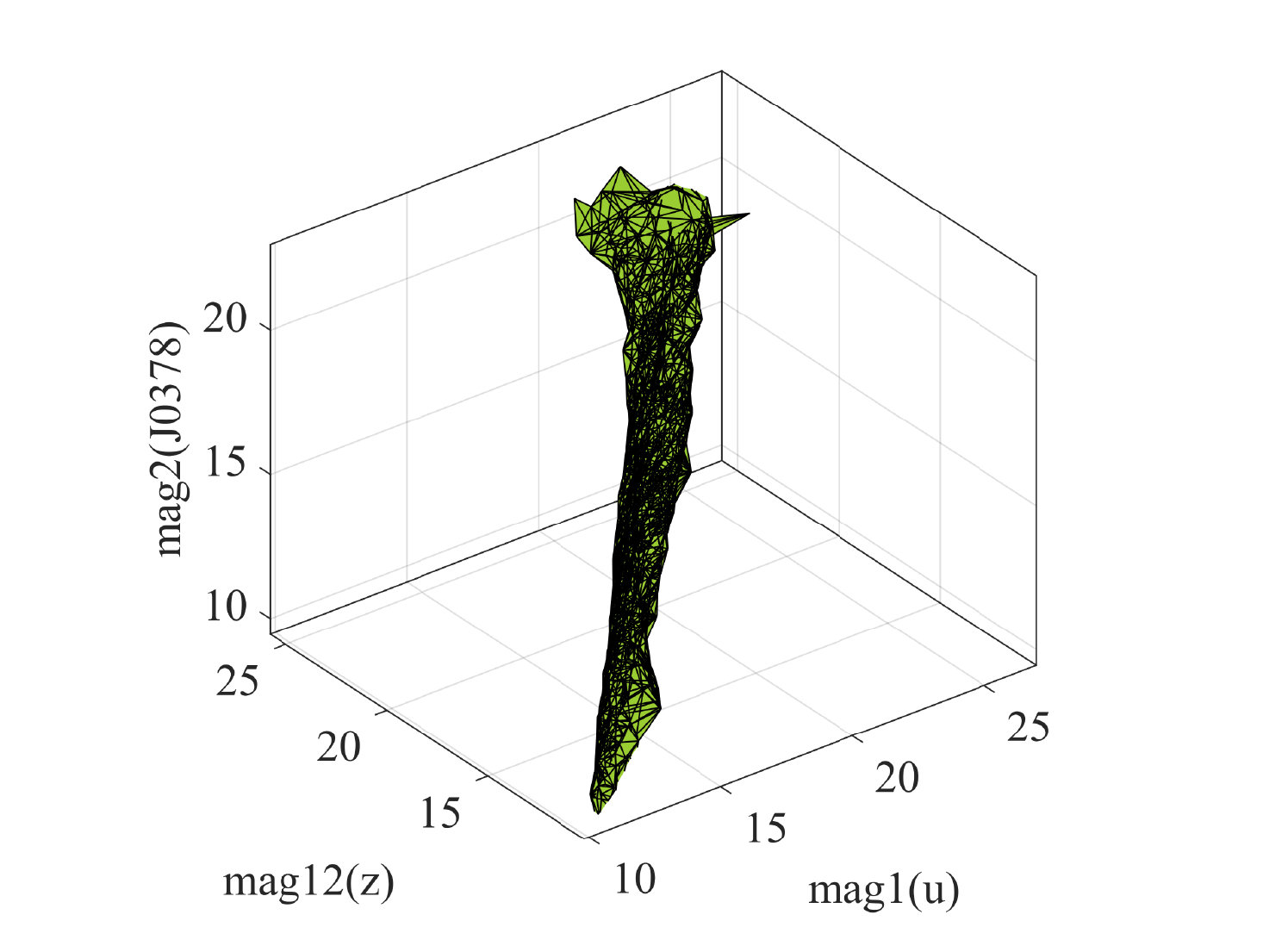}
    \caption{Last two contours for extrapolation constraining.}
\end{figure*}

\FloatBarrier

\section{Magnitude distributions}
\label{appb}
We present the magnitude distributions for each class, magnitude, and for both samples and interpolations. The red line indicates STAR, the green is for GALAXY, and the blue is for QSO. 

\begin{figure*}
\centering 
\includegraphics[width= \textwidth]{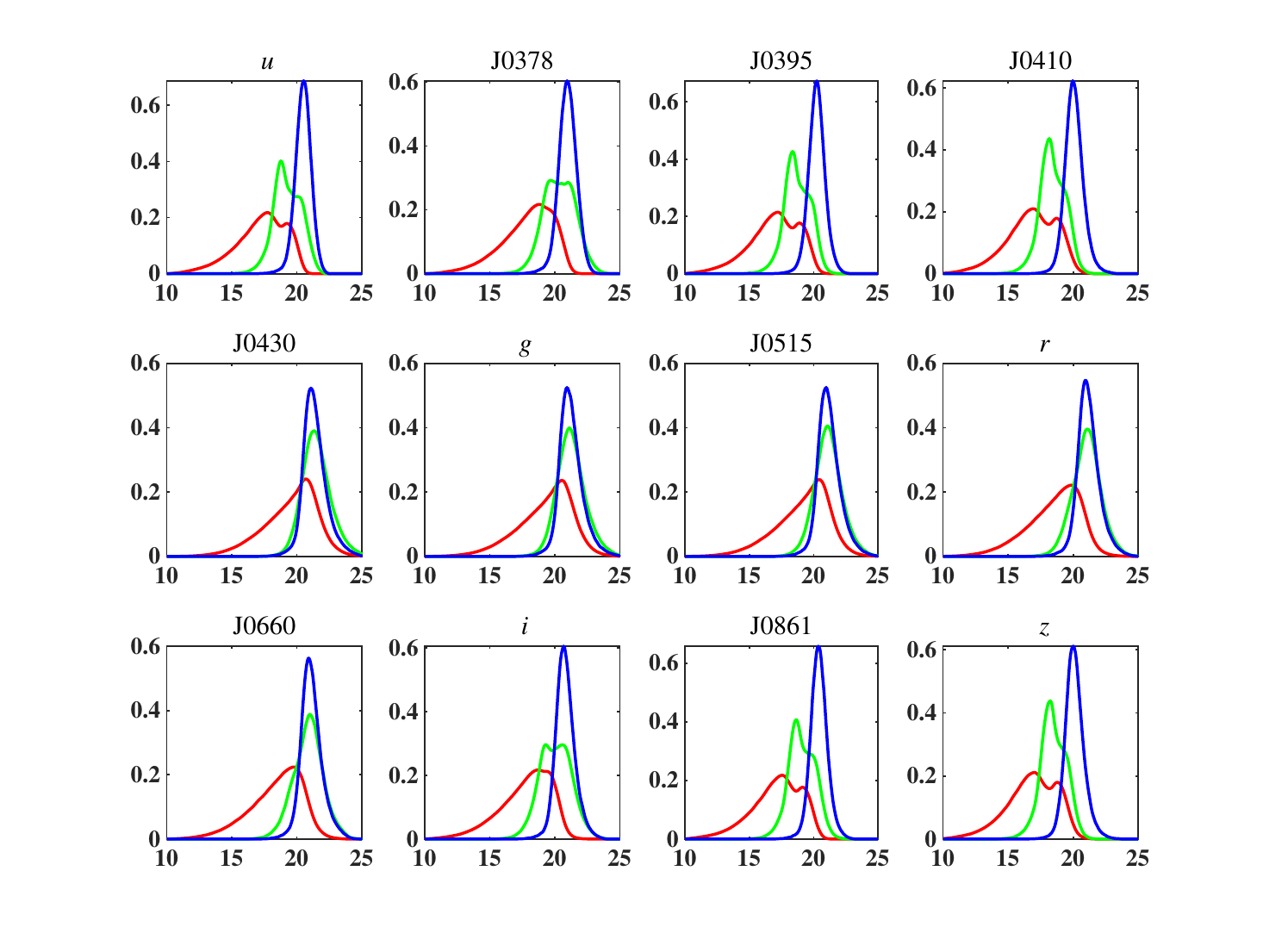}
\caption{Magnitude distributions for the interpolation objects. The STARs are red, the GALAXYs are green, and the QSOs are blue. The x-axis shows the magnitude, and the y-axis shows the probability \label{magb1}}
\end{figure*}

\begin{figure*}
\centering 
\includegraphics[width= \textwidth]{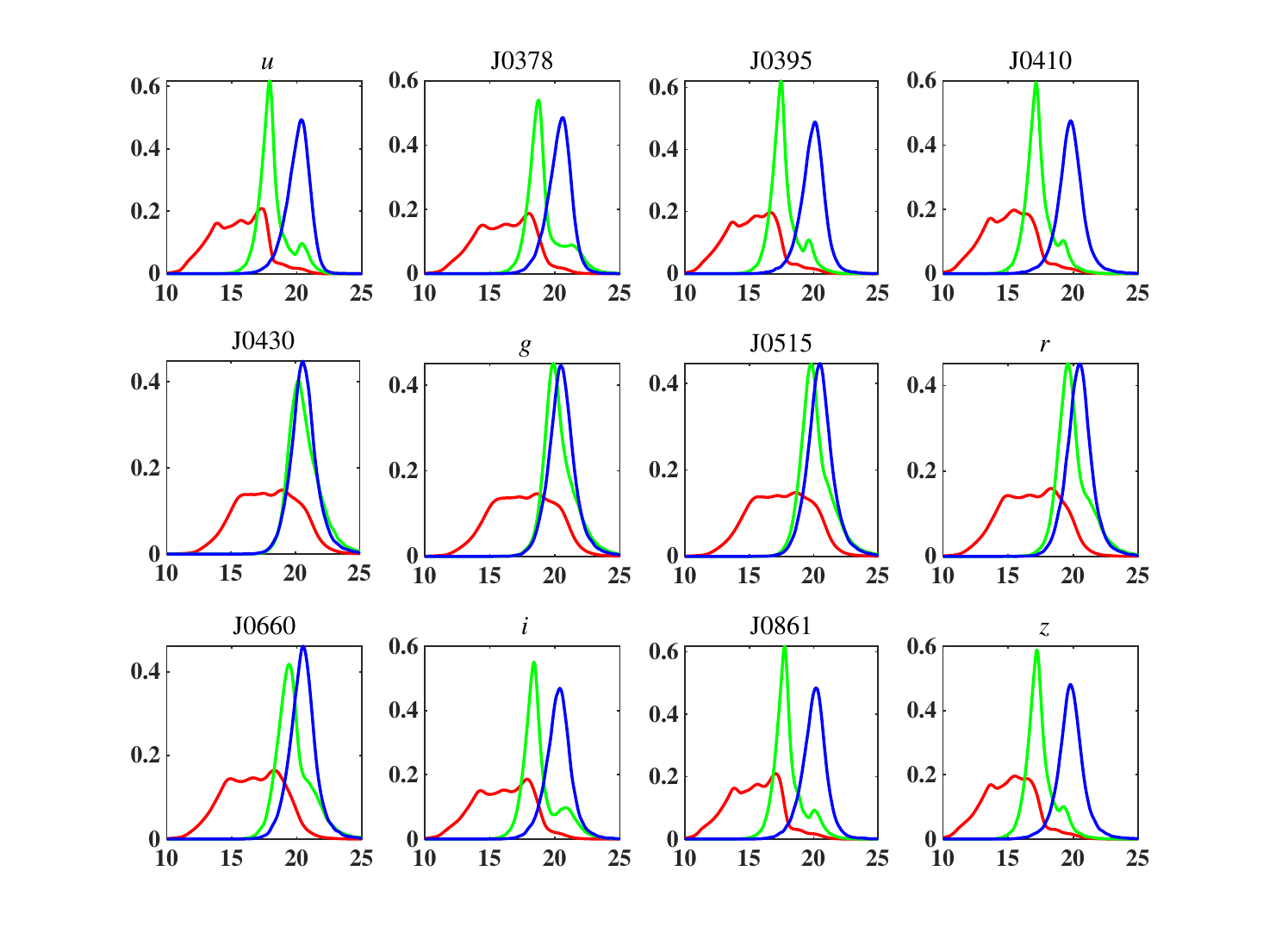}
\caption{Magnitude distribution for sample objects. The axes and line colors are the same as the interpolations.}
\end{figure*}

\begin{figure*}
    \centering 
    \includegraphics[width=0.45\textwidth]{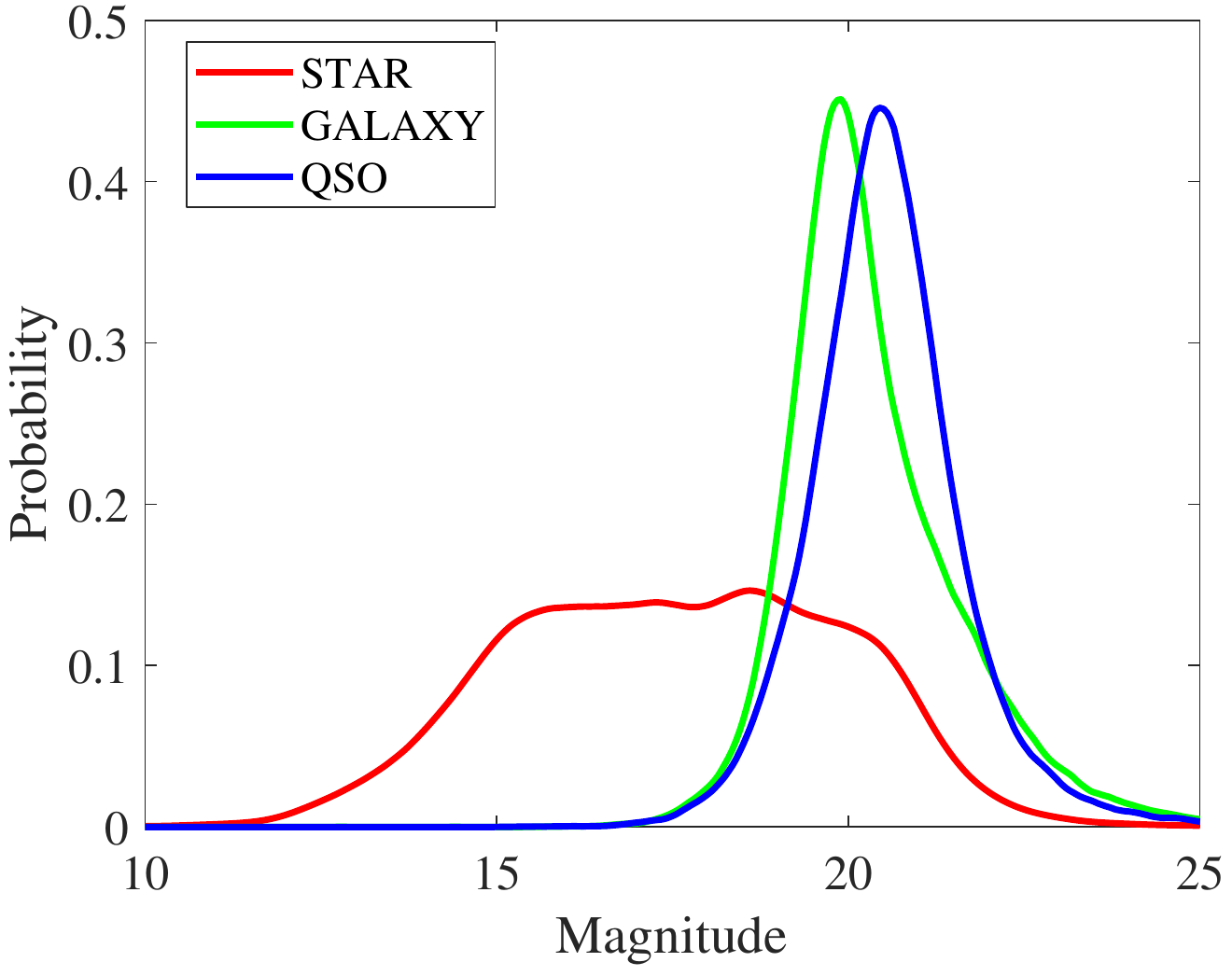}
    \includegraphics[width=0.45\textwidth]{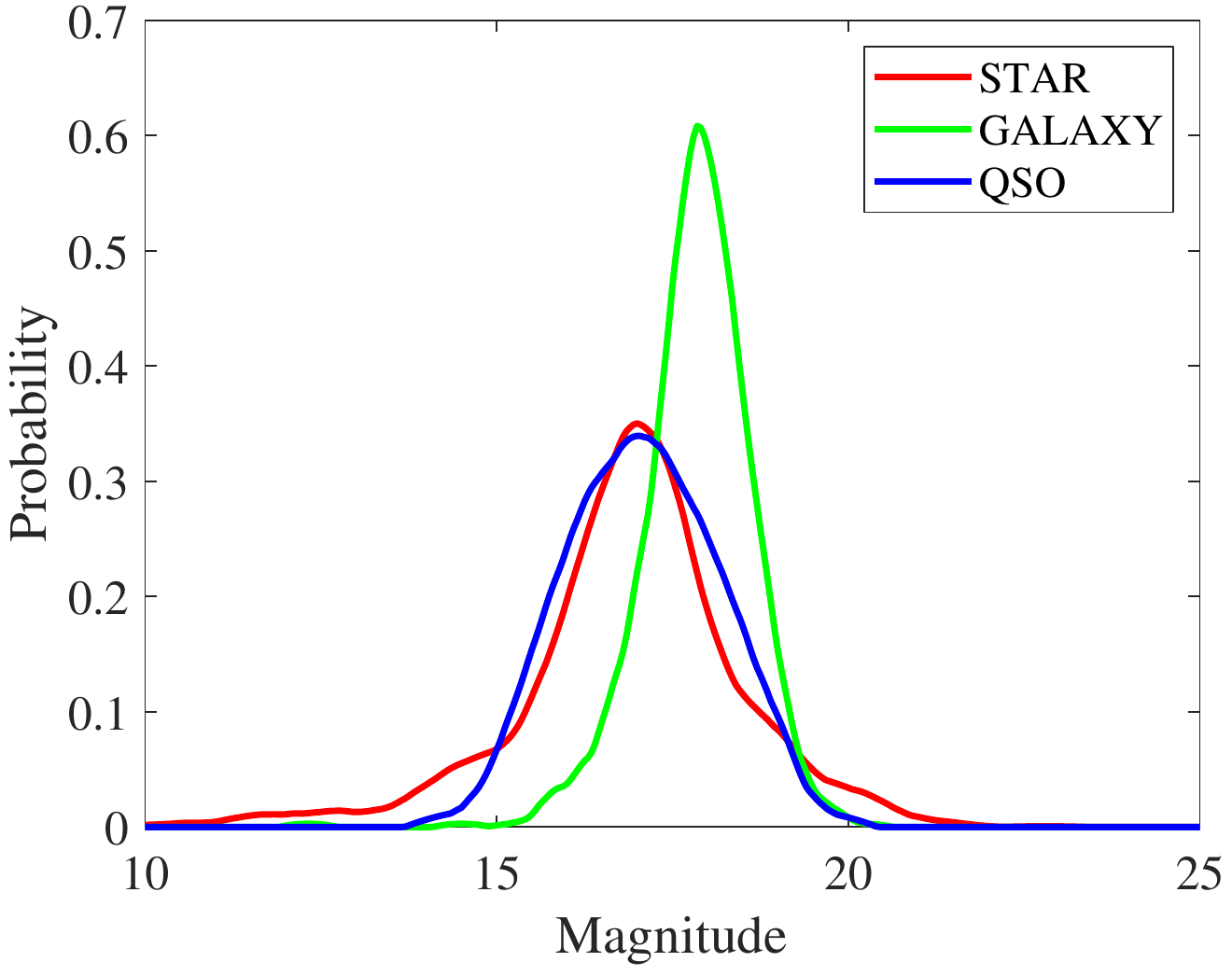}
    \includegraphics[width=0.45\textwidth]{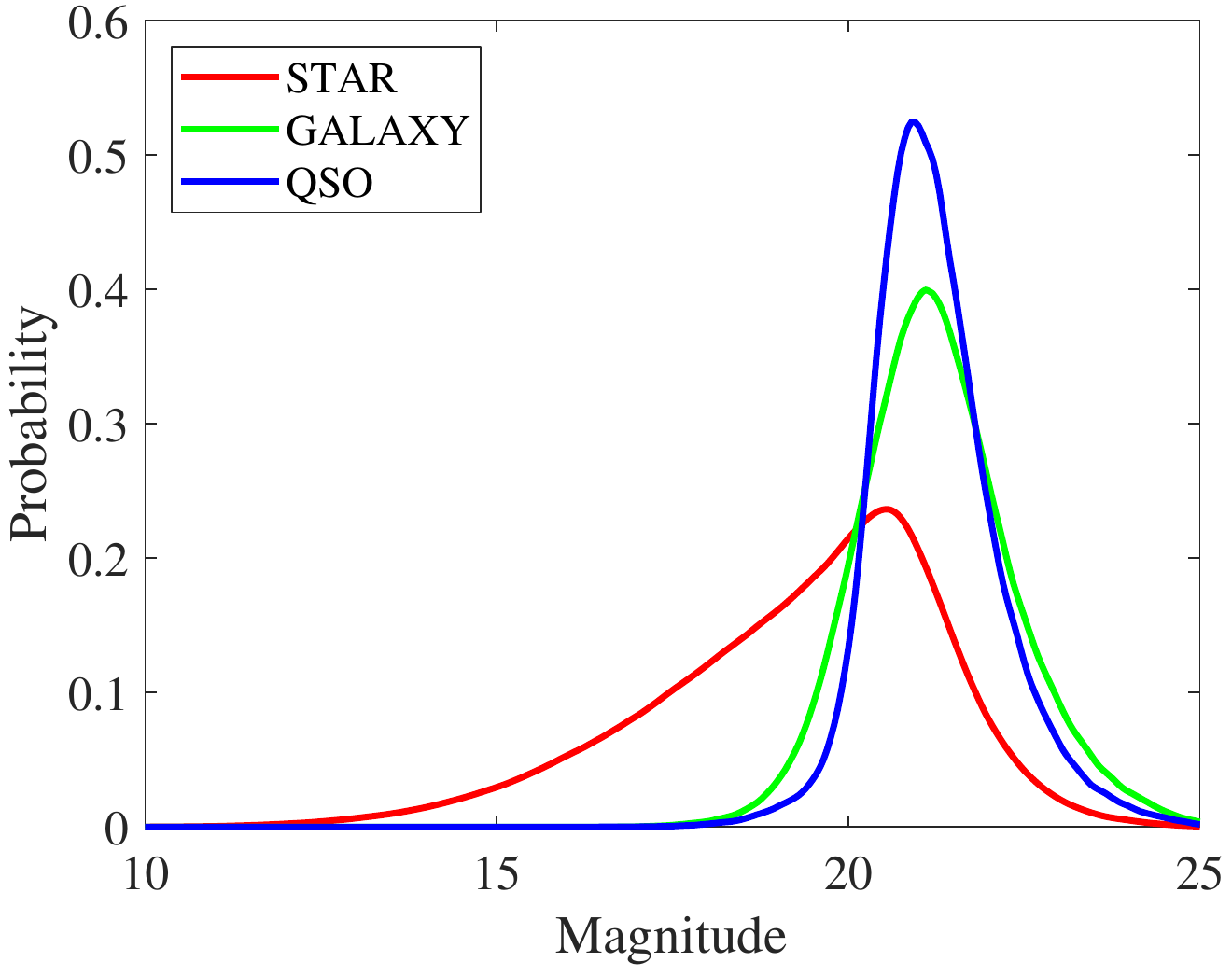}
    \includegraphics[width=0.45\textwidth]{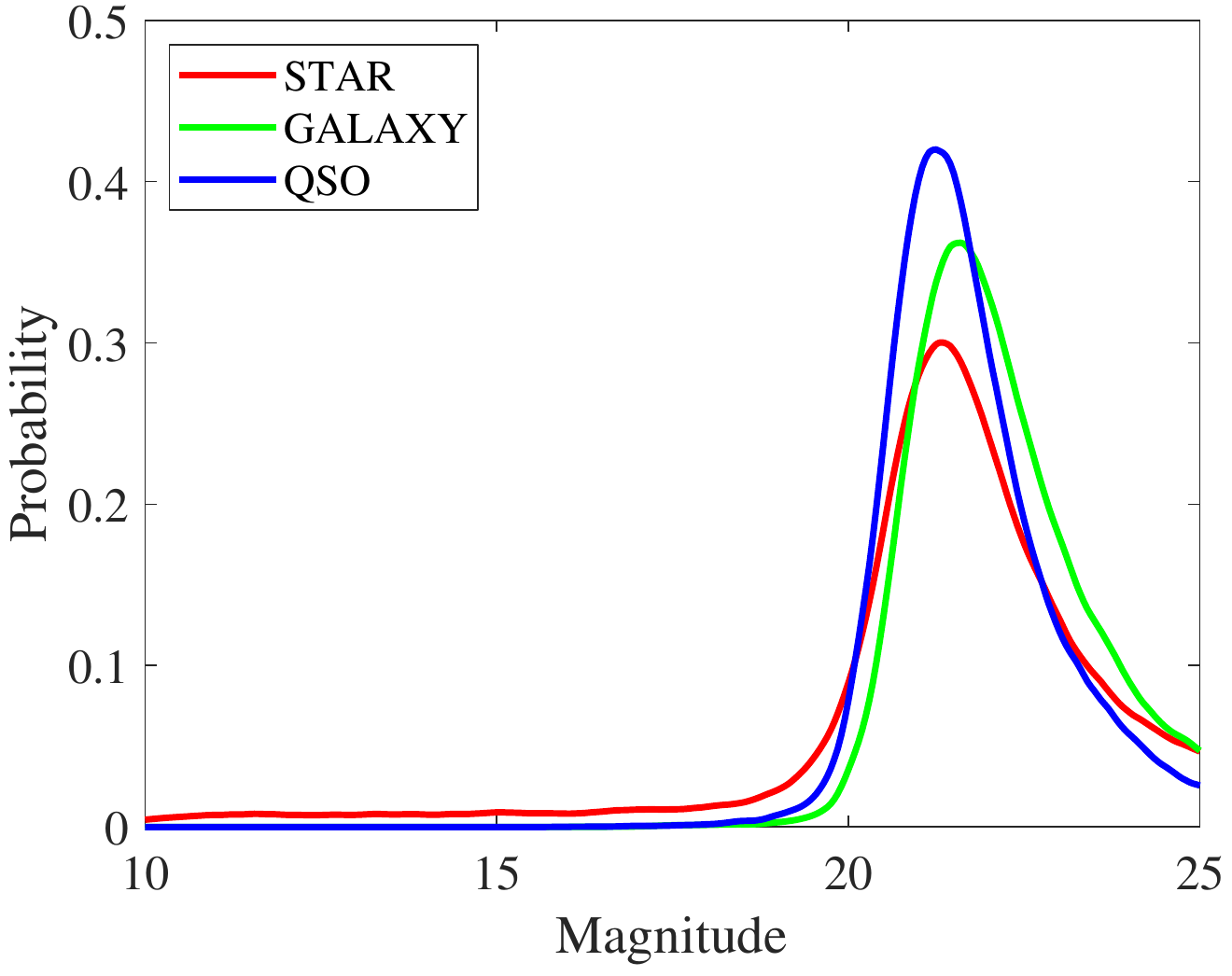}
    \includegraphics[width=0.45\textwidth]{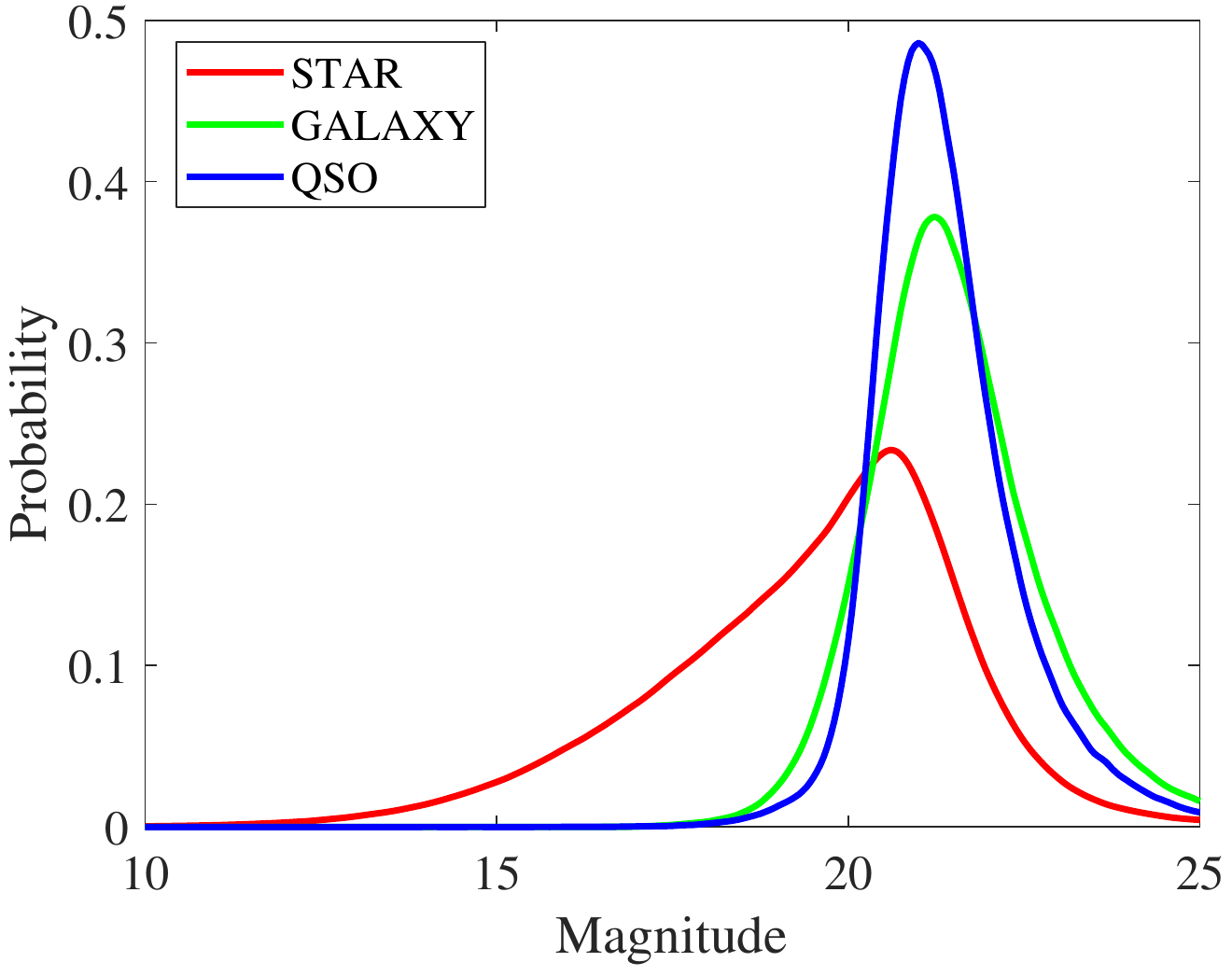}
    \caption{Magnitude distributions in the g-band of our training. The top panels show the sample set and blind test set from left to right. The middle panels show the interpolation and extrapolation objects, and the bottom panel presents the J-PLUS catalog distribution.}
\end{figure*}

\begin{figure*}
    \centering 
    \includegraphics[width=0.45\textwidth]{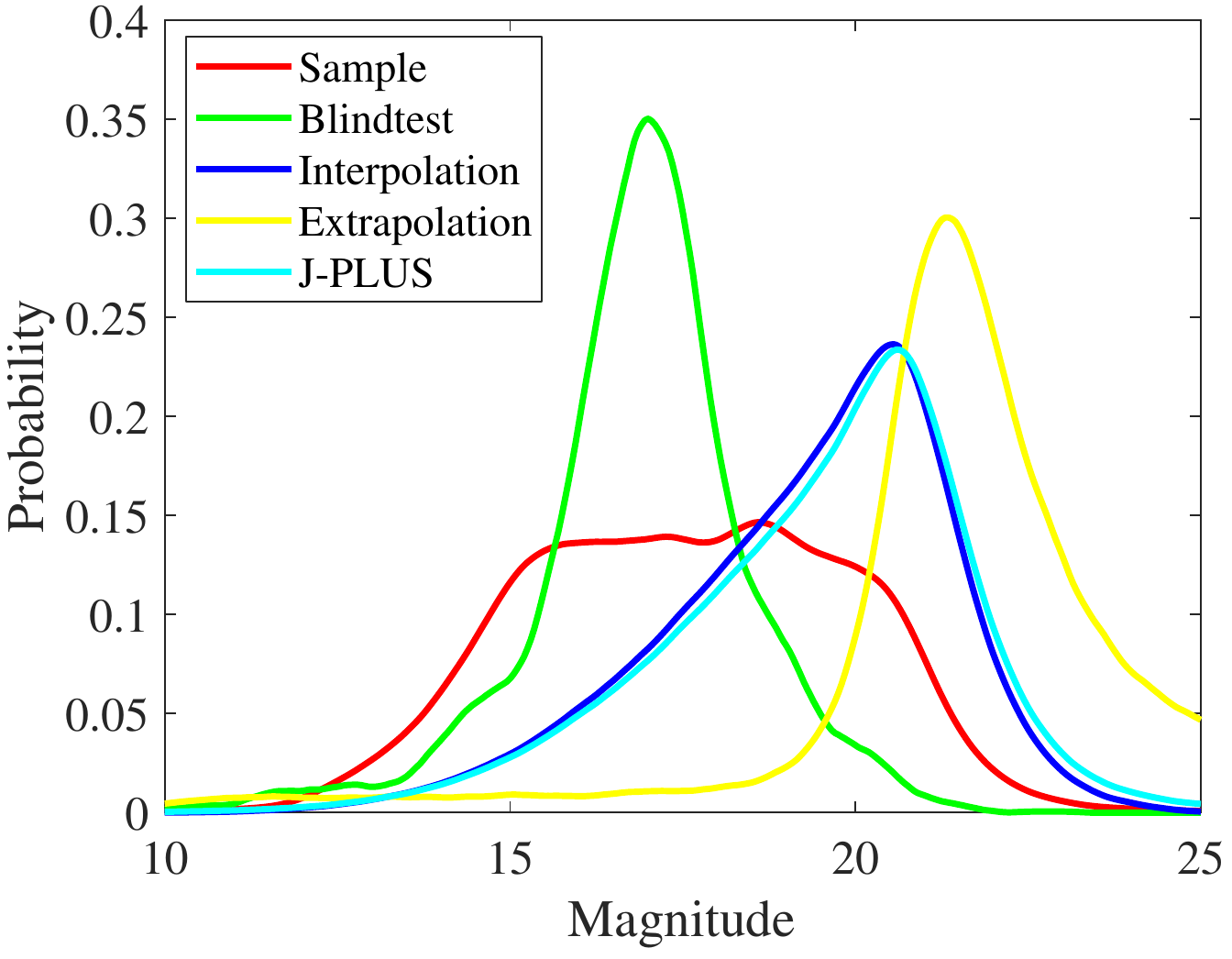}
    \includegraphics[width=0.45\textwidth]{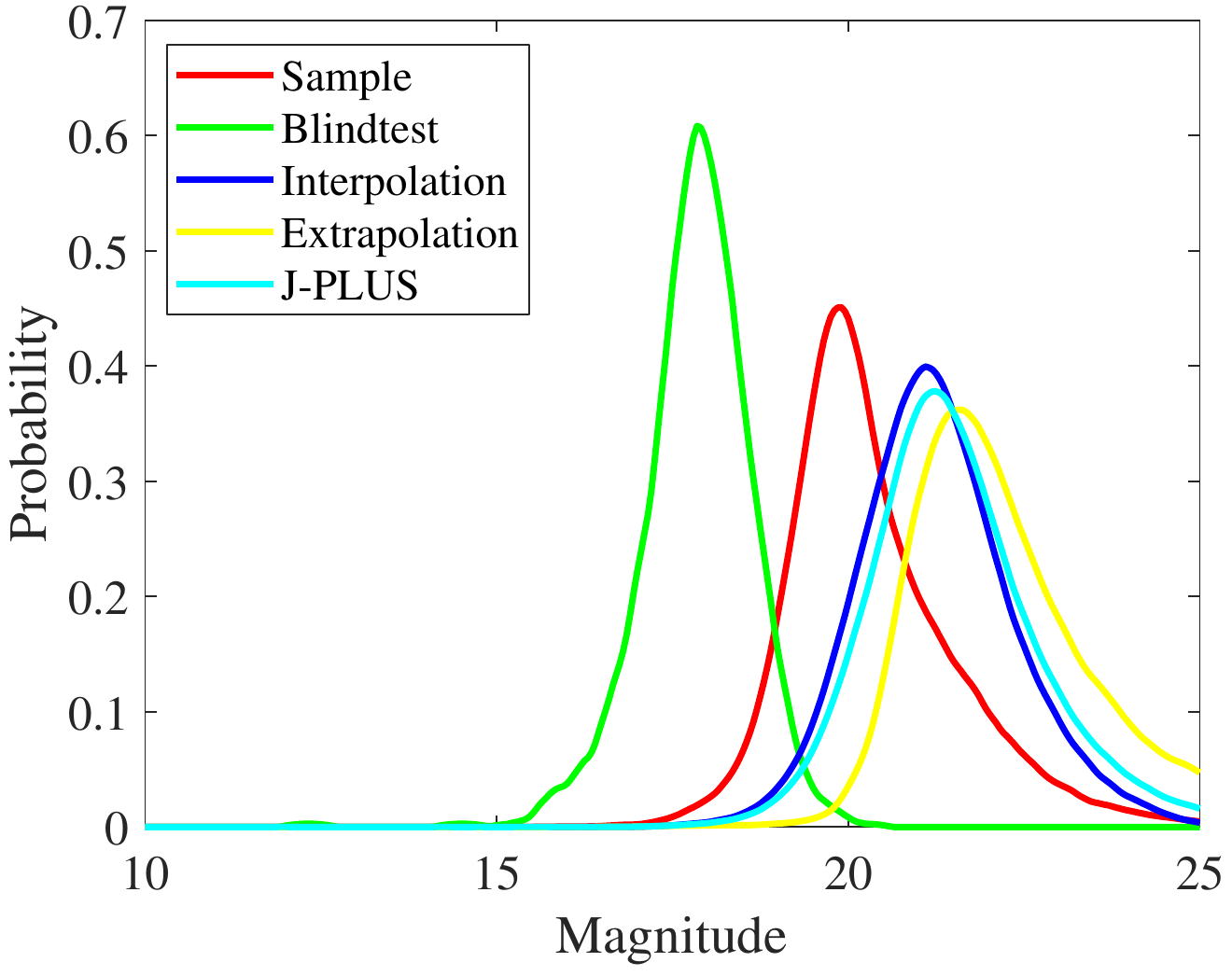}
    \includegraphics[width=0.45\textwidth]{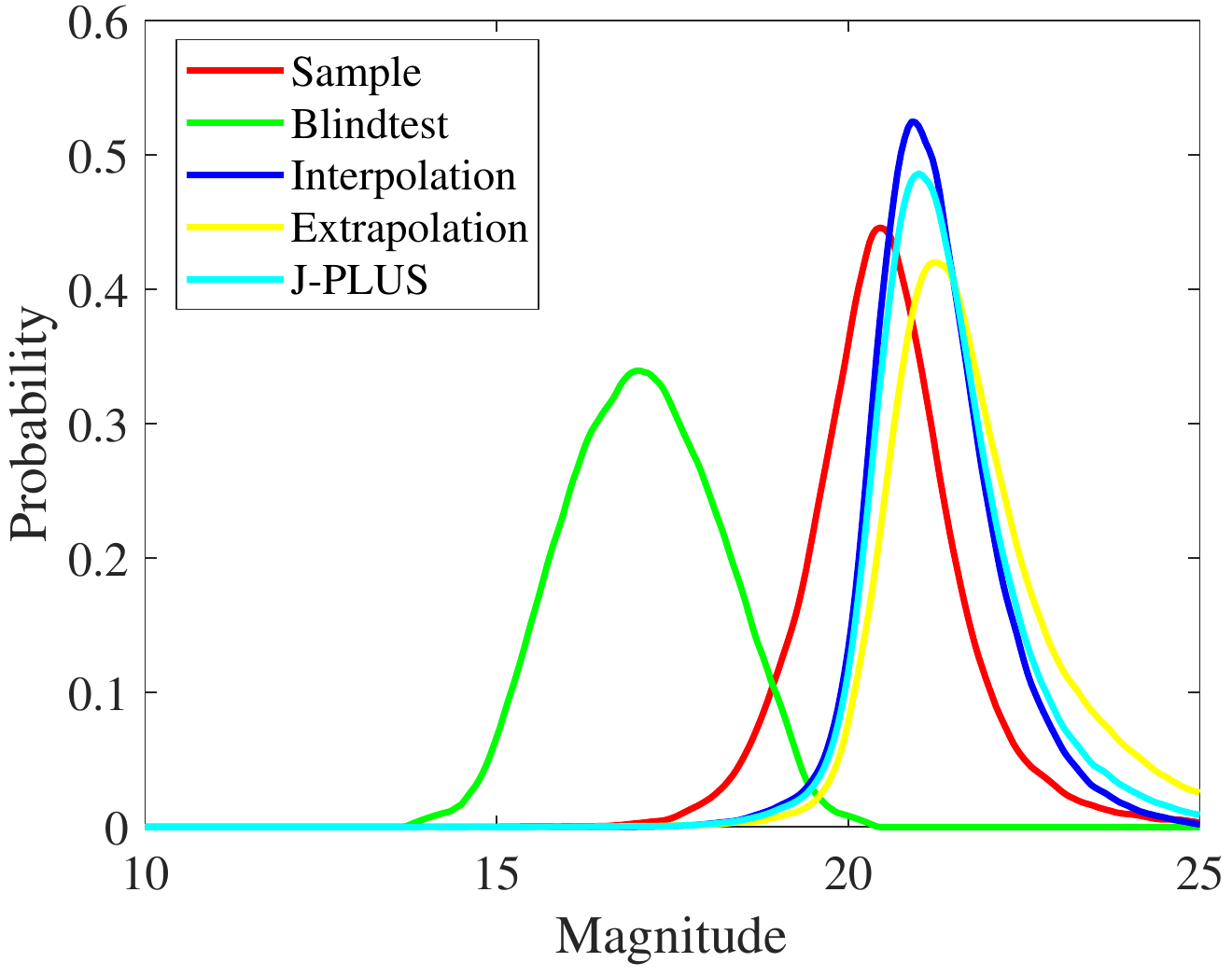}
    \caption{Magnitude distributions in the g-band of each label. The three panels show the label STAR, GALAXY, and QSO from the top left to the bottom, respectively.}
\end{figure*}

\FloatBarrier

\section{Sample of our training}
\label{appc}
We present the training sample in Table \ref{sampletab}, and the subclasses of STARs are included. The overlap of the stars in the sample is presented in Table \ref{sampleols}. The overlap of galaxies between SDSS DR16 and LAMOST DR7 is 9,871. The QSO overlaps between each catalog are 4,593 for VV13 and SDSS DR16, 1,339 for VV13 and LAMOST DR7, and 3,802 for SDSS DR16 and LAMOST DR7. Though the overlapping is so enormous for the LAMOST catalogs, there is still one independent object in the A-, F-, G-, and K-type star catalog. This catalog can provide more information. Also, there are 6,147 and 108 independent objects in APOGEE and VV13.

\begin{table*}
\centering 
\caption{Sample set \label{sampletab}}
\begin{tabular}{cccccc}
\hline \hline \noalign{\smallskip}
ID & R.A. & Dec. & class & subclass & catalog \\ 
\noalign{\smallskip}
\hline 
\noalign{\smallskip}
25998-16 & 117.49152 & 39.45784 & STAR & G8 & LAMOST M \\
25998-143 & 117.58813 & 39.46421 & STAR & M3 & LAMOST A- F- G- K- \\
25998-309 & 116.08787 & 39.47680 & STAR & F7 & LAMOST \\
25998-495 & 117.13277 & 39.48703 & QSO & &      SDSS \\
25998-942 & 116.99474 & 39.50345 & GALAXY & & SDSS\\
25998-8981 & 116.47530 & 39.83560 & STAR & G2 & LAMOST M \\
25998-9909 & 116.52973 & 39.82853 & STAR & & APOGEE\\
26036-5884 & 145.42144 & 30.01548 & STAR & A2V & LAMOST A- F- G- K-\\
26025-6501 & 125.43704 & 30.12984 & QSO & & VV13 \\
26025-7265 & 124.35765 & 30.17180 & STAR & & SDSS \\
\hline
\end{tabular}
\tablefoot{The first four columns are the same as in Table \ref{table1}. The "subclass" is labeled from the LAMOST DR7 catalog, indicating the subclass of stars. The blank in the subclass means that the subclass is missing or it is not a star. The column "catalog" shows the origin catalog, where SDSS means SDSS DR16 and LAMOST means LAMOST DR7.}
\end{table*}

\begin{table*}
\centering 
\caption{Sample overlap of STARs \label{sampleols}}
\begin{tabular}{l|ccccc}
\hline \hline \noalign{\smallskip}
Catalog & APOGEE & LAMOST & AFGK & LA & LM \\ 
\noalign{\smallskip}
\hline 
\noalign{\smallskip}
SDSS & 91 & 7,266 & 3,993 & 960 & 333 \\
APOGEE &  & 7,564 & 6,419 & 69 & 294  \\ 
LAMOST &   &   & 212,114 & 5,145 & 2,5604 \\ 
AFGK &   &   &   & 1,408 & 421 \\ 
LA &   &   &   & & 6 \\ 
\hline
\end{tabular}
\tablefoot{The overlap of each catalog in the class STAR. SDSS stands for SDSS DR16, and LAMOST stands for LAMOST DR7. AFGK, LA, and LM are the LAMOST A-, F-, G-, and K-type star, LAMOST A-star, and LAMOST M-star catalog, respectively.  }
\end{table*}

\FloatBarrier

\end{appendix}

\end{document}